\documentclass[english]{article}
\usepackage{newcent}
\usepackage{avant}
\renewcommand{\familydefault}{\rmdefault}
\usepackage[T1]{fontenc}
\usepackage[latin9]{inputenc}
\usepackage[a4paper]{geometry}
\usepackage{color}
\usepackage{array}
\usepackage{wrapfig}
\usepackage{textcomp}
\usepackage{bbding}
\usepackage{mathrsfs}
\usepackage{multirow}
\usepackage{amsmath}
\usepackage{amssymb}
\usepackage{wasysym}
\usepackage{graphicx}
\usepackage{setspace}
\usepackage{esint}
\usepackage[numbers]{natbib}

\makeatletter

\newcommand{\noun}[1]{\textsc{#1}}
\providecommand{\tabularnewline}{\\}

\newcommand{\lyxaddress}[1]{
\par {\raggedright #1
\vspace{1.4em}
\noindent\par}
}

\@ifundefined{date}{}{\date{}}
\usepackage{sublabel} 
\usepackage{setspace} 
\usepackage{amsfonts}
\usepackage{amssymb}
\usepackage{wasysym} 
\usepackage{latexsym} 
\usepackage{graphicx} 
\usepackage{epstopdf}
\usepackage{multirow} 
\usepackage{multicol} 
\usepackage{color}
\usepackage{colortbl}
\usepackage{subfig}
\usepackage{subfloat}
\usepackage{fixltx2e}
\usepackage{fancybox} 
\usepackage{euscript} 
\usepackage{exscale} 
\usepackage{pifont} 
\usepackage{array}
\usepackage{afterpage}
\usepackage{bm}
\usepackage{makeidx}
\makeindex
\usepackage{charter}
\usepackage{everysel}
\usepackage{keyval}
\usepackage{ragged2e}
\usepackage{perpage}
\usepackage{scalefnt}
\usepackage{ifpdf}
\usepackage{arydshln}
\date{}
\definecolor{R}{rgb}{1,0.5,0.5}
\definecolor{G}{rgb}{0.7,1,0.5}
\definecolor{B}{rgb}{0.78,0.88,1}
\definecolor{Y}{rgb}{1,0.7,0}
\definecolor{M}{rgb}{1,0.75,0.95}
\definecolor{GY}{rgb}{0.91,0.98,0.51}
\definecolor{Gold}{rgb}{1,0.6,0}

\AtBeginDocument{

}

\makeatother

\begin{document}

\title{\textsf{\textbf{\huge Adventures Beyond Reductionism.\\The Remarkable
Unfolding of\\Complex Holism: Sustainability Science}}}

\author{{\normalsize A. Sengupta}}

\maketitle

\lyxaddress{\begin{center}
Institute for Complex Holism, Kolkata, INDIA\\\textbf{E-mail:} osegu@iitk.ac.in
\par\end{center}}
\begin{abstract}
\noindent \textquotedblleft If it could be demonstrated that any complex
organ existed, which could not possibly have been formed by numerous
successive, slight modifications, my theory would absolutely break
down\textquotedblright . Can Darwinian random mutations and selection
generate biological complexity and holism? In this paper we argue
that the \textquotedblleft wonderful but not enough\textquotedblright{}
tools of linear reductionism cannot lead to chaos and hence to complexity
and holism, but with ChaNoXity this seems indeed plausible, even likely.
Based on the Pump-Engine realism of mutually interacting supply and
demand --- demand institutes supply that fuels demand --- we demonstrate
that the \textquotedblleft supply\textquotedblright{} of symmetry
breaking Darwinian genetic variation, in direct conflict with the
symmetry inducing \textquotedblleft demand\textquotedblright{} of
natural selection, defines the antagonistic arrows of the real and
negative worlds. Working in this competitively collaborating nonlinear
mode, these opposites generate the homeostasy of holistic life. Protein
folding, mitosis, meiosis, hydrophobicity and other ingredients have
their respective expressions in this paradigm; nucleotide substitution,
gene duplication-divergence, HGT, stress-induced mutations, antibiotic
resistance, Lamarckism would appear to fit in naturally in this complexity
paradigm defined through emergence of novelty and self-organization. 

With obvious departures from mainstream reductionism, this can have
far reaching implications in the Darwinian and nano medicine of genetic
diseases and disorders. 

Our goal is to chart a roadmap of adventure beyond (neo)-Darwinian
reductionism.\end{abstract}
\begin{description}
\item [{Keywords:}] Chanoxity; Reductionism; Darwinian Holism; Self-organization
and Emergence; Demand, Supply, Logistic. 
\end{description}

\section{Introduction: Beyond Reductionism}

Biological systems are complex holistic systems: thermodynamically
open and far-from-equilibrium, self organizing, emergent. The normal
tools of Newtonian analysis structured around linear reductionism
fail to address these issues, just as classical Newtonian mechanics
failed to embrace the microscopic, necessitating the quantum revolution
100 years ago. The ``inadequate'' reductionist tools of linear mathematics,
where a composite whole is diagnosed as a sum of its parts, work so
long as its foundational \textquotedblleft normal\textquotedblright ,
isolated, near-equilibrium --- rather than \textquotedblleft extreme\textquotedblright /\textquotedblleft revolutionary\textquotedblright ,
open, far-from-equilibrium, stressed --- conditions are met. Increasingly,
it is being felt that most of the important manifestations of nature
and life display holistic behaviour which is the philosophy that parts
of a whole cannot exist and be understood except in the context of
the entirety: wholes generate interdependent, interacting, effects
that are qualitatively different from what can be induced by the parts
on their own. Complex self-organizing systems evolve on emergent feedback
mechanisms and processes that \textquotedblleft interact with themselves
and produce themselves from themselves\textquotedblright , they are
\textquotedblleft more than the sum of their parts\textquotedblright .
Holism simply put, is the thesis that parts of a whole cannot exist
and be understood except in their relation to the whole; complex systems
cannot dismantle into their parts without destroying themselves. The
cybernetic system being analysed is involved in a closed loop where
action by the system causes some change in its environment and that
change is fed to the system via feedback information that causes the
system to adapt to these new conditions --- the system's changes affect
its behavior. This ``circular causal'' relationship is necessary
and sufficient for the cybernetic perspective of competitive-collaboration
that forms the basis of life in Nature. 

A remarkable example of the \textit{competitive-collaboration} of
the parts is the open source/free software dialectics, developed essentially
by an independent, dispersed community of individuals. Wikipedia as
an exceptional phenomenon of this collaboration, along with the Linux
operating system, are noteworthy manifestations of the power and reality
of self-organizing emergent systems: the ``dependencies'' of software
packages --- that are depended on by others --- and the resulting
entangled web in its totality comprising the success of the system.
How are these bottom-up community expressions of ``peer-reviewed
science'' --- with bugs, security holes, and deviations from standards
having to pass through peer-review evaluation of the system (author)
in dynamic equilibrium of competitive-collaboration with the reviewing
environment --- able to ``outperform a stupendously rich company
that can afford to employ very smart people and give them all the
resources they need? Here is a posible answer: Complexity. Open source
is a way of building complex things'' \citep{Naughton2006}, not
orchestrated in the main, by any super-intelligence. 

(Neo)Darwinian microevolution defined as change in allele frequencies,
is a two-step, mutually independent linear reductionist process of
any genetic change small or large in a population inherited over several
generations. For an event to be considered evolutionary, changes have
to occur at the genetic level of a population and be passed on from
one generation to the next. This means that the genes, or more precisely,
the alleles in the population change and are passed on through the
phenotypes of the population. The first stage of symmetry breaking,
random, infinitely small, heritable mutations of genetic variations,
is followed by symmetry generating natural selection of fixation of
beneficial changes as its principal motive force: natural selection
acts to preserve and accumulate minor advantageous genetic mutations.
Darwinism is a linearly smooth, gradual, continuous process, bereft
of ``surprises'' and ``unpredictability''. Can the Darwinian paradigm
of variation-selection-retention explain complexity and holism: ``If
it could be demonstrated that any complex organ existed, which could
not possibly have been formed by numerous, succesive slight modifications,
my theory would absolutely break down'' observed Darwin \citep{Darwin1859}.
In reality, symmetry-inducing ``demand'' institutes a symmetry-breaking
``supply'' which in turn fuels the ``demand'' in a feedback, interacting
loop, essential for complexity and holism \citep{Sengupta2010-a,Sengupta2010-b}:
evolutionary pressures act on the whole organism, not on single genes,
and genes can have different effects depending on the other genes
present. ``A gene is never visible to natural selection, and in the
genotype, it is always in the context with other genes, and the interaction
with those other genes make a particular gene either more favorable
or less favorable'' \citep{Mayr2001}. As in economic holism \citep{Sengupta2010-c},
this implies that the product of mutually antagonistic mutation and
selection evolve in time generating complex, emergent structures.
Somatic mutation, sexual genetic recombination, gene flow and horizontal
gene transfer increase variation while natural selection and random
genetic drift decrease available free energy (exergy) that represents
the ``price''/``cost'' of maintaining the bi-directional feedback
mechanism of complexity and ``life''. Biological evolution also
includes macroevolution of all life being connected that can be traced
back to one common ancestor. In this work we are principally concerned
with the foundations of microevolution. 

Nature abhors gradients: according to the Second Law when a system
is displaced from thermodynamic equilibrium, Nature tries to restore
it by destroying the gradient. For large departures, if the system
is unable to return to the old configuration, a new steady equilibrium
state is sought by more efficient management through pattern formation
of emergent phenomena and structures characteristic of complexity.
Symmetry breaking, in particular breaking of equivalences leading
to  partitioning of the space, is Nature's way of introducing patterns,
structures and complexity in an originally structureless and symmetric
system. Symmetry and equivalence-breaking phase transitions is how
Nature emerges holistically via nonlinear competitive-collaboration,
\textit{not} by adopting alternative evolutionary bifurcating routes
linearly. 

What lies beyond reductionism, at the heart of complex holism? The
emergence of complex systems invalidates reductionist approaches in
the understanding of open, far-from-equilibrium, hierarchical systems.
Nonlinear complexity does not, however, violate any of the familiar
reductionist analytic tools, applicable in their respective linear
domains: for neighbouring or contiguous hierarchical levels, a reductionist
approach can be expected to provide meaningful results. Thermodynamic
expansion and long-range gravitational contraction obeying the virial
theorem generate bi-directional positive-negative feedback loops of
negative (absolute) temperatures, specific heat, entropy and distances
in the black hole \citep{Sengupta2010-c}. Its manifestation of gravitational
attraction institutes the immunity of survival against eventual second-law
entropic implosive cold death.

\textit{Darwin's theory of natural selection}\textsf{\textbf{}}%
\footnote{For a spirited dissection of the inadequacies of Darwinism and Neo-Darwinism
see\textsf{\textbf{ }}Carsten Herrmann-Pillath, \textit{The Concept
of Information and the Problem of Holism Vs. Atomism in Biological
and Economic Uses of Universal Darwinism} (August 24, 2007) at SSRN:
http://ssrn.com/abstract=1009787. \textit{Universal Darwinism} refers
to any of several concepts which apply the ideas and theories of Darwinism
beyond their original scope of organic evolution on Earth.%
} states that if there are (i) variation and variety in a given population
in an environment, (ii) differential reproduction\textbf{ }and survival
of individual members of the new generation, and (iii) inheritance
of this variation by the next generation with \textit{random} modifications,
then evolution by natural selection follows with the new generation,
more adapted to the environment, passing on its characteristics to
the next. The central idea is that a species evolves because natural
selection acts on small heritable variations in the members of the
species: those adapting better to their environment tend to leave
more progeny and transmit their characters, while those less able
to adapt leave fewer progeny or die out, so that in the course of
generations there is a progressive tendency in the species to a greater
degree of adaptation. 

The genotype or genome, the genetic constitution in every cell of
the organism, is the storehouse of the genetic blueprint in the DNA.
Phenotype, the characteristics manifested by an organism, is the end
product created by the organism that emerges through execution of
the instructions in the genotype and is subjected to the battle for
survival; the genotype, however, is the storehouse of accumulated
evolutionary benefits of succeeding generations. The phenotypes compete,
and the fittest among them have a higher chance of exchanging genes
among themselves. 

\textit{Neo-Darwinism}, the modern version of Darwinism, is a synthesis
the means of transmittal of genetic information from one generation
to the next responsible for variations. Neo-Darwinism postulates that
natural selection acts on the heritable genetic variations in alleles
of genes in populations, ultimately caused by variation in the order
of bases in the nucleotides in genes. Mutations (especially random
copying errors in DNA) mainly contribute to these genetic variations,
the raw material for natural selection. Since genetic characteristics
are not entirely identical among individuals in a population, genes
of individuals with characteristics that enable them to reproduce
successfully tend to survive at the expense of genes that tend to
fail. This feature of \textit{natural selection at the gene level}
with consequences at the organism or \textit{phenotype level}, is
not a random process. Gene flow --- the movement of genes from one
population to another --- is another important contributor to genetic
variation, and sexual recombination of chromosomes leading to independent
assortment of new gene combinations into a population is a third source
of genetic variation.

\section{Nonlocality, Entanglement and Holism. The Transactional Interpretation
of Quantum Mechanics\textsf{ }\textmd{\citep{Cramer1986-88}}}

\begin{spacing}{0.9}
\noindent \begin{flushright}
\textsf{\textsl{\small \FiveStarOpenCircled{} }}\textsl{An experiment
is an active intervention into the course of Nature. We set up this
or that experiment to see how Nature reacts. If from such a description
we can further distill a model of a free-standing ``reality'' independent
of our interventions then so much the better. Classical physics is
the ultimate example of such a model. However, there is no logical
necessity for a realistic worldview to always be obtainable. If the
world is such that we can never identify a reality independent of
our experimental activity, then we must be prepared for that, too.}\textsf{\textsl{\small{}
}}\textsc{\hfill{}}\textsf{\textbf{\small Fuchs and Peres}}\textsc{
\citep{Fuchs2000}}
\par\end{flushright}
\end{spacing}

\noindent \begin{flushright}
\textsf{\textsl{\small \FiveStarOpenCircled{}}}\textsl{\small{} }\textsl{It
has been suggested that quantum phenomena exhibit a characteristic
holism or nonseparability, and that this distinguishes quantum from
classical physics. The puzzling statistics that arise from measurements
on entangled quantum systems demonstrate, or are explicable in terms
of, holism or nonseparability rather than any problematic action at
a distance.}\textsc{\hfill{}}\textsf{\textbf{\small Stanford Encyclopedia
of Philosophy}}
\par\end{flushright}{\small \par}

\bigskip{}

\noindent In the stressed, far-from-equilibrium, ``revolutionary''
world of today, the inadequate tools of normal science appear to have
indeed run their course. As it had in the classical-quantum transition
some 100 years back. Today, however, the hazards are far greater,
the reductionism-holism revolution being more fundamental, with reductionism
possibly a first-order linear representation of holism. We are clearly
``horribly stuck'' in almost all facets of human endeavour --- cultural
diversity, social inclusion, economic justifiability, political balance
--- with little understanding of what might possibly be behind this
monumental betrayal. 

The single most distinguishing feature of quantum from classical is
the notion of \textit{nonlocality}. Quantum nonlocality embodies the
paradox of quantum entanglement in which measurements on spatially
separated quantum systems instantaneously influence each other violating
local realism, the philosophy that changes in one physical system
can have no immediate effect on another spatially separated system.
This ``local realistic'' view of nature asserts that events separated
in time and space can be correlated at most through speed-of-light
contact --- no influence can travel faster than this maximum. Quantum
nonlocality implies that these foundations of classical Newtonian
physics are not inviolable, that there is a principle of holistic,
faster-than-light, interconnectedness across spacelike or \textit{negative
timelike }intervals. 

The mathematical formalism of quantum mechanics has not faced any
serious challenge since its inception, although its interpretation
continues to remain shrouded in mystery and dogged by controversy.
Nonlocality, the paradoxical source of this mystery, puzzles rationality.
Various \textit{interpretations} of quantum mechanics have been advanced
to understand this famous EPR paradox of which the \textit{Copenhagen
Interpretation}  is admittedly the most well-known. According to this
doctrine, $\Psi(r,t)$ is a mathematical representation of ``our
knowledge of the system'' manifested through the measurement 
\begin{equation}
\left\langle O\right\rangle =\int\Psi O\Psi^{*}d\tau.\label{eq: Copenhagen}
\end{equation}

\noindent This is the only admissible exposition of the physical significance
of the retarted solutions $\Psi(r,t)\simeq e^{i(kr-\omega t)}$ of
positive energy of the Schrodinger Equation to the future, $\Psi^{*}(r,t)\simeq e^{-i(kr-\omega t)}$
being the advanced solutions of negative energy to the past, of the
complex conjugate time-reversed equation. Copenhagen interpretation
admits no other significance to the state vector $\left|\Psi\right\rangle $;
specifically contrary to normal usage \textit{it is not a physically
functional entity of space and time,} it is rather an encoded mathematical
message of our knowledge of nature. In the \textit{collapse} of $\left|\Psi\right\rangle $
following a measurement process, the implied change must occur simultaneously
at all locations described by the state vector at that instant: a
physical wave would necessitate instantaneous transmission if the
``knowledge'' alternative were to be abandoned. 

Decoherence \citep{Schlosshauer2004} the mechanism by which open
quantum systems interact with their environment leading to spontaneous
suppression of interference and appearance of classicality --- utilized
principally to explain the measurement problem involving transition
from the quantum world of superpositions to the definiteness of the
classical objectivity --- partial tracing over the environment of
the total density operator produces an ``environment selected''
basis in which the reduced density is diagonal. This irreversible
decay of the off-diagonal terms is the basis of decoherence that effectively
bypasses ``collapse'' of the state on measurement to one of its
eigenstates. This ``derivation of the classical world from quantum-mechanical
principles'' however only succeeds in bypassing the real issue because
the non-diagonal terms are specifically responsible for heterozygosity
and non-locality; this is to be compared with nonlinearly-induced
emergence of complex patterns and structures through the multifunctional
graphical convergence route \citep{Sengupta2003}. Multiplicities
inherent in this mode, liberated from the strictures of linear superposition
and reductionism, allow interpretation of objectivity and definiteness
as in classical probabilistic systems through a judicious application
of the axiom of choice: to define a choice function is to conduct
an experiment, \citep{Sengupta2010-b}.

The above relationship of $\Psi$ and $\Psi^{*}$ leads to the of
the \textit{Transactional Interpretation} \citep{Cramer1986-88} in
which the retarded, physical \textit{offer wave} from an emitter elicits
an advanced, physical \textit{confirmation wave} from the absorber,
interacting with each other to complete the ``handshake'' \textit{transaction}
of an explicitly nonlocal character. The future absorber influencing
the past emitter focuses on the \textit{compound of ``knowledge''
and ``ignorance''.} According to Cramer\citep{Cramer1986-88} ``The
root of the (non-transactional) inconsistencies lies in the implicit
assumption of Copenhagen interpretation that the state vector collapse
occurs at a particular instant at which a particular measurement is
made and 'knowledge' is gained, that before this instant the state
vector is in its full uncollapsed state, and that there can be a well-defined
'before' and 'after' in the collapse description. In the transactional
interpretation the collapse of the development of the transaction,
is atemporal and thus avoids the contradictions and inconsistencies
implicit in any time-localized state-vector collapse.'' 

The Copenhagen interpretation (\ref{eq: Copenhagen}) regards quantum
mechanics to be intrinsically about awareness, observations, and measurements
emanating from a unitary evolution of the Schrodinger equation with
little on what it is ontologically or what infact it seeks to describe.
According to this philosophy, the wavefunction is simply an auxiliary
mathematical tool devoid of any physical significance, whose only
import lies in its ability to generate the probabilities $\Psi^{*}\Psi$
compactly, representing our knowledge of the preparation and subsequent
evolution of a physical system; only experimental results lie in the
perview of physical theories and ontological questions are invalid.
The $\left|\Psi\right\rangle $ function has only a ``symbolic''
significance in associating expectation values with dynamic variables
and does not represent anything real, the imaginary component in the
state variable forbidding a pictorial represention of the real world.
The dynamics of the Schrodinger equation describes how the observer's
knowledge of the system changes as a function of time. 

Quantum nonlocality is technically expressed in terms of\textsf{\textbf{
}}\textit{entanglement}\textsf{\textbf{.}} Any 
\[
\left|\Psi\right\rangle _{SE}=\sum_{i,j}\alpha_{ij}\left|\phi_{i}\right\rangle \otimes\left|\psi_{j}\right\rangle 
\]
in the tensor product space $\mathcal{H}_{S}\otimes\mathcal{H}_{E}$
that cannot be factored as a product of its component parts $\left\{ \left|\phi_{i}\right\rangle \right\} \in\mathcal{H}_{S}$
and $\left\{ \left|\psi_{j}\right\rangle \right\} \in\mathcal{H}_{E}$,
that is $\left|\Psi\right\rangle _{SE}\ne\left|\phi\right\rangle \otimes\left|\psi\right\rangle $,
is said to be \textbf{\textit{entangled }}(nonlocal); $\left|\Psi\right\rangle _{SE}$
is \textbf{\textit{unentangled}}\textbf{ }(separable) if it is factorisable
into the components. Thus, choose orthonormal bases $\{\left|\uparrow\right\rangle _{S},\left|\downarrow\right\rangle _{S}\}$
and $\{\left|\uparrow\right\rangle _{E},\left|\downarrow\right\rangle _{E}\}$
in $\mathcal{H}_{S}$ and $\mathcal{H}_{E}$ so that $\mathcal{H}_{SE}$
is spanned by the vectors $\left|\uparrow\right\rangle _{S}\left|\uparrow\right\rangle _{E}$,
$\left|\uparrow\right\rangle _{S}\left|\downarrow\right\rangle _{E}$,
$\left|\downarrow\right\rangle _{S}\left|\uparrow\right\rangle _{E}$,
and $\left|\downarrow\right\rangle _{S}\left|\downarrow\right\rangle _{E}$.
Then for the qubits %
\footnote{Unlike a classical bit\emph{ }which must be either of the two possible
values ``on'' $\left|\uparrow\right\rangle $ or ``off'' $\left|\downarrow\right\rangle $,
the \textit{qu}(antum)\textit{bit} can be either $\left|\uparrow\right\rangle $,
or $\left|\downarrow\right\rangle $, or a superposition $\alpha\left|\uparrow\right\rangle +\beta\left|\downarrow\right\rangle $
$\left(\mbox{with }\alpha^{2}+\beta^{2}=1\right)$ of both in the
two-dimensional Hilbert space spanned by $\left|\uparrow\right\rangle $
and $\left|\downarrow\right\rangle $. %
} $\left|\Phi\right\rangle _{S}=\sigma_{1}\left|\uparrow\right\rangle _{S}+\sigma_{2}\left|\downarrow\right\rangle _{S}$
and $\left|\Upsilon\right\rangle _{E}=\varepsilon_{1}\left|\uparrow\right\rangle _{E}+\varepsilon_{2}\left|\downarrow\right\rangle _{E}$
\sublabon{equation}
\begin{eqnarray}
\left|\Psi\right\rangle _{SE} & = & \sigma_{1}\varepsilon_{1}\left|\uparrow\uparrow\right\rangle +\sigma_{1}\varepsilon_{2}\left|\uparrow\downarrow\right\rangle +\sigma_{2}\varepsilon_{1}\left|\downarrow\uparrow\right\rangle +\sigma_{2}\varepsilon_{2}\left|\downarrow\downarrow\right\rangle \nonumber \\
 & = & \left|\Phi\right\rangle _{S}\otimes\left|\Upsilon\right\rangle _{E}\label{eq: NonLocal(a)}
\end{eqnarray}
is a separable state, whereas 
\begin{eqnarray}
\left|\Phi_{\pm}\right\rangle _{SE} & = & \alpha_{1}\left|\uparrow\uparrow\right\rangle \pm\alpha_{2}\left|\downarrow\downarrow\right\rangle \ne\left|\Phi\right\rangle _{S}\otimes\left|\Upsilon\right\rangle _{E}\label{eq: Bell(b)}\\
\left|\Psi_{\pm}\right\rangle _{SE} & = & \beta_{1}\left|\uparrow\downarrow\right\rangle \pm\beta_{2}\left|\downarrow\uparrow\right\rangle \ne\left|\Phi\right\rangle _{S}\otimes\left|\Upsilon\right\rangle _{E}\label{eq: Bell(c)}
\end{eqnarray}
\sublaboff{equation}are nonseparable, entangled, Bell states.%
\footnote{A 2-qubit state is separable iff $ad=bc$ for $\left|\Psi\right\rangle =a\left|\uparrow\uparrow\right\rangle +b\left|\uparrow\downarrow\right\rangle +c\left|\downarrow\uparrow\right\rangle +d\left|\downarrow\downarrow\right\rangle $.%
} 

An entangled state does not define vectors in the individual factor
spaces $\mathcal{H}_{S}$ and $\mathcal{H}_{E}$ unless the state
is actually unentangled. For physically separated $S$ and $E$, a
measurement outcome of $\left|\uparrow\right\rangle $ on $S$ implies
that any subsequent measurement on $E$ in the same basis will always
yield $\left|\uparrow\right\rangle $. If $\left|\downarrow\right\rangle $
occurs in $S$, then $E$ will be guaranteed to return $\left|\downarrow\right\rangle $;
hence system $\left|E\right\rangle $ has been altered by local random
operations on $\left|S\right\rangle $. This non-local puzzle of entangled
quantum states --- the orthodox Copenhagen doctrine maintains that
neither of the particles possess any definite position or momentum
before they are measured --- is resolved by bestowing quantum mechanics
with non-local properties determined by Bell's inequality. In this
sense entanglements induced by \textit{iterations} of nonlinear \textit{separable}
systems like the generalized logistic qubit of the product of supply
$(\downarrow)$ and demand $(\uparrow)$ functions (see Sec. \ref{sub: General Logistic}),
are destined to be far more complex than that of the partial Bell
states Eqs. (\ref{eq: Bell(b)}, \textit{c}). 

For an expository initiation to the exciting new world of \textit{quantum
biology, }see for example \citep{Ball2011}.

\section{ChaNoXity: Pump-Engine Realism of the Participatory Universe\textsf{
}\textmd{{[}\citealp{Sengupta2010-c}{]}}\vspace{-0.15in}
}

\noindent \begin{flushright}
\textsl{\FiveStarOpenCircled{} Let us see how we humans use the second
law for our purposes. Whenever we run any kind of engine, we're using
the second law for our benefit: Taking energy inside of substances
that tend to spread out, but can't because of (the activation energy)
$E_{a}$, giving it the necessary energy, having the diffusing energy
in the form of hot expanding gases push a piston that turns crankshafts,
gears and wheels, with the exhaust gases, still fairly hot, but no
longer available for any more piston-pushing in this engine.\\Did
we beat the second law? No way. But by using the second law --- taking
the energy from spontaneous ``downhill'' reactions and transferring
much of it to force a nonspontaneous process to go ``uphill'' energy-wise
and make something --- we got what we wanted. $\cdots$ \\Living
creatures are essentially energy processing systems that cannot function
unless a multitude of ``molecular machines'', biochemical cycles,
operate synchronically in using energy to oppose second law predictions.
All of the thousands of biochemical systems that run our bodies are
maintained and regulated by feedback subsystems, many composed of
complex substances. Most of the compounds in the feedback systems
are also synthesized internally by thermodynamically nonspontaneous
reactions, effected by utilizing energy ultimately transferred from
the metabolism of food. When these feedback subsystems fail --- due
to inadequate energy inflow, malfunction from critical errors in synthesis,
the presence of toxins or competing agents such as bacteria or viruses
--- dysfunction, illness, or death results: energy can no longer be
processed to carry out the many reactions we need for life that are
contrary to the direction predicted by the second law.\hfill{}}\textsf{\textbf{\small Lambert}}
\citep{Lambert2007}
\par\end{flushright}

\subsection{The Logistic Nonlinear Qubit: competitive-collaboration of Supply
and Demand\textsf{ }}

\noindent The logistic difference equation $x_{t+1}=\lambda x_{t}(1-x_{t})$
representing nonlinear interaction between individualistic supply
$x$ and collaborative demand $1-x$ is the starting point in our
study of complexity and holism, \citep{Sengupta2006,Sengupta2010-c,Sengupta2010-a,Sengupta2010-b},
with $\lambda$ is an environment parameter. In the context of population
dynamics $\lambda$ referring to ``an intrinsic reproductive rate
of the average fecundity of an individual'' \citep{Camazine2001}
corresponds to the \textit{mean} \textit{fitness }of the organism;
here fecundity is the number of gametes and progeny zygotes produced
by the parents surviving to adulthood. The gametic lifecycle of sexually
reproduced organisms with differential survival and reproduction of
genotypes leading to selection can then be summarized as follows \citep{Hamilton2009} 

\noindent \sublabon{table}
\begin{table}[!tbh]
\begin{centering}
\begin{tabular}{c|c|c|c|c|}
\cline{2-3} 
 & \multicolumn{2}{c|}{\textcolor{red}{\small }%
\begin{tabular}{ccccccc}
S & T & A & G & E &  & III\tabularnewline
\end{tabular}} & \multicolumn{1}{c}{} & \multicolumn{1}{c}{}\tabularnewline
\cline{2-4} 
 & \textcolor{red}{\small ``Bottle-neck'' \citep{Dawkins2006}} & \multicolumn{2}{c|}{\textcolor{blue}{\small Mitotic}} & \multicolumn{1}{c}{}\tabularnewline
\cline{5-5} 
\multirow{1}{*}{} & {\small \cellcolor{R}}\textbf{ ZYGOTE} & %
\begin{tabular}{c}
\textcolor{blue}{\small $t+1\left\Vert \right.$Emergence:}\tabularnewline
\textbf{\textcolor{blue}{\small FETUS (Uterus)}}\tabularnewline
\end{tabular} & \multirow{2}{*}{%
\begin{tabular}{c}
\textcolor{blue}{\small Self-}\tabularnewline
\textcolor{blue}{\small organization:}\tabularnewline
\multirow{1}{*}{\textbf{\textcolor{blue}{\small ADULT}}}\tabularnewline
\end{tabular}} & %
\begin{tabular}{c}
\tabularnewline
S\tabularnewline
\end{tabular}\tabularnewline
\cline{1-3} 
\multicolumn{1}{|c|}{%
\begin{tabular}{c}
S\tabularnewline
T\tabularnewline
\end{tabular}} & \textcolor{green}{\small Fertilization $\left\Vert \right.t$} & $\overset{\textrm{Viability}}{\longrightarrow}$ &  & %
\begin{tabular}{c}
T\tabularnewline
A\tabularnewline
\end{tabular}\tabularnewline
\cline{2-2} \cline{4-4} 
\multicolumn{1}{|c|}{%
\begin{tabular}{c}
A\tabularnewline
G\tabularnewline
\end{tabular}} & \multicolumn{1}{c}{$\underset{\textrm{{Compatibility}}}{\uparrow}$} & \multicolumn{1}{c}{\textbf{\textsc{Selection:$\,\lambda$}}} & $\overset{\textrm{{Sexual}}}{\underset{\textrm{{Survival}}}{\downarrow}}$ & %
\begin{tabular}{c}
G\tabularnewline
E\tabularnewline
\end{tabular}\tabularnewline
\cline{2-2} \cline{4-4} 
\multicolumn{1}{|c|}{%
\begin{tabular}{c}
E\tabularnewline
II\tabularnewline
\end{tabular}} & \textcolor{green}{\small Sex} & $\underset{\textrm{\textrm{Gametic}}}{\overset{\textrm{Fecundity}}{\longleftarrow}}$ & \multirow{2}{*}{%
\begin{tabular}{c}
\textcolor{blue}{\small Mitosis:}\tabularnewline
\textbf{\textcolor{blue}{\small PARENTS}}\tabularnewline
\end{tabular}} & %
\begin{tabular}{c}
\tabularnewline
IV\tabularnewline
\end{tabular}\tabularnewline
\cline{1-3} 
 & \textbf{\textcolor{green}{GAMETE}} & \textcolor{green}{\small Meiosis: Gonads} &  & \tabularnewline
\cline{2-5} 
 & \multicolumn{2}{c|}{\textcolor{red}{\small }%
\begin{tabular}{ccccccc}
S & T & A & G & E &  & I\tabularnewline
\end{tabular}} & \multicolumn{1}{c}{} & \multicolumn{1}{c}{}\tabularnewline
\cline{2-3} 
\end{tabular}
\par\end{centering}

\caption{{\small \label{tab: life-cycle}Gametic life-cycle of organisms. The
basic characteristic that distinguishes meiosis from mitosis is the
cross-over of homologous chromosomes resulting in the production of
sperm and egg in the gonads. The cycle can be sequenced into four
steps leading to the notion of an }\textit{\small extended meiosis}{\small{}
represented by Stages I, II and III of meiotic gonads to mitotic emergence,
Sec. \ref{sub: Meiosis-NegW}.}}
\end{table}

\noindent It is asumed that in the absence of limiting factors, $\lambda x$
the population in a succeeding generation, is the positive feedback
that is effectively regulated by the negative feedback of depletions
$(1-x)$: the zygotic transformation $t\mapsto t+1$ corresponds to
the logistic map. 

The adversaries \textit{individualistic supply} and \textit{collectivistic
demand} collaborate nonlinearly to generate life in a win-win game
where no participant wins and none lose. This self-organizing, emergent
system working on a positive-negative bi-directional $\rightleftarrows$
feedback loop, adjusts itself to the environmental conditions it finds
itself in leading to a homeostasis of ``inevitable limitations, compromises
and trade-offs'' that neither of the participants, working alone
and independently, can achieve. Both adversaries have equal stake
in the complexity of holistic life and participate as equal partners,
each competing with its opponent for its own collaborative good. In
this capital-culture contest \citep{Sengupta2010-c} of self-organization
and emergence, one of the contestants assumes a dissipative passive
(recessive) role of an ``offerer'' that elicits an active ``confirmation''
from a concentrating (dominant) opponent leading to a handshake ``transaction''
of an explicitly non-local character: the cause and effect entangle
in generating a two-phase complex mixture, with the bottlenecked life-cycle
making ``possible the equivalent of going back to the drawing board''
because ``really radical changes can be achieved only by throwing
away the previous design and starting afresh''. The replicating concentrator
uses the offerer as a vehicular ``tool by which it levers itself
into the next generation'' \citep{Dawkins2006}: it needs the vehicle
not only to express itself meaningfully and purposefully despite its
bewildering multitude of interactive interactions, more importantly
the finite and complicated nature of the vehicle acts as a necessary
physical impediment in inhibiting the cancerous growth of uncontrolled
replication, with the environment of an ``extended phenotype'' resolving
the ``paradox of the two ways of looking at life''. Non-reductionist
graphical convergence in an extended multifunctional space is almost
a natural corollary to this unconventional view of the phenotype of
the replicator-interactor antagonistic collaboration embracing the
unconventionality of HGT.

While biologic life is supply regulated depending principally on the
resources available, an individualistic supply economy --- like a
non-democratic form of political governance --- can lead to severe
collective stresses. Working in the two-spin mode of individualism
$(\downarrow)$ and collectivism $(\uparrow)$ represented respectively
by the increasing-positive and decreasing-negative slopes of the logistic
map, the feedback loop of self-organization and emergence achieve
in one nonlinear step the dual functions of inducing the resource
of scarce order in an universal backdrop of pervasive disorder. Thus
if deaths were absent with only the ordering component $x$ available,
there would be no cutoff to the explosive growth of fitness; likewise
with only $1-x$ present the vanishing unfit, steadily eroded through
selection, would lead to the eventual extinction of the population. 

The logistic map through its nonlinear handshake of these interactive
opponents achieves the remarkably wondrous transaction of inclusive
holism.

\subsection{\label{sub: near-to-equilibrium}Alleles and Genotypes. Confrontation
of Opposites: A Generalization}

\noindent As the commentary in the preceding Section indicates quantum
nonlocality, properly interpreted, shares with chanoxity the fundamental
characteristic of collaboration between competing adversaries that
defines holistic entanglement. We have however argued elsewhere \citep{Sengupta2010-c,Sengupta2010-a,Sengupta2010-b},
that quantum nonlocality and complex holism are not the same: quantum
mechanics is a linear theory while complex chanoxity is violently
nonlinear. In the linear setting, multipartite systems in $2^{N}$-dimensional
tensor products $\mathcal{H}_{1}\otimes\cdots\otimes\mathcal{H}_{N}$
of 2-dimensional spin states, correspond to the $2^{N}$ dimensional
space of unstable fixed points of chanoxity \citep{Sengupta2006}.
This formal equivalence while clearly demonstrating how holism emerges
in $2^{N}$-cycle complex systems, also focuses on the significant
differences between complex holism and linear non-locality which can
eventually be traced to the constraints imposed by reductionism, also
Sec. \ref{sub: Fixed-Periodic}. The converged holism of complex ``entanglement''
reflects the fact that the subsystems have combined nonlinearly to
form an emergent, self-organized system that cannot be decoupled without
destroying its structure; quantum nonlocality and the notion of partial
tracing for obtaining properties of individual components from the
whole is not restricted by this defining property of complex holism. 

To establish the perspective of our considerations that follow, consider
the iconic Mendel monohybrid and dihybrid pea-plant experiments as
an example. The monohybrid case is summarized in the Punnett diagram
of Fig. \ref{fig: Punnett-1}: the alleles%
\footnote{\textbf{Alleles} are the alternate forms of the two genes of an organisms
located on chromosomes that control each heritable characteristic
or \textit{trait}, one contributed by the female and the other by
the male. When gametes develop during meiosis --- a process of cell
division that specifically produces the sex cells --- each gamete
receives only one of these alleles. %
} $\mathbf{T}$ and $\mathbf{t}$ formally correspond to the ``spin''
strategies $\left(\downarrow\right)$ and $\left(\uparrow\right)$,
and the genotypes of the entire set of genes in a cell, organism or
individual, correspond to the respective stable states {\large $\bullet$.
}Of the two alleles for every trait, we take the chromosome contributed
by the female $(\mathbf{T})$ in her ova as \textit{dominant} and
the male allele $(\mathbf{t})$ in his sperm as \textit{recessive.
}Joined together in fertilization, there are thus three possible genotypes
for each characteristic trait: $\mathbf{TT}$ homozygous dominant
with $\mathbf{Tt}$/$\mathbf{tT}$ heterozygous of the same phenotype,
and $\mathbf{tt}$ homozygous recessive expressed in a different phenotype.
\sublabon{figure}
\begin{figure}[!tbh]
\begin{centering}
{\renewcommand{\arraystretch}{1.25} %
\begin{tabular}{|c||c|c|}
\hline 
{\large \male }$\left(\uparrow\right)$ allele  & \multirow{2}{*}{$\mathbf{t}(0(\downarrow))$} & \multirow{2}{*}{$\mathbf{t}(1(\uparrow))$}\tabularnewline
\cline{1-1} 
{\large \female }$\left(\downarrow\right)$ allele &  & \tabularnewline
\hline 
\hline 
${\color{green}{\color{red}{\normalcolor \mathbf{T}(0(\downarrow))}}}$ & ${\color{green}\mathbf{{\normalcolor Tt}}}$ & ${\color{green}\mathbf{{\normalcolor Tt}}}$\tabularnewline
\hline 
${\color{green}{\normalcolor \mathbf{T}(1(\uparrow))}}$ & ${\color{green}\mathbf{{\normalcolor Tt}}}$ & ${\color{red}\mathbf{{\normalcolor Tt}}}$\tabularnewline
\hline 
\multicolumn{3}{|c|}{Homozygous tall parents $\overset{\mbox{F}_{1}}{\longrightarrow}$ }\tabularnewline
\multicolumn{3}{|c|}{All heterozygous $\mathbf{Tt}$ tall daughters}\tabularnewline
\hline 
\multicolumn{3}{|c|}{(i)}\tabularnewline
\hline 
\end{tabular}%
\begin{tabular}{c}
$\longrightarrow$\tabularnewline
\end{tabular}%
\begin{tabular}{|c||c|c|}
\hline 
{\large \male }$\left(\uparrow\right)$ allele & \multirow{2}{*}{$\mathbf{T}({\color{red}{\normalcolor 0(\downarrow))}}$} & \multirow{2}{*}{${\normalcolor \mathbf{{\color{green}{\normalcolor t}}}}(1(\uparrow))$}\tabularnewline
\cline{1-1} 
{\large \female }$\left(\downarrow\right)$ allele &  & \tabularnewline
\hline 
\hline 
\textcolor{red}{${\color{green}{\color{red}{\normalcolor \mathbf{T}(0(\downarrow))}}}$} & \cellcolor{B}${\color{red}\mathbf{{\normalcolor TT}}}$ & \cellcolor{B}${\color{green}\mathbf{{\normalcolor Tt}}}$\tabularnewline
\hline 
${\color{green}{\normalcolor \mathbf{t}(1(\uparrow))}}$ & \cellcolor{B}${\color{green}\mathbf{{\normalcolor tT}}}$ & \cellcolor{G}${\color{green}\mathbf{{\normalcolor tt}}}$\tabularnewline
\hline 
\multicolumn{3}{|c|}{Heterozygous tall parents $\overset{\mbox{F}_{2}}{\longrightarrow}$ }\tabularnewline
\multicolumn{3}{|c|}{1 $\mathbf{TT}$ tall, 2 $\mathbf{Tt}$ tall, 1 $\mathbf{tt}$ short}\tabularnewline
\multicolumn{3}{|c|}{daughters: $3:1$ ratio}\tabularnewline
\hline 
\multicolumn{3}{|c|}{(ii)}\tabularnewline
\hline 
\end{tabular}}
\par\end{centering}

\caption{\label{fig: Punnett-1}\textbf{\small Mendel and Meiosis - 1.}{\small{}
In this linear setting, the parental alleles }\textbf{\small T}{\small{}
and }\textbf{\small t}{\small{} produce the only heterozygous genotypes
}\textbf{\small Tt}{\small{} of tall plants which segregate into gametes
with half carrying one of the allele and the other half the other
allele, correspond to the nonlinear entanglements of $2{}^{2}$ cycle.
Note that $\left(\left\downarrow \right\downarrow \right)$ is a Nash
equilibrium. The three possible F$_{1}$ gamete genotypes --- $\mathbf{TT}$
(homozygous dominant), $\mathbf{Tt}$ (heterozygous), and $\mathbf{tt}$
(homozygous recessive) --- in $3:1$ ratio, arise from successful
$\male-\female$ mating. }}
\end{figure}

Our use of the terms ``dominant'' and ``recessive'' for the female
and male alleles in a diploid cell is at variance with the simplest
form of allelic interaction formulated by Mendel where the phenotypic
effect of one allele completely masks that of the other in heterozygous
combinations when the phenotype produced by the two alleles is identical
to that produced by the homozygous genotype of the dominant member.
In the nonlinear case of holistic evolution regulated by interactive
feedbacks between ``demand'' and ``supply'', the symmetry breaking,
gravity-stimulated,%
\footnote{\textbf{\emph{Gravity}} is the thermodynamic legacy of the negative
world $\mathbb{W}_{-}$ in $\mathbb{W}_{+}$ generating the characteristic
dissipation-concentration, two-phase $(\left\uparrow \right\downarrow )$
signature of complexity and holism \citep{Sengupta2010-c}. %
} ordering supply-pump moderates the second law demand-engine of symmetry
inducing dissipation, and dominance represents the overriding entropy
modulation by the negative world $\mathbb{W}_{-}$ on $\mathbb{W}_{+}$
manifested through the ordered structures of Nature, see Sec. \ref{sec: Yang-Yin}. 

The choice of the female --- rather than the male --- as the dominant
of the complementary pair is based on our understanding that the uterus
symbolizes the receptacle of biological order in mammals: it is here
that the second law of dissipation appears to have been completely
defeated by order-inducing gravitational coalescence with its roots
in $\mathbb{W}_{-}$. In this negative multifunctional dual where
``anti-second law'' requires heat to flow spontaneously from lower
to higher temperatures with positive temperature gradient along increasing
temperatures, the engine and pump interchange their roles with ordering
compresssion of the system by the environment --- rather than the
entropic expansion against it of $\mathbb{W}_{+}$ --- being the natural
direction in $\mathbb{W}_{-}$. This postulates the rather startling
hypothesis that the female of a species is possibly the most significant
direct link of $\mathbb{W}_{+}$ with its source of negativity anchored
in $\mathbb{W}_{-}$ --- she is the ``capital'' supplier-in-adversary
in competitive-collaboration of unity-in-diversity with male cultural
demand.%
\footnote{\label{fn: capital-culture}``Capital'' and ``Culture'' are technical
terms introduced in \citep{Sengupta2010-c}. 

\textbf{Capital:} A factor of production not significantly consumed,
which is not wanted for itself but for its ability to help in producing
other goods; any form of wealth capable of being employed in the production
of more wealth. In a fundamental sense, capital consists of any produced
thing that can enhance a person's power to perform useful work. http://en.wikipedia.org/wiki/Capital. 

\emph{Capital} represents free-energy, exergy, information and individualism;
the genotypic supply in a multicellutar organism and \emph{markets}
in human society. On its own, capital is as insatiated as egg without
sperm.

\textbf{Culture:} The set of shared attitudes, values, goals, and
practices that characterizes an institution, organization or group;
the social production and transmission of an integrated pattern of
human knowledge, belief, and behavior that depends upon the capacity
for symbolic thought and social learning. http://en.wikipedia.org/wiki/Culture. 

According to Edward Tylor, culture \textquotedblleft is that complex
whole of knowledge, belief, art, morals, law, custom, and any other
capabilities and habits acquired by man as a member of society\textquotedblright . 

\emph{Culture} represents entropic dissipation and collective cooperation;
an organism's phenotypic demand, the \emph{state} in human society.
On its own it is as impotent as sperm without egg.%
} 

\begin{figure}[!tbh]
\noindent \begin{centering}
$\hspace{1in}$\includegraphics[scale=0.5]{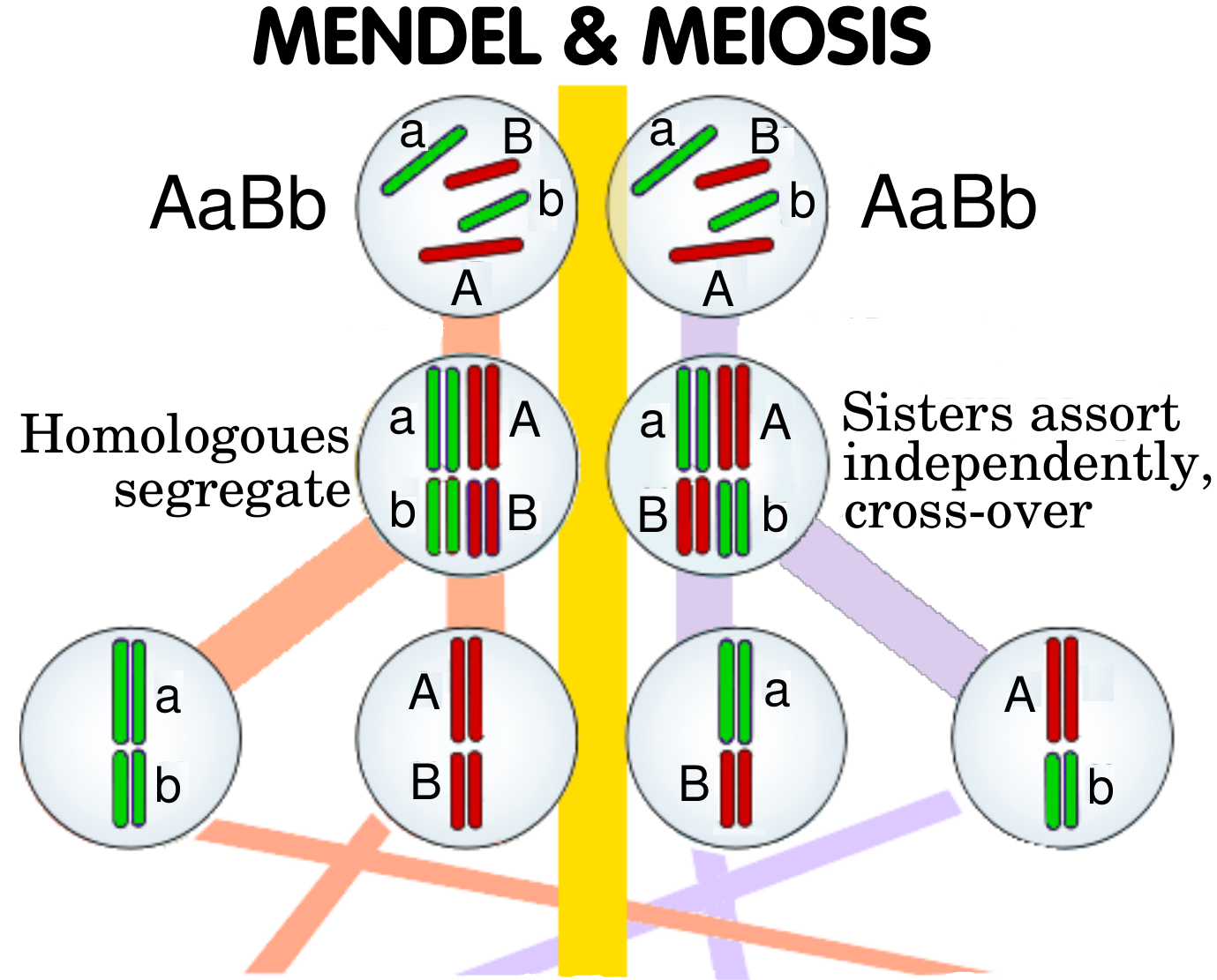}
\par\end{centering}

\begin{centering}
{\renewcommand{\arraystretch}{1.4}%
\begin{tabular}{|c||cc|cc|}
\hline 
{\large \female }$\left(\downarrow\right)$\textbackslash{}{\large \male }$\left(\uparrow\right)$ & 00($\mathbf{AB}$) & 01($\mathbf{Ab}$) & 10$\mathbf{(aB)}$ & 11$\mathbf{{\normalcolor (}ab{\normalcolor )}}$\tabularnewline
\hline 
\hline 
00($\mathbf{AB}$) & \cellcolor{B}0000 & \cellcolor{G}0001 & \cellcolor{Y}0100 & \cellcolor{R}0101\tabularnewline
01($\mathbf{Ab}$) & \cellcolor{G}0010 & \cellcolor{G}0011 & \cellcolor{R}0110 & \cellcolor{R}0111\tabularnewline
\hline 
10($\mathbf{aB}$) & \cellcolor{Y}1000 & \cellcolor{R}1001 & \cellcolor{Y}1100 & \cellcolor{R}1101\tabularnewline
11($\mathbf{ab}$) & \cellcolor{R}1010 & \cellcolor{R}1011 & \cellcolor{R}1110 & \cellcolor{R}1111\tabularnewline
\hline 
\end{tabular}}
\par\end{centering}

\caption{\label{fig: Mendel}\textbf{\small Mendel and Meiosis - 3.} {\small The
duplicated sister chromatids of Fig. \ref{fig: Punnett-1} perform
as \textquotedblleft two independent traits\textquotedblright{} A,
B with the traits in (a, b) and (A, B) \textquotedblleft crossing
over\textquotedblright{} to generate the Male-Female collaborations
$\left({\normalcolor \left\uparrow ,\right\downarrow }\right)$ of
(a, B) and $\left(\left\downarrow ,\right\uparrow \right)$ of (A,
b). Gamete compatibility of the traits lead to successful fusing of
the gametes and formation of zygotes with genotypes shown. }}
\end{figure}

An observation of immense significance is that the diploid-cross matrix
$\mathbf{\mathsf{C}}$ for two independent traits $\mathbf{\mathsf{A}}$
and $\mathbf{\mathsf{B}}$ is infact a tensor product of the monohybrid
F1 matrices of Fig. \ref{fig: Punnett-1}. Thus $\mathbf{\mathsf{C}}=\mathbf{\mathsf{A}}\otimes\mathbf{\mathsf{B}}$
in Fig. \ref{fig: Punnett-2} for the $\mathbf{\mathsf{A}}$ and $\mathbf{\mathsf{B}}$
traits generates the product $\mathbf{AaBb\otimes AaBb}$ of Fig.
\ref{fig: Mendel}, noting that the order of the alleles is immaterial
here. This suggests a common base for quantum nonlocality and Mendelian
inheritance which, as we have argued earlier \citep{Sengupta2010-a,Sengupta2010-b},
is a linear representation of complex holism. 
\begin{figure}[!tbh]
\noindent \begin{centering}
{\renewcommand{\arraystretch}{1.25}%
\begin{tabular}{|c||c|c|}
\hline 
$\mathbf{\mathsf{A}}$ & $\mathbf{A}$ & $\mathbf{a}$\tabularnewline
\hline 
\hline 
$\mathbf{A}$ & $\mathbf{AA}$ & $\mathbf{Aa}$\tabularnewline
\hline 
$\mathbf{a}$ & $\mathbf{aA}$ & $\mathbf{aa}$\tabularnewline
\hline 
\end{tabular}$\;\;\bigotimes\;\;$%
\begin{tabular}{|c||c|c|}
\hline 
$\mathbf{\mathsf{B}}$ & $\mathbf{B}$ & $\mathbf{b}$\tabularnewline
\hline 
\hline 
$\mathbf{B}$ & $\mathbf{BB}$ & $\mathbf{Bb}$\tabularnewline
\hline 
$\mathbf{b}$ & $\mathbf{bB}$ & $\mathbf{bb}$\tabularnewline
\hline 
\end{tabular}$\;\;=\;\;$%
\begin{tabular}{c}
\begin{tabular}{|cc|cc|}
\hline 
\cellcolor{R}$\mathbf{AABB}$ & \cellcolor{R}$\mathbf{AABb}$ & \cellcolor{B}$\mathbf{AaBB}$ & \cellcolor{B}$\mathbf{AaBb}$\tabularnewline
\cellcolor{R}$\mathbf{AAbB}$ & \cellcolor{R}$\mathbf{AAbb}$ & \cellcolor{B}$\mathbf{AabB}$ & \cellcolor{B}$\mathbf{Aabb}$\tabularnewline
\hline 
\cellcolor{Y}$\mathbf{aABB}$ & \cellcolor{Y}$\mathbf{aABb}$ & \cellcolor{G}$\mathbf{aaBB}$ & \cellcolor{G}$\mathbf{aaBb}$\tabularnewline
\cellcolor{Y}$\mathbf{aAbB}$ & \cellcolor{Y}$\mathbf{aAbb}$ & \cellcolor{G}$\mathbf{aabB}$ & \cellcolor{G}$\mathbf{aabb}$\tabularnewline
\hline 
\multicolumn{4}{|c|}{(iii)}\tabularnewline
\hline 
\end{tabular}\tabularnewline
\end{tabular}}
\par\end{centering}

\caption{\label{fig: Punnett-2}\textbf{\small Mendel and Meiosis - 2. }{\small The
diploid cross of two independent traits is the tensor product $\otimes$
of the individual traits. This suggests reductionist similarities
with quantum entanglement in the organizational assembly of an increasing
number of components into a composite whole. Taking $\mathsf{A}=\mathsf{B}$
as the valid tensor product of a matrix with itself, leads to the
situation where each of the emergent periodic point homologous units
like $\{\mathbf{AAbb}\parallel\mathbf{aAbb}\}$ and $\{\mathbf{AaBB}\parallel\mathbf{aaBB}\}$
of individual genotypes can be considered as organs of a particular
trait. The $(3:1)\times(3:1)$ cross of $\mathsf{A}\otimes\mathsf{B}$
generates the degeneracy in the arrangement of alleles into 4 groups:
$(0000)$, $(0001,0010,0011)$, $(0100,1000,1100)$, and $(0101,0110,0111,1001,1101,1010,1011,1110,1111)$
corresponding to $\female$ contributions of both traits from both
parents, $\female$ contributions of any one trait from both parents,
and the remainder, leading to the classical 1:3:3:9 ratio of genotypes.
Clearly, this follows the classical linear and reductionist sense
of ``dominant'' and ``recessive'' traits in heterozygosity, with
little intermingling as mandated in holism.}}
\end{figure}

Newton and Darwin fundamentally constructed two different types of
reductionist worlds. ``Newton's universe was stationary, cycling
without change through all eternity, perfectly knowable and completely
predictable. In Darwin's world, history mattered. The shape of the
future depended on the outcome of past events. No elegent equations
could predict the future of even a single organism, because chance
itself is inherent in life. Newton and Darwin created two utterly
different conceptions of Nature: one for lifeless objects, the other
for living things; one for stability, the other for change'' \citep{Rothschild1995}.
Nonetheless, it is not enough in life for effect to simply depend
on cause; reciprocally cause is simultaneously influenced by the effect
it produces. Advanced and retarded waves acting together in harmony
defines reality, one without the other is incomplete as Eq. (\ref{eq: Copenhagen})
and Eq. (\ref{eq: rigged}) below explicitly demonstrate. 

\sublaboff{figure}

\section{\label{sec: Yang-Yin}Yang-Yinism of Darwinian Evolution }

\begin{flushright}
\textsl{\FiveStarOpenCircled{} This (transactional) model describes
any quantum event as a ``handshake'' executed through an exchange
of advanced and retarded waves. Any emission process makes ``advanced''
waves $e^{i\omega t}$ on an equal basis with ordinary ``retarded''
waves $e^{-i\omega t}$. Both advanced and retarded waves are valid
solutions of the electromagnetic wave equation, but in conventional
electrodynamics the advanced solutions are rejected as unphysical
or acausal. This advanced-retarded handshake is the basis for the
transactional interpretation of quantum mechanics. It is a two-way
contract between the future and the past. The transaction is explicitly
nonlocal because the future is, in a limited way, affecting the past.
}\textsf{\textsl{\small \hfill{}}}\textsf{\textbf{\small John J.
Crammer}}\textsf{\textsl{\small{} }}\citep{Cramer1986-88}
\par\end{flushright}

\noindent In the above we have advocated a unified framework for the
evolutionary dynamics of Nature of which the linear depictions of
quantum non-locality, biological Punnett squares, and economic payoffs
and equilibria are different manifestations. Complexity results from
the interaction between parts of a system such that it manifests properties
not carried by, or dictated by, individual components: complexity
resides in the interactive competitive-collaboration between the parts
and the properties of a complex system are said to ``emerge, without
any guiding hand''. Competitive-collaboration, as opposed to reductionism,
in the context of this characterization means that the interdependent
parts retain their individual attributes, with each contributing to
the whole in its own characteristic fashion within a framework of
dynamically emerging inclusive globality of the whole. Although the
properties of the whole are generated by the parts, the individual
units acting independently on their own cannot account for the collective
behaviour of the total; a complex system is an assembly of many interdependent
parts, interacting with each other through competitive nonlinear collaboration
leading to self organized, emergent inclusiveness. 

Our earlier contention \citep{Sengupta2010-c,Sengupta2010-a,Sengupta2010-b}
that \textit{complex holism represents a stronger form of entanglement
than quantum nonlocality} is further corroborated in the present study:
Punnett squares and payoff matrices represent definite forms of equilibria
although itertated prisoner's dilemma is often used as a pedagogical
example of how collaboration emerges from confrontation. The $4\times4$
matrix of dihybrid cross continues to be meaningful for a single trait
under tensor product $\mathsf{R}\otimes\mathsf{R}$ of $\mathsf{R}$
with itself characterizing evolutionary iteration. The 9:~3:~3:~1
F$_{2}$ ratio is not evolutionary holistic, only a linear non-dynamic
expression of this omnipresence. 

Mathematically, inverse and direct limits denoted by $\underleftarrow{\lim}$
and $\underrightarrow{\lim}$ \citep{Dugundji1966} constitute a rationale
for the simultaneous existence opposing directional arrows that follow
from very general considerations. Thus reality of the union of a family
of nested sets entails the existence of their intersection, and conversely.
In the context of Hilbert spaces, these limits taken together specialize
to the extended rigged Hilbert space $\Psi\subset\mathcal{H}\subset\Psi^{\times}$

\begin{equation}
\begin{array}{c}
\begin{array}{rcl}
\begin{array}{rc}
{\color{red}\longleftarrow} & \begin{array}{l}
{\color{red}\mbox{Entropy decreasing}}\\
{\color{red}\mbox{Replicator}}
\end{array}\end{array} &  & {\color{green}\textrm{Collaborative}(\uparrow)}\hspace{0.77in}{\color{red}\textrm{Exergy}\rightarrow}\\
--------------- & ---- & -----------------\\
{\color{green}{\color{green}\underleftarrow{\lim}\mathcal{H}^{k}}\triangleq\Psi=\bigcap_{k}\mathcal{H}^{k}\subset\cdots\subset\mathcal{H}^{1}} & \subset\mathcal{H}\subset & {\color{red}\mathcal{H}_{-1}\subset\cdots\subset\bigcup_{k}\mathcal{H}_{-k}=\Psi^{\times}\triangleq\,{\color{red}\underrightarrow{\lim}\mathcal{H}_{-k}}}\\
--------------- & ---- & -----------------\\
{\color{green}\leftarrow\mbox{Entropy}}\hspace{0.4in}{\color{red}\mbox{Competitive}(\downarrow)} &  & \begin{array}{cc}
{\color{green}\begin{array}{r}
\mbox{Entropy increasing}\\
\mbox{Interactor}
\end{array}} & {\color{green}\longrightarrow}\end{array}
\end{array}\end{array}\label{eq: rigged}
\end{equation}

\noindent with $\Psi$ the space of physical states prepared in actual
experiments and $\Psi^{\times}$ antilinear functionals on $\Psi$
that associates with each state a real number interpreted as the result
of measurements on the state --- the spaces of test functions $\Psi$
and distributions $\Psi^{\times}$ enlarge the Hilbert space $\mathcal{H}$
to the rigged space $(\Psi,\mathcal{H},\Psi^{\times})$. If $\{X_{k}\}_{k\in\mathbb{Z}_{+}}$
is an increasing family of subsets of $X$ and $\eta_{mn}\!:X_{m}\rightarrow X_{n}$
is the inclusion map for $m\leq n$, then $\underrightarrow{\lim}X_{k}=\bigcup X_{k}$\textcolor{blue}{{}
}corresponds to the entropy decreasing direct iterates $f^{i}$ of
the logistic map; for $\{X^{k}\}_{k\in\mathbb{Z}_{+}}$ a decreasing
family of subsets of $X$ with $\pi^{nm}\!:X^{n}\rightarrow X^{m}$
the inclusion map, $\underleftarrow{\lim}X^{k}=\bigcap X^{k}$ represents
the entropy increasing inverse iterates $f^{-i}$. As topological
spaces, direct and inverse limits carry the final and initial topologies
with respect to their respective canonical morphisms which are identified
in Sec. \ref{sub: General Logistic}. Equation (\ref{eq: rigged})
can be visualized as a pyramid with a direct $\underrightarrow{\lim}$
base of concentrative compression and an inverse $\underleftarrow{\lim}$
tip of dissipative expansion. The homeostasis $\mathcal{H}$ is a
dynamical mixture of these extremes, with the entropy increasing engine
acting in the domain of the pump to increase collective cooperation
and the entropy decreasing pump operating in that of the engine to
boost individual selfishness. 

The forward and backward iterates of $f$ collectively define graphical
convergence leading to the multifunctional homeostasis \citep{Sengupta2003}
of nonlocality and entanglements reflecting limitations, compromises
and tradeoffs, Fig. \ref{fig: ChaNoXity-a}, \textit{b. }Both the
Copenhagen interpretation Eq. (\ref{eq: Copenhagen}) and transactional
interpretation constitute illustrations of these limits.

\subsection{Cohabitation of Opposites }

\begin{figure}[!tbh]
\noindent \begin{centering}
\textsf{\textbf{\large THE UNFOLDING OF SUSTAINABLE HOLISM\medskip{}
}}
\par\end{centering}{\large \par}

\noindent \begin{centering}
\begin{tabular}{|c|}
\hline 
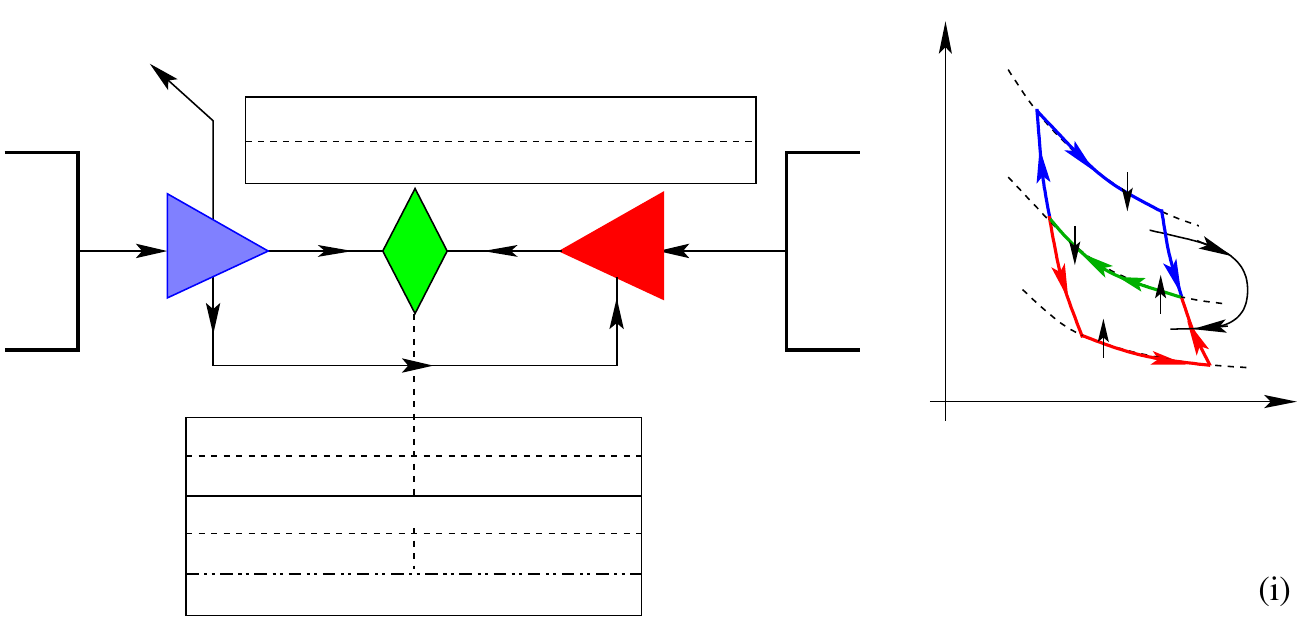\tabularnewline
{\small Human development and environmental conservation must be integrated
to respond reactively,}\tabularnewline
{\small rather than proactively, by choice if societal metabolism
is to be attained, sustained, propagated.}\tabularnewline
\hline 
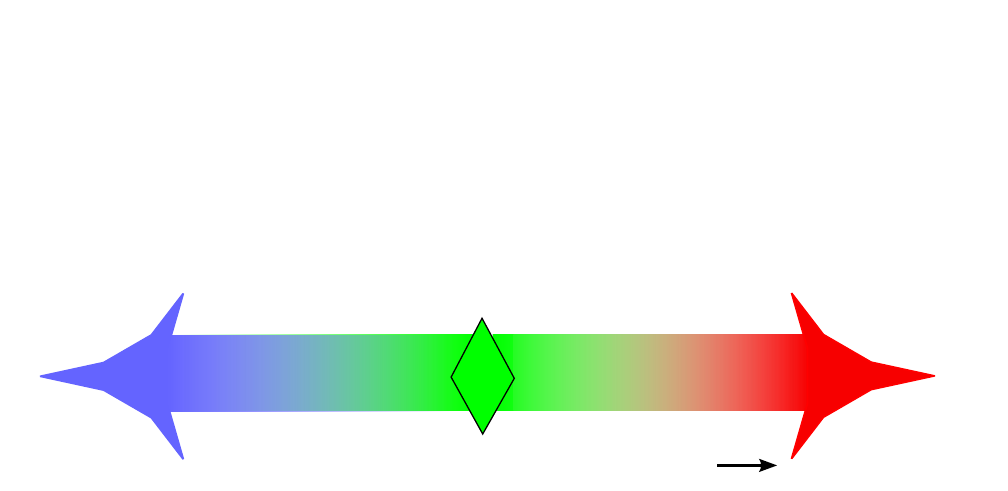\tabularnewline
\hline 
{\small Produce is a firm's phenotype, technology its ``push'' genotype,
and cultural consumer selection of}\tabularnewline
{\small produce a ``pull'' on the genotype: Economics should be
the study of the }\textbf{\emph{\small social}}\emph{\small{} relations
and proce-}\tabularnewline
\emph{\small sses}{\small{} governing production, distribution and
exchange of the requisites of life \citep{Hodgson1993} on the push-pull}\tabularnewline
{\small boundary $T$ by ``taking the product to the customer''
while ``getting him to come to you'', rather}\tabularnewline
{\small than the persuit of material wealth of unlimited wants by
scarce means. The ``productive base'' of}\tabularnewline
{\small the anabolic pump confronts the ``consumptive demand'' of
catabolic engine \citep{Dasgupta2007} ensuring the z}\tabularnewline
{\small collaborative and sustainably enduring dialectics encoded
in $\iota q=(1-\iota)Q$.\hspace{1in}(ii)}\tabularnewline
\hline 
\end{tabular}
\par\end{centering}

\noindent \caption{\textbf{\small \label{fig: ChaNoXity-a}A Blueprint for Sustainable
Holism: Universal Darwinism \citep{Corning2005}.} {\small Reduction
of the dynamics of opposites to an equivalent Pump-Engine thermodynamic
system; $W_{\textrm{rev}}=\eta_{E}Q_{h}$, $W(T)=\eta Q_{h}=\eta_{E}/(1-\eta_{E})Q=\left(T_{h}/T\right)\eta_{E}Q$,
where $\eta_{E}=1-\left(T_{c}/T_{h}\right)$, $\eta=1-\left(T/T_{h}\right)$
are thermodynamic efficiencies of the respective engines. The collaborative
confrontation of $Q(T)\triangleq Q_{h}-W(T)=Q_{h}-[1-\iota(T)]W_{\textrm{{rev}}}=\left(T/T_{h}\right)Q_{h}$
and $q(T)=\left[\left(1-\iota\left(T\right)\right)/\iota\left(T\right)\right]Q$,
permits the interpretation of $Q$ as \textquotedblleft demand\textquotedblright{}
that is met by the \textquotedblleft supply\textquotedblright{} $\iota q$
in a bidirectional feedback loop that sustains, and is sustained by
each other, in the context of the whole. The entangled characterization
$(\left\uparrow \right\uparrow )\oplus(\left\uparrow \right\downarrow )\oplus(\left\downarrow \right\uparrow )\oplus(\left\downarrow \right\downarrow )$
of the holistic temperature $T$ is to be compared with the non-locality
of Eqs. (\ref{eq: NonLocal(a)}, }\textit{\small b}{\small , }\textit{\small c}{\small ).
Heterozygosity of an organism, as considered here, is a continuous
parameter determined by the homeostasy of inheritance $T$.}}
\vspace{-0.2in}
\end{figure}

Figure \ref{fig: ChaNoXity-a}, \textit{b} summarizes our current
understanding of complex holism. The fundamental issue is the existence
of a negative world $\mathbb{W}_{-}\triangleq\mbox{Multi}(X)$ for
every real world $\mathbb{W}_{+}\triangleq\mbox{map}(X)$  defined
by \sublabon{equation}
\begin{equation}
\mathbb{W}_{-}\triangleq\{\mathsf{w}\!:\{w\}\,{\textstyle \bigoplus}\,\{\mathsf{w}\}=\emptyset\};\label{eq: matter-neg(a)}
\end{equation}
$\mathbb{W}_{-}$ is the negative, or exclusion, set of $\mathbb{W}_{+}$%
\footnote{Notice that this definition is meaningless if restricted to $\mathbb{W}_{+}$
or $\mathbb{W}_{-}$ alone; it makes sense, in the manner defined
here, only in relation to the pair $(\mathbb{W}_{+},\mathbb{W}_{-})$. %
}. Here, the usual topological treatment of pointwise convergence in
function space is generalized to generate the boundary $\textrm{Multi}_{\parallel}(X)$
between $\textrm{map}(X)$ and $\textrm{multi}(X)$ where $\textrm{map}(X)$
and $\textrm{multi}(X)$ are respectively proper functional and non-functional
subsets of all the correspondences $\mbox{Multi}(X)$ on $X$ 
\begin{equation}
\mbox{Multi}(X)=\mbox{map}(X)\,{\textstyle \bigcup}\,\mbox{Multi}_{\parallel}(X)\,{\textstyle \bigcup}\,\mbox{multi}(X)\label{eq: Multi}
\end{equation}
This generalization defines neighbourhoods of $f\in\textrm{map}(X)$
to consist of those functional relations in $\textrm{Multi}(X)$ whose
images at any point $x\in X$ lies not only arbitrarily close to $f(x)$
but whose inverse images at $y=f(x)$ contain points arbitrarily close
to $x$: the graph of $f$ lies not only close to $f(x)$ at $x$
in $V$, but must additionally be such that $f^{-}(y)$ has at least
branch in $U$ about $x$, with $f$ constrained to cling to $f$
as the number of points on its graph increases with convergence. Unlike
for simple pointwise convergence, no gaps in the graph of the limit
is permitted not only on the domain of $f$ but in its range too;
this \emph{topological extension} of the function space map$(X)$
to Multi$(X)$ is fundamental in our treatment that allows non-functional
limits as possibilities, and constitutes the bedrock of non-reductionist
holistic sustainability. 

Hence for all $A\subseteq\mathbb{W}_{+}$ there exists a neg(ative)
set $\mathsf{A}\subseteq\mathbb{W}_{-}$ associated with (generated
by) $A$ that satisfies \vspace{-0.2in}

\begin{eqnarray}
A\:{\textstyle \bigoplus}\:\mathsf{G} & \!\!\!\triangleq\!\!\! & A-G,\quad G\leftrightarrow\mathsf{G}\nonumber \\
A\:{\textstyle \bigoplus}\:\mathsf{A} & \!\!\!=\!\!\! & \emptyset.\label{eq: matter-neg(b)}
\end{eqnarray}
\sublaboff{equation}The pair $(A,\mathsf{A})$ act as relative discipliners
of each other in ``undoing'', ``controlling'', ``stabilizing'' the
other. This induces a state of dynamic homeostasis in $\mathbb{W}_{+}$
of limitations, compromises and trade-offs that permits out-of-equilibrium
complex composites of a system and its environment to coexist despite
the privileged omnipresence of the Second Law. The evolutionary process
ceases when the opposing influences in $\mathbb{W}_{+}$ and its moderator
$\mathbb{W}_{-}$ balance. $\mathbb{W}_{-}$ is the source of \textit{all
}creativity\textit{ }in $\mathbb{W}_{+}$, that however can natively
support only sterile dissipation: through the induction of gravity
--- its ordering signature --- in $\mathbb{W}_{+}$, the negative
world $\mathbb{W}_{-}$ is indeed the progenitor of Schroedinger's
neg-entropy. On its own, capital is as destructively dissipative as
its parent gravity, leading to the only eventuality of crashes and
catastrophic heat death. 

\begin{figure}[!tbh]
\noindent \begin{centering}
{\renewcommand{\arraystretch}{1.3}%
\begin{tabular}{c||c||c}
\multicolumn{3}{c}{%
\begin{tabular}{|c||c|}
\hline 
\multicolumn{2}{|>{\centering}p{5.93in}|}{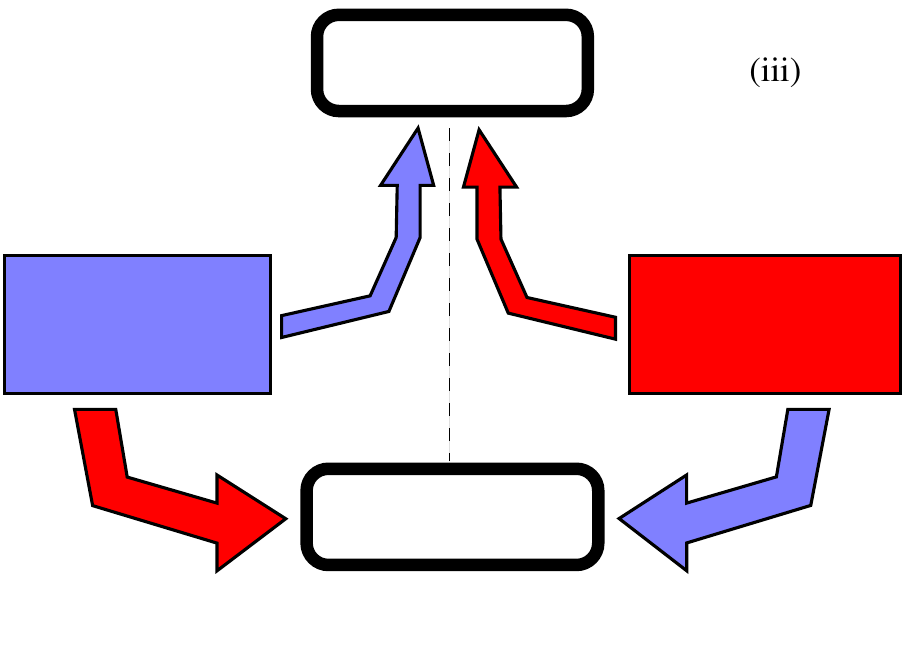}\tabularnewline
\hline 
\end{tabular}}\tabularnewline
\multicolumn{3}{c}{%
\begin{tabular}{|>{\centering}p{2.1in}|c|>{\centering}p{2.1in}|}
\multicolumn{1}{>{\centering}p{2.1in}|}{} & \multirow{11}{*}{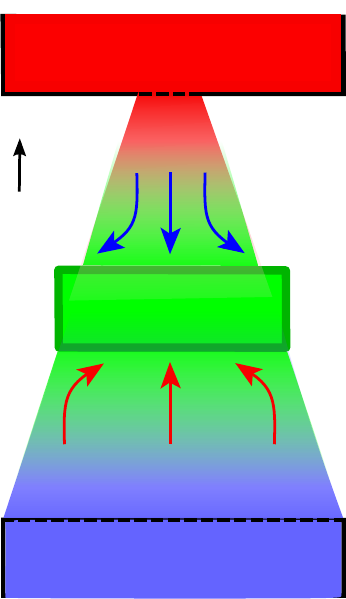} & \multicolumn{1}{>{\centering}p{2.1in}}{}\tabularnewline
\multicolumn{1}{>{\centering}p{2.1in}|}{} &  & \multicolumn{1}{>{\centering}p{2.1in}}{}\tabularnewline
\multicolumn{1}{>{\centering}p{2.1in}|}{} &  & \multicolumn{1}{>{\centering}p{2.1in}}{}\tabularnewline
\cline{1-1} \cline{3-3} 
${\color{green}{\normalcolor \underrightarrow{\mbox{\textsc{{\normalcolor Forward-Inverse\,\ arrow}}}}}}$ &  & \textcolor{red}{${\normalcolor \underleftarrow{\mbox{\textsc{{\normalcolor Backward-Direct\,\ arrow}}}}}$}\tabularnewline
\textsc{$\mathbb{W}_{+}$:} Natural Selection &  & \textsc{$\mathbb{W}_{-}$: }Somatic Mutation\tabularnewline
\cline{1-1} \cline{3-3} 
Top-down Engine $E$ &  & Bottom-up Pump $P$\tabularnewline
\cline{1-1} \cline{3-3} 
Disorder: Entropy increasing  &  & Order: Entropy decreasing\tabularnewline
\cline{1-1} \cline{3-3} 
Dissipative: Self-organization &  & Concentrative: Emergence\tabularnewline
\cline{1-1} \cline{3-3} 
Collective: Cooperative &  & Individualistic: Competitive\tabularnewline
\cline{1-1} \cline{3-3} 
${\color{green}{\normalcolor \underrightarrow{\mbox{\textsc{Altruist Culture}}(\uparrow)\!:\mbox{Phenotype}}}}$ &  & \textcolor{red}{${\normalcolor \underleftarrow{\mbox{\textsc{Selfish Capital}}(\downarrow)\!:\mbox{Genotype}}}$}\tabularnewline
\cline{1-1} \cline{3-3} 
Offer, Interactor &  & Confirmation, Replicator\tabularnewline
\cline{2-2} 
\multicolumn{3}{|c|}{\textsc{Handshake, Inheritance}: Sustainable synthetic cohabitation
of opposites, $T$}\tabularnewline
\multicolumn{3}{|c|}{{\small ``Growth by itself is not enough. High productive activity
and widespread poverty can coexist, and can }}\tabularnewline
\multicolumn{3}{|c|}{{\small endanger the environment. Hence sustainable development requires
that societies meet human needs both}}\tabularnewline
\multicolumn{3}{|c|}{{\small by increasing productive potential and by ensuring equitable
opportunities for all'' \citep{Brundtland1987}\hspace{0.85in}(iv)}}\tabularnewline
\hline 
\end{tabular}}\tabularnewline
\multicolumn{3}{c}{%
\begin{tabular}{|c|}
\textbf{\small Definition-1.}{\small{} An open thermodynamic system
of many interdependent organs is }\textbf{\emph{\small complex}}{\small{}
if it is in synthetic}\tabularnewline
{\small competitive cohabitation with its induced negative dual in
a hierarchical two-phase homeostasy of collective}\tabularnewline
{\small top-down, dissipative, self-organizing, entropy-increasing,
phenotypical interactor ``culture'' and individualis-}\tabularnewline
{\small tic bottom-up, concentrative, emergent, entropy-decreasing,
genotypical replicator ``capital'', coordinated and }\tabularnewline
{\small mediated by environment.\hspace{4.25in}(v)}\tabularnewline
\hline 
\textbf{\small Definition-2.}{\small{} For an environment $(T_{c},T_{h})$,
the equity-growth homeostasis $T$ defines}\textit{\small{} }\textbf{\textit{\small sustainable
holism, }}\tabularnewline
{\small and conduct of the capital-culture environment in maintaining
$\lambda\in(3,\lambda_{*})$ within the window of a particular}\tabularnewline
{\small{} $2^{N}$-cycle constitutes}\textbf{\textit{\small{} sustainable
development}}\textit{\small .}{\small \hspace{3.1in}(vi)}\tabularnewline
\hline 
\textbf{\small Definition-3.}{\small{} }\textbf{\textit{\small Life}}{\small{}
is the sustainable inheritance of entropy-exergy antagonism of the
cooperative-individual-}\tabularnewline
{\small ism of Nature.\hspace{4.85in}(vii)}\tabularnewline
\hline 
\end{tabular}}\tabularnewline
\end{tabular}}
\par\end{centering}

\caption{{\small \label{fig: ChaNoXity-b}}\textbf{\small{} ChaNoXity, The New
Science of Complex Holism}{\small{} }\textbf{\small \citep{Sengupta2010-c}:
Sustainable Inheritance.}{\small{} Universal Darwinism is a generalization
of classical Darwinism that applies with ``essential and auxiliary
explanations specific to each scientific domain'' to all open evolving
systems sharing the common attribute of variation-inheritance-selection.
Hodgson and Knudsen \citep{Hodgson2004} uses this as the defining
attribute of universal, generalized Darwinism, distinguished by self-replication. }}
\end{figure}

\sublabon{figure}
This $\mathbb{W}_{+}-\mathbb{W}_{-}$ dualism is formalized in the
Engine$(\mathbb{W}_{+})$~$\leftrightarrow$~Pump$(\mathbb{W}_{-})$
bidirectional, positive-negative of Fig. \ref{fig: ChaNoXity-a}.
An \textit{engine} $E$ needs fuel to deliver; this is its ``demand''
in exchange for an ``offer'' of non-entropic, constructive work.
Complex systems achieve this by establishing an autocatalytic positive
feedback mechanism in the form of a \textit{pump $P$ }that\textit{
}counters the\textit{ }symmetrization of $E$ though symmetry breaking
inducement of structures, compare Eq. (\ref{eq: rigged}). Complex
systems therefore realize an induced homeostasis of the entropy and
free energy adversaries, in the context of a given available enthalpy
of maximum Carnot work $W_{\textrm{{rev}}}$. For macroscopic processes,
the entropy increase due to $E$ must \textit{eventually} dominate
its decrease due to $P$ of increasing exergy, simply because $\mathbb{W}_{+}$
is administered by the law of increasing entropy. The replicator-interactor
dynamics \citep{Dawkins2006} of out-of-equilibrium systems is formalized
through the interactor \textit{firm} of cultural selection as vehicles
for the \textit{habits }and\textit{ routines ---} the later as collective
analogues of habits of individuals --- of replicators of technological
innovation, Fig. \ref{fig: ChaNoXity-b}. A firm is the entropic expression
of exergic routines. However Nelson \citep{Nelson2007}, in a more
general and divergent point of view on the evolution of entities like
science, technology, and business organizations that he terms human
\textquoteleft culture\textquoteright{} signifying ``a collective
phenomenon affecting the way that individuals within a society think
and act'', believes that ``attempts to force the details of cultural
evolution into a framework (of) biological evolution, in particular
to assume that close analogues to entities like genes and processes
like the dynamics of inclusive fitness, are generally misconceived
and counterproductive''. Note that in our setting, altruist culture
together with its antagonistic partner the selfish gene, comprise
the basis for evolution under the overall supervision of the environment.

\sublaboff{figure}In justification of the natural existence and applicability
of a backward arrow in establishing the dynamics of Nature --- as
for example in the physical reality of an advanced confirmation wave
from an absorber in the future influencing the retarded offer of an
emitter in the past --- we recall the salient features of a topological
argument of \citep{Sengupta2010-a} that allows a contrapuntal source
of ``negative entropy'' to coexit with the natural entropic signature
of the Second Law; for the details the original reference should be
consulted. The origin of the unusual ``negative entropy'' lies in
the inducement of an \textit{exclusion topology }\citep{Murdeshwar1990}
in $\mathbb{W}_{+}$ due to the $\mathbb{W}_{+}-\mathbb{W}_{-}$ interaction.\textit{
}For any subset $A\subseteq X$, while the normal $A$-\textit{inclusion
topology on} $X$ comprises $\emptyset$ and every superset of $A$
containg $A$, the abnormal $A$-\textit{exclusion topology} consists
of all subsets of $X-A$ that exclude $A$. Thus $A$ is open in the
inclusion topology and closed in the exclusion, and generally \textit{every
open set of one is closed in the other}. For $x\in X$, the $x$-\textit{inclusion}
open neighbourhoods comprises all non-empty supersets of $\{x\}$;
the $x$-\textit{exclusion} neighbourhoods are non-empty open subsets
of $\mathcal{P}(X-\{x\})$ \textit{exclusive} of $\{x\}$. The abnormal
exclusion topology is in a sense ``negative'' of the normal inclusion
topology: whereas the neighbourhood of a point in the inclusion must
always contain the point, the exclusion neighbourhood never contains
its defining point. This rather novel property endows the exclusion
topology with the remarkable attribute that while any sequence converges
to the defining point in its own topology, only the eventually constant
$\{v,v,v,\cdots\}$ converges to a $v\neq w$ --- all directions with
respect to the defining point in this unusual topology are infact
equivalent. 

For a sequence converging to $w_{*}\in\mathbb{W}_{+}$, there exist
according to Eqs. (\ref{eq: matter-neg(a)},\textit{ }\emph{c}) an
increasing sequence of negelements $(\mathsf{w}_{i})_{i\geq0}\rightarrow\mathsf{w}_{*}\in\mathbb{W}_{-}$
in the $\mathsf{w}_{*}$-inclusion topology generated by the $\mathbb{W}_{-}$-images
of the neighbourhood system of $\mathbb{W}_{+}$. Since the only manifestation
of neg-sets in the observable world is their influence on $\mathbb{W}_{+}$,
the $\mathbb{W}_{-}$ sequence converges in $\mathbb{W}_{-}$ if and
only if $(w_{i})_{i\ge0}\rightarrow w_{*}\in\mathbb{W}_{+}$, which
means that the $\mathsf{w}_{*}$-inclusion arrow in $\mathbb{W}_{-}$
induces, through its interaction with $\mathbb{W}_{+}$, an $w_{0}$-exclusion
arrow in $\mathbb{W}_{+}$ opposing the inclusion arrow converging
to $w_{*}$. The inclusion subspace topology is the natural initial
topology on inverse limits. 

The effect of $(\mathsf{w}_{i})_{i\geq0}\in\mathbb{W}_{-}$ on $\mathbb{W}_{+}$
is to regulate the evolution of this forward arrow to an effective
state of stasis of dynamical equilibrium. The existence of a negelement
$\mathsf{w\in W}$ for every $w\in\mathbb{W}_{+}$, by Eqs. (\ref{eq: matter-neg(a)},\textit{
c}) requires all forward arrows in $\mathbb{W}_{+}$ to have a matching
forward arrow in $\mathbb{W}_{-}$ that  appears backward when viewed
in $\mathbb{W}_{+}$. It is this opposing complimentary dualistic
nature of the apparently backward-$\mathbb{W}_{-}$ sequences on $\mathbb{W}_{+}$
--- responsible by (\ref{eq: matter-neg(b)}) for moderating the normal
uni-directional evolution in $\mathbb{W}_{+}$ --- that establishes
a stasis of dynamical balance between the opposing forces generated
in the composite of a compound system and its environment. The conjugation
operation of changing a retarded wave $\Psi$ to its complex conjugate
advanced state $\Psi^{*}$ is an example of the inhibitory action
of $\mathbb{W}_{-}$ on $\mathbb{W}_{+}$. Obviously, such evolutionary
processes cease when the opposing influences in $\mathbb{W}_{+}$
due to itself and its moderator $\mathbb{W}_{-}$ achieve holistic
balance, marking a state of dynamic equilibrium. An additional support
of these arguments is provided by the inverse and direct limits referred
earlier, see Ref. \citep{Sengupta2010-a} for more. 

\begin{figure}[!tbh]
\begin{centering}
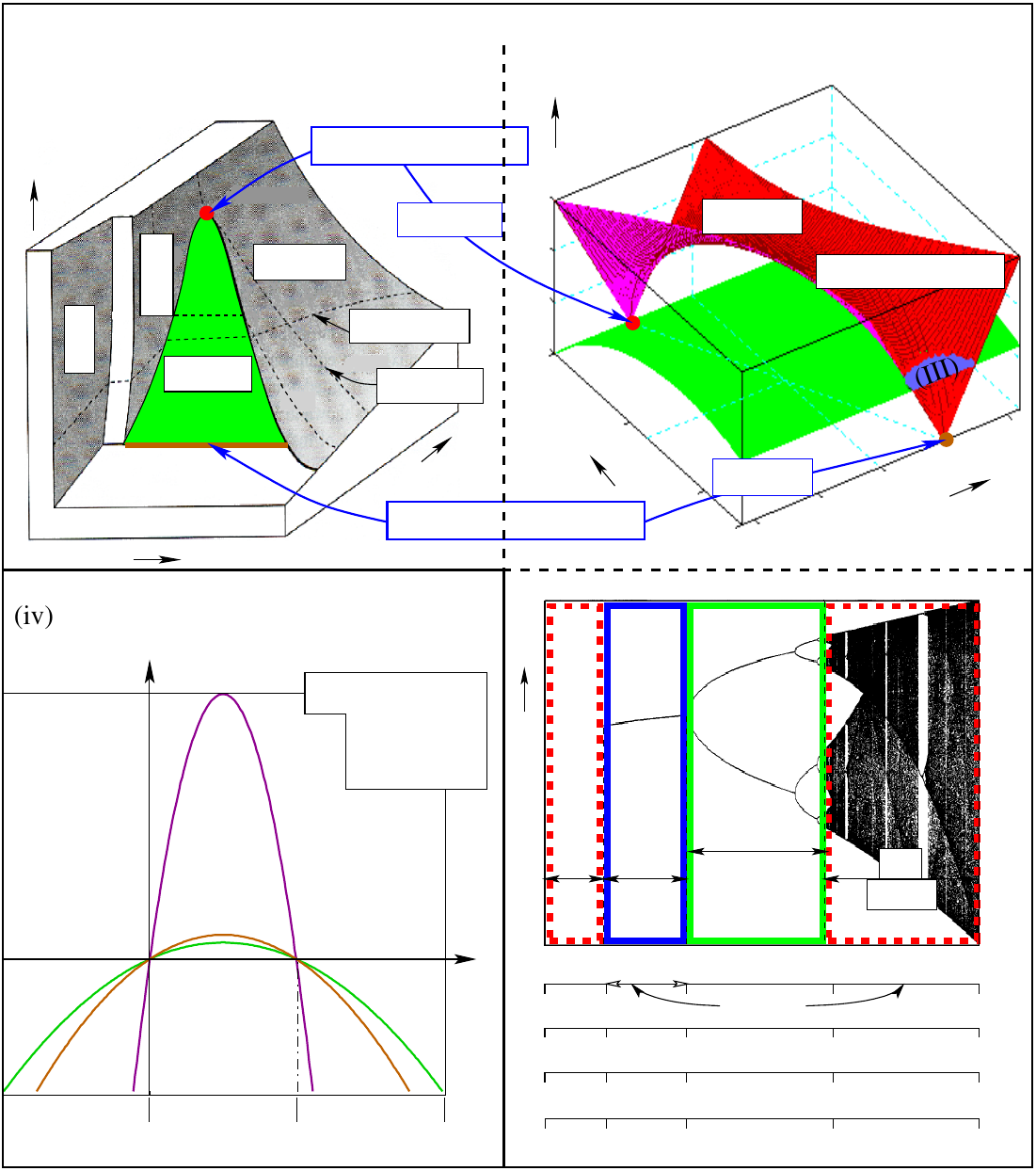
\par\end{centering}

\caption{{\small \label{fig: 2-phase}The 2-phase complex $\iota=\alpha$ region,
(ii) with }\textit{\small critical point}{\small{} $T_{c}=T_{h}$ at
$\alpha_{-}=-1$, yielding to $\alpha$-dependent $\alpha=\eta\zeta$
at low $T_{c}$. The }\textit{\small triple point}{\small{} $\alpha_{+}=1$,
$T_{c}=0$ is approachable only through this route. Compared to the
normal transition of (i), self-organization in (ii) occurs for $\alpha=Pv=\mbox{const}$.
$T_{+}-T_{-}:=(T_{h}-T_{c})\sqrt{T_{h}^{2}+4T_{c}T_{h}}/(2T_{h}-T_{c})$
at $\iota=\alpha$ is taken as an indicator of first-order-second-order
transition; $\iota\alpha>0$ in (I)/(III) defines $\mathbb{W}_{+}$,
$\iota\alpha<0$ in (II)/(IV) for $\mathbb{W}_{-}$. $\iota_{c}=-T_{c}/(T_{h}-T_{c})$.
BB: Big Bang, BH: Black Hole. }}
\end{figure}

The space of functions $A=\mbox{map}(X)$ is an open subspace of of
$Y=\mbox{Multi}(X)$ in the topology of pointwise biconvergence \citep{Sengupta2003};
hence Cl$(\mbox{multi}(X))$ is closed in the normal inclusion topology,
but open referred to the exclusion. The adversity of these components
of $\mbox{Multi}$ induces a common boundary of the constant multifunction
as indicated in Fig. \ref{fig: 2-phase}(c). Relative to the derived
set $\mbox{Der}(A)\triangleq\{y\in Y\!:(\forall N\in\mathcal{N}_{y})(N{\textstyle \cap}(A-\{y\})\neq\emptyset)\}=\{y\in Y\!:(\exists\textrm{ a net }\zeta\rightarrow y\textrm{ in }A-\{y\})\}$,
a subset $A\subseteq Y$ can be classified into the three types \medskip{}

1.\emph{ }\textbf{\emph{Altruist:}}\textbf{ }$\mbox{Der}(A)\subseteq Y-A$, 

2.\emph{ }\textbf{\emph{Cooperative:}} $(\mbox{Der}(A)\cap A\neq\emptyset)\wedge(\mbox{Der}(A)\cap(X-A)\neq\emptyset)$,

3.\emph{ }\textbf{\emph{Selfish:}} $\mbox{Der}(A)\subseteq A$. \medskip{}

This categorization adapted from Sengupta \citep{Sengupta2003}, illustrates
that of the nine possibilities the Cooperative-Cooperative entry (2,2)
represents the give-and-take dynamic homeostasy born of the collective-individualism
of complex holism. The exclusion topology of $\mbox{multi}(X)$ and
the inclusion topology of $\mbox{map}(X)$ conspire in the spirit
of mutual sustainability of complexity to evolve $(2,3)\rightarrow(2,2)$. 

The positive-negative feedback responsible for this transition is
related to the topologies of $\mbox{map}(X)$ and $\mbox{multi}(X)$
in Eq. (\ref{eq: Multi}), the dynamics of convergence being determined
by the families of open sets. In this sense the solution
\begin{equation}
x=f^{-}(y),\quad f\!:(X,\mathcal{U})\rightarrow(Y,\mathcal{V})\label{eq: ill-posed}
\end{equation}
of the ill-posed problem $f(x)=y$ defined by a non-injective, non-surjective
function $f$ --- with $f^{-}f(U\in\mathcal{U}):=\mbox{sat}(U)$ and
$ff^{-}(V\in\mathcal{V})=f(X)\cap V:=\mbox{comp}(V)$ reflecting the
ill-posedness of Eq. (\ref{eq: ill-posed}), neither being identities
on their respective spaces --- as a bidirectional dynamical system
with $y$ determining $x$ that in turn defines $y$, reflects the
homeostasy of the nonlinear evolution. The tools of initial and final
topologies are specifically tailored for a situation like this, with
the coarsest initial topology on the domain and finest final topology
on the range specifying the pre-image and image continuity of $f$
respectively. 

\textsl{\emph{For }}\emph{$e\!:X\rightarrow(Y,\mathcal{V})$, }\textsl{\emph{the}}\emph{
}preimage\emph{ }\textsl{\emph{or}}\emph{ }initial topology of\emph{
$X$} generated by\emph{ }\textsl{$e$}\emph{ }and $\mathcal{V}$\emph{
}\textsl{\emph{is \sublabon{equation}}}\emph{ }
\begin{equation}
\textrm{IT}\{e;\mathcal{V}\}\overset{\textrm{def}}{=}\{U\subseteq X\!:U=e^{-}(V)\textrm{ for }V\in\mathcal{V}_{\textrm{comp}}\},\label{eq: IT}
\end{equation}

\noindent \textsl{\emph{while for $q\!:(X,\mathcal{U})\rightarrow Y$,
the}}\emph{ }image\emph{ }\textsl{\emph{or}}\emph{ }final topology
of\emph{ $Y$ }generated by $\mathcal{U}$ and\emph{ }\textsl{$q$}\emph{
}\textsl{\emph{is}} 
\begin{equation}
\textrm{FT}\{\mathcal{U};q\}\overset{\textrm{def}}{=}\{V\subseteq Y\!:U=q^{-}(V)\mbox{ for }U\in\mathcal{U}_{\textrm{sat}}\}\label{eq: FT}
\end{equation}

\noindent where $\mathcal{U}_{\textrm{sat}}$, $\mathcal{V}_{\textrm{comp}}$
are the saturations $U_{\textrm{sat}}=\{\textrm{sat}(U)\!:U\in\mathcal{U}\}$
of the open sets of $X$ and the components $V_{\textrm{comp}}=\{\textrm{comp}(V)\!:V\in\mathcal{V}\}$
of the open sets of $Y$ whenever these are also open in $X$ and
$Y$; plainly, $\mathcal{U}_{\textrm{sat}}\subseteq\mathcal{U}$ and
$\mathcal{V}_{\textrm{comp}}\subseteq\mathcal{V}$. Hence the topology
of $(X,\textrm{IT}\{e;\mathcal{V}\})$ consists of, and only of, the
$e$-saturations of all the open sets of $X$, and the open sets of
$(Y,\textrm{FT}\{\mathcal{U};q\})$ are the $q$-images in $Y$ (and
not just in $q(X)$) of all the $q$-saturated open sets of $X$. 

Combining these equations for the problem (\ref{eq: ill-posed}),
yields $U\in\mathcal{U}_{\textrm{sat}}=\mathcal{U}$ and $V\in\mathcal{V}_{\textrm{comp}}=\mathcal{V}$
simultaneously from the initial-final problem 
\begin{equation}
\mbox{IFT}\{\mathcal{U};f;\mathcal{V}\}=\{(U\subseteq X,V\subseteq Y\!:U=f^{-}(V))\}\label{eq: IFT}
\end{equation}
\sublaboff{equation}which clearly reduces to a homeomorphism for
bijective $f$ with a well-defined inverse $f^{-1}$. As should be
transparently evident from (\ref{eq: IFT}), the initial-final problem
is one of competitive-collaboration with the finest topology $\mathcal{U}$
on $X$ representing the decreasing entropy of coalescence, concentration
and confirmation initiated by the inverse problem, and the coarsest
topology $\mathcal{V}$ of increasing entropic dispersion, dissipation
and offer on $Y$ due to the direct problem. Thus in Fig. \ref{fig: mitosis},
the fixed points of $f^{2N}$, $N=1,2,3$ are equivalent (periodic
points) in the sense that, in (ii) for example, $00\sim01\sim10\sim11$
as they all lie on the chain-dotted $2^{N}$-periodic cycle of the
figure. Taking the collection of these periodic points to represent
a maximal saturated open set of $X$, the $f$-images $f_{13}$ and
$f_{24}$ fragments $f_{12}$, to be further dissipated into $f_{15}$,
$f_{26}$, $f_{37}$, $f_{48}$. The associated induced dispersion
in $\mathbb{W}_{-}\mbox{-multi}$ on the ordinate competing collaboratively
with the concentration on the $\mathbb{W}_{+}\mbox{-map}$ of the
abscissa completes this prognosis of Eq. (\ref{eq: IFT}), with the
$V\subseteq Y$ of (\ref{eq: IFT}) considered closed in $Y$ and
their complements open in $\mbox{multi}(X)$ that induces an additional
exclusion topology on $\mbox{map}(X)$.

The following passage from Greaves \citep{Greaves2007} provides a
medical interpretation of the Engine-Pump bidirectionality of Fig.
\ref{fig: ChaNoXity-a},\textit{b} with reference to cancer (and other
chronic diseases). ``Intrinsic vulnerability to cancer must be counterintuitive
to anyone who views our bodies as the product of purposeful design
or engineering. Darwinian medicine provides the opposite view: the
blind process through which we have emerged carries with it inevitable
limitations, compromises and trade-offs. The reality is that for accidental
or biologically sound adaptive reasons, we have historically programmed
falliability. Covert tumours arise constantly, reflecting our intrinsic
vulnerability, and each and every one of us harbours mutant clones
with malignant potential. Clinical cancer rates would be even worse
if it were not for the fact that cancer clone emergence is relatively
inefficient evolutionary process, subject to many constraints or bottlenecks.
Perhaps only 1\% of the covert pre-malignant clone ever acquire the
necessary additional or complimentary mutations required for graduation
to malignancy''. Needless to say, the biologically sound natural
entropic adversity of the self-organizing Engine keeps this exergic
cancerous emergence caused by $P$ in holistic check, even with the
possible attendant mismatch of an extravagant lifestyle. Nevertheless,
with the passage of time the cosmic determinism of second law asserts
itself, the inhibiting influence of $P$ diminishes and eventually
vanishes. Coupled with possible lifestyle extravaganza, this may leave
large-$\chi$ holistic organisms like \textit{homo sapiens} with unpaired
excess of accumulated $\mathbb{W}_{-}$-emergence of broken symmetry,
no longer amenable to the failed entropic self-organization of $\mathbb{W}_{+}$
--- leading to a genetic predisposition and intrinsic vulnerability
to malignancy. Benign tumor growth appears as a natural corollary,
subvertible under normal circumstances through the much more effortless
entropic dissipative failure of normal organs. 

Define the equilibrium holistic state of homeostatic Engine-Pump adaptability
by the equation of state $Pv=f(T)$ of the participatory universe
\sublabon{equation}

\begin{eqnarray}
\alpha(T):=\eta\zeta=\left(\frac{T_{h}-T}{T-T_{c}}\right)\left({\displaystyle \frac{T}{T_{h}}}\right)\triangleq\frac{q(T)}{Q_{h}} & \!\!\!=\!\!\! & \frac{q(T)}{Q(T)}\left(\frac{T}{T_{h}}\right),\label{eq: alpha(ii)}\\
 & \!\!\!=\!\!\! & \left(\frac{1-\iota(T)}{\iota(T)}\right)\left(\frac{T}{T_{h}}\right);\label{eq: alpha(i)}
\end{eqnarray}
\sublaboff{equation}$Q(T)\triangleq Q_{h}-W(T)=Q_{h}-[1-\iota(T)]W_{\textrm{{rev}}}=Q_{h}\left(\frac{T}{T_{h}}\right)$,
$Q+q=\left(\frac{T_{h}-T_{c}}{T_{h}}\right)\left(\frac{T}{T-T_{c}}\right)Q_{h}=\eta_{E}\zeta Q_{h}$,
where $\zeta\leftrightarrow P=0$ at $T=0$ and $\eta\leftrightarrow v$
as the product of the efficiency $\eta$ of a reversible engine and
the coefficient of performance $\zeta$ of a reversible pump, see
Fig. \ref{fig: ChaNoXity-a}. Then \sublabon{equation}
\begin{eqnarray}
{\color{magenta}{\color{magenta}{\normalcolor T_{\pm}(\alpha)}}} & = & \frac{1}{2}\left[(1-\alpha)T_{h}\pm\sqrt{(1-\alpha)^{2}T_{h}^{2}+4\alpha T_{c}T_{h}}\right]\label{eq: T_pm(a)}\\
 & = & \begin{cases}
((1-\alpha),\,0)=(0,0)_{\alpha=1}, & \tau=0\\
(1,\,-\alpha)=(1,1)_{\alpha=-1}, & \tau=1
\end{cases}\label{eq: T_pm(b)}
\end{eqnarray}
\sublaboff{equation}

\noindent for any value of adaptation $\alpha$. The homeostatic balancing
condition 
\begin{equation}
\iota(T)=\alpha(T),\label{eq: iota=00003Dalpha}
\end{equation}
--- where the thermodynamic irreversibility \sublabon{equation}
\begin{equation}
\iota=\frac{T-T_{c}}{T_{h}-T_{c}},\quad\alpha~\mbox{const},\label{eq: 2-phase_irrever(a)}
\end{equation}
is formally equivalent to the quality of a two-phase mixture%
\footnote{Irreversibility can be interpreted \citep{Sengupta2010-c} as the
\textit{quality} of a two phase liquid-vapour mixture of selfish individualism
and altruist collectivism. Holism is a two-phase mixture of ``capital''
and ``culture'', interlocked in a feedback confrontation, as are
the more than 200 organs in the human body, acting for and on behalf
of the whole with the individual organs surviving no longer than the
body with the later only their collective phenotypic expression. Collectivism
is not an after-thought by-product of individualism: it lies beyond
the passivity of the ``invisible hand'' as an equal player of the
win-win ``game of life''. %
} 
\begin{equation}
x=\frac{v-v_{f}}{v_{g}-v_{f}},\quad T~\mbox{const},\label{eq: 2-phase_irrever(b)}
\end{equation}
\sublaboff{equation}both $T$ and $v$ being thermodynamic intensive
properties with $v_{f}\le v\le v_{g}$ --- defines the most appropriate
equilibrium complexity conditions \sublabon{equation}
\begin{eqnarray}
\theta_{\pm} & = & \frac{(1+\tau)\pm(1-\tau)\sqrt{1+4\tau}}{2(2-\tau)},\quad\tau=\frac{T_{c}}{T_{h}},~\theta=\frac{T}{T_{h}}\label{eq: T_plus(a)}\\
 & = & \begin{cases}
(0.5,0), & \tau=0\\
(1,1), & \tau=1,
\end{cases}\label{eq: T_plus(b)}
\end{eqnarray}
\begin{eqnarray}
\iota_{\pm} & = & \frac{1-2\tau\pm\sqrt{1+4\tau}}{2(2-\tau)}\label{eq: iota(+/-)}\\
 & = & \begin{cases}
(0.5,0), & \tau=0\\
\pm\frac{1}{2}(\sqrt{5}\mp1), & \tau=1
\end{cases}
\end{eqnarray}
\sublaboff{equation}that determines the irreversibility of the interaction
as the complex holistic state $T_{+}$, such that any tendency to
revert back to the original condition (small $\iota$: predominance
of pump $P$) implies large $E$-$P$ adaptability $\alpha$ inviting
$E$-opposition and the homeostasy of Eq. (\ref{eq: iota=00003Dalpha}).
Note that at $T_{c}=0$, $T_{\text{\textminus}}=T_{c}$ while at $T_{c}=T_{h}$,
$T_{+}=T_{\text{\textminus}}=T_{c}$. Biologically,  irreversibility
$\iota$ corresponds to the selection coefficient $s$ of the fraction
of a transformed resource (difference in temperature, specific volume)
in terms of the available quantity, where the favoured template phenotype
represents the reversible state $\iota=0$ of no change. Observe that
$\iota$ represents entropic degrading fraction $\mathbb{W}_{+}/(\mathbb{W}_{+}\oplus\mathbb{W}_{-})$,
and $(1-\iota)$ corresponds to the exergic reciprocal $\mathbb{W}_{-}/(\mathbb{W}_{+}\oplus\mathbb{W}_{-})$
of life's collaborative antagonism which accordingly represents the
entropic decrease in biological fitness. The quasi-static value $\iota=0$
at no loss typifies the maximum theoretical capacity/potential of
the organism; of this, life's inevitable wear-and-tear in staying
alive entails the cost of a resulting fitness. Complexity is a holistic
expression of \emph{individual selfishness} and \emph{collective altruism}
of concentration and dissipation; to reflect this duality we propose
the following measure 
\begin{eqnarray}
\sigma_{\textrm{{C}}} & = & \sqrt{\iota(1-\iota)}\label{eq: CIndex}
\end{eqnarray}
of complexity where $\iota$ is as in Eq. (\ref{eq: iota(+/-)}).
It is to be noted that this $\mathbb{W}_{+}$-entropic irreversibility
is indeed larger than 1/2 as it perhaps ought to be, but is not so
large as to relegate the growth contribution due to $\mathbb{W}_{-}$
to insignificance.

Of fundamental importance is the fact that the roots Eq. (\ref{eq: T_pm(a)})
form continuous curves in the regions $0\le T_{c}\le T_{h}$ --- meaningful
only for $T_{h}\rightarrow+\infty$ --- bifurcating as individual
holistic components at $\alpha_{\pm}=\pm1$: at these values the continuous
curves disengage from each other as separate linear entities before
``collaborating'' once again in generating the profiles $T_{\pm}$
in the complex region $(T_{c},T_{h})$. These adaptations of the engine-pump
are substantive in the sense that the specific $\alpha$-values denote
physical changes in the global behaviour; they are the critical and
triple points in Fig. \ref{fig: 2-phase}. The two-phase complex surface
denoted by $\alpha=\iota$ is to be distinguished from the general
$Pv$ region $\alpha=\eta\zeta$. Since the participatory universe
satisfies a more involved nonlinear equation (\ref{eq: alpha(ii)})
compared to the simple linear relationship of an ideal gas, diagram
\ref{fig: 2-phase}(ii) is more involved than the corresponding (i),
with the transition at the triple point $\alpha_{+}=1$ displaying
very definite distinctive features. While (ii) clearly establishes
that the triple point cannot be accessed from the $\iota=\alpha$
surface and requires a detour through the general $\alpha=\eta\zeta$,
it also offers a fresh insight on the origin of the insular nature
of the absolute zero $T=0$. 

Given a $(T_{c},T_{h})$ i.e. the environment, the holistic homeostasis
$T_{+}\triangleq T$ defines \textit{sustainability,} and management
of the triad capital-environment-culture in the sense implied by Fig.
\ref{fig: 2-phase}(c) in maintaining $\lambda\in(3,\lambda_{*})$
within the defining window of a particular $2^{N}$-cycle, constitutes
\textit{sustainable development,} Fig. \ref{fig: mitosis}. As has
been noted ``what sustainability is, what its goals should be, and
how these goals are to be achieved are open to interpretations'',
with an accepted definition of the term remaining ``elusive because
it is expected to achieve many things''; ``in essence, sustainable
development is a process of change in which the exploitation of resources,
the direction of investments, the orientation of technological development,
and institutional change are all in harmony and enhance both current
and future potential to meet human needs and aspirations'' \citep{Brundtland1987}.
The idea of sustainable development, considered a self-contradictory
oxymoron because \textit{development} appears antithetical to environmental
\textit{conservation,} is merely an expression of the inevitability
of the entropic consequence of the second law even as ``sustainability
is improving the quality of human life while living within the carrying
capacity of supporting eco-systems'' \citep{IUNC1991}. The resolution
of this paradox of adversarial and antagonistic collaboration of opposites
is based on the tools of complex holism summarized in Fig. \ref{fig: ChaNoXity-a}:
we believe that an adequate understanding of the dialectics involved
is essential for a proper formulation of policies and programmes in
the stressed, far-from-equilibrium reality that lies beyond the simplicity
of linear reductionism.

\subsection{Two-Phase Mixture of Individualism and Collectivism}

Equation (\ref{eq: alpha(ii)}) and Fig. \ref{fig: 2-phase}(ii) show
that the 2-phase individualistic-collective ``liquid-vapour'' region
$\iota=\alpha$ is distinguished by the imposed constancy of $\alpha$
--- and hence of the product $Pv$ --- just as $P$ and $T$ separately
remain constant in Fig. \ref{fig: 2-phase}(i). At the critical point
$v_{f}=v_{g}$ for passage to second order phase transition, $T_{c}=T_{h}$
requires $T_{+}=T_{-}$ which according to Eqs. (\ref{eq: T_pm(b)})
and (\ref{eq: T_plus(b)}) can happen only at $\alpha_{-}=-1$ corresponding
to the $(P_{\textrm{cr}},T_{\textrm{cr}})$ of figure (i). At the
other unique adaptability of $\alpha_{+}=1$ at $T_{c}\rightarrow0$,
the system passes into region (IV) from (III) just as (I) passes into
(II) as $T_{c}\rightarrow T_{h}$ at $\alpha_{-}$. Observe from Eq.
(\ref{eq: T_pm(a)}) that \sublabon{equation} 
\begin{equation}
(T_{c}\rightarrow0)\Longleftrightarrow(T_{h}\rightarrow\infty)\label{eq: reciprocal_a}
\end{equation}
allows the self-organizing complex phase-mixture of collaboration
and competition to maintain its state $T$ as the condition of homeostatic
equilibrium%
\footnote{Does the melting of the arctic icebergs and the recent severe blizards
in Europe and USA indicate the veracity of Eq. \ref{eq: reciprocal_a}?%
} when $(T_{+},T_{-})=(0,0)_{\alpha_{+}=1}$. Simultaneously however,
because $T_{c}<T_{h}$,
\begin{equation}
(T_{c}\rightarrow T_{h})\Rightarrow(T_{h}\rightarrow\infty)\label{eq: reciprocal_b}
\end{equation}
 implies from Eq. (\ref{eq: T_pm(b)}) that $(T_{+},T_{-})=(T_{h},T_{h})_{\alpha_{-}=-1}$
is also true. Hence 

\begin{equation}
(\alpha_{+})_{T_{c}=0}\sim(\alpha_{-})_{T_{c}=T_{h}}\label{eq: recirocal_c}
\end{equation}
generates the equivalence 
\begin{equation}
(T_{+}-T_{-})_{\alpha_{+},\, T_{c}=0}=(T_{+}-T_{-})_{\alpha_{-},\, T_{c}=T_{h}}\label{eq: reciprocal_d}
\end{equation}
\sublaboff{equation}providing an interpretation of the simultaneous
validity of Eqs. (\ref{eq: reciprocal_a}, \textit{b}). The limiting
consideration (\ref{eq: reciprocal_a}) leaves us with two regions:
(I) characterized by $\iota\alpha>0$ of the complex real world $\mathbb{W}_{+}$
and (IV) of $\iota\alpha<0$ of the negative world $\mathbb{W}_{-}$.
The three phases of matter of solid, liquid and gas of our perception
manifests only in $\mathbb{W}_{+}$, the negative world not admitting
this distinction is a miscible concentrate in all proportions. The
reciprocal implications (\ref{eq: reciprocal_a}-\textit{ d}) at the
big-bang degenerate singularity $\alpha_{+}=+1$ at $t=0$ \citep{Sengupta2010-a},
instantaneously causes the birth of the $(\mathbb{W}_{+},\mathbb{W}_{-})$
duality at some \textit{unique admissible} value of $\alpha$ for
$0<T_{c}<T_{h}$ and complexity criterion $\iota=\alpha$, breaking
the equivalence $\alpha_{_{+}}\sim\alpha_{-}$ of Eq. (\ref{eq: recirocal_c}). 

The correspondence between dynamics of the engine-pump system and
the logistic map $\lambda x(1-x)$, with the competitive backward-direct
iterates $f^{i}(x)$ corresponding to the ``pump'' $\mathbb{W}_{-}$
and the collaborative, forward-inverse iterates $f^{-i}(x)$ to the
``engine'' $\mathbb{W}_{+}$ --- assured by the exclusion-inclusion
topologies and direct-inverse limits --- constitutes the basis of
our analysis. Note from Fig. \ref{fig: 2-phase} that the two-phase
complex region $\lambda\in(3,\lambda_{*})$, $T\in(T_{c},T_{h})$,
$\iota\in(0,1)$ is the outward manifestation of the tension between
the regions (I), (III) of $\iota\alpha>0$ on the one hand and (II),
(IV) of $\iota\alpha<0$ on the other: observe that at the environment
$T_{c}=(0,T_{h})$ the two worlds merge at $\alpha_{\pm}=\pm1$ bifurcating
as individual components for $0<T_{c}<T_{h}$. The logistic map ---
and its possible generalizations --- with its rising and falling branches
denoted $\female\left(\downarrow\right)$ and $\male\left(\uparrow\right)$
constitutes a perfect example of an elementary \textit{nonlinear qubit},
not represented as a (complex) linear combination: nonlinear combinations
of the branches generate the evolving structures, as do the computational
base $(1\;0)^{\textrm{T}}$ and $(0\;1)^{\textrm{T}}$ for the linear
qubit. This qubit can be prepared efficiently by its defining nonlinear,
non-invertible, functional representation, made to interact with the
environment through discrete non-unitary time evolutionary iterations,
with the final (homeostatic) equilibrium ``measured'' and recorded
through its resulting complex structures. 

\begin{figure}[!tbh]
\begin{centering}
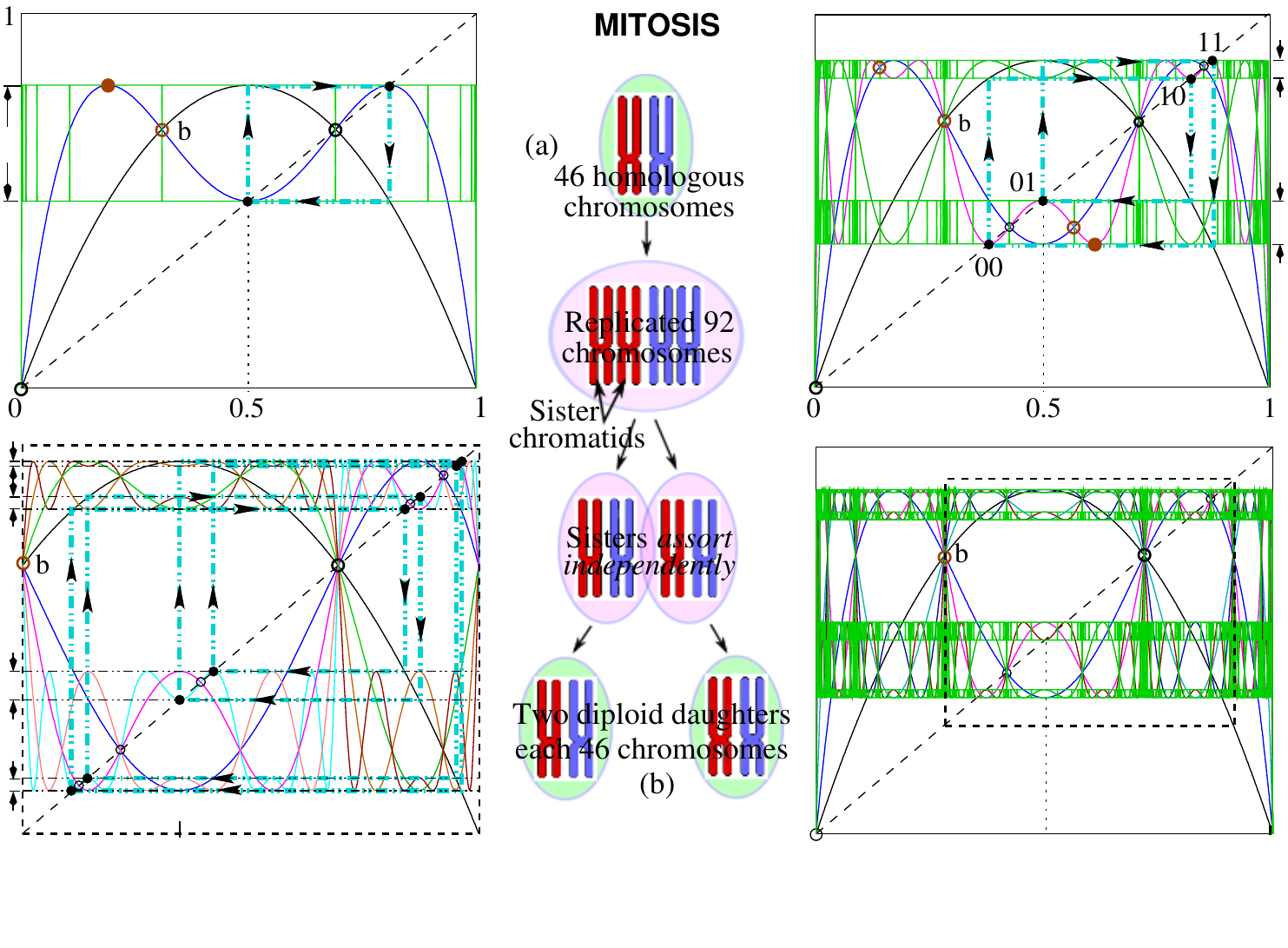
\par\end{centering}

\caption{{\small \label{fig: mitosis}}\textbf{\small Mitosis and Complex Evolution.}{\small{}
The effective nonlinearity $0\le\chi\le1$ of the logistic }\textit{\small nonlinear
qubit}{\small{} $f_{\lambda}(x)=\lambda x(1-x)$ in the representation
$f_{\lambda}(x)=x^{1\text{\textminus}\chi}$ increases with $\lambda$,
as the system becomes more holistic with an larger number of interacting
parts of unstable fixed points shown unfilled, the stable filled points
being the interacting, interdependent, components of the evolved pattern.
The resulting holistic patterns of one, two $\cdots$ are entangled
manifestations of these observables, none of which can be independently
manipulated outside of the collaborative whole. Collaboration of the
direct iterates $f_{\lambda}^{i}(x)$ of individualism and the inverse
iterates $f_{\lambda}^{-i}(x)$ of collectivism leads to homeostasis
of the graphically converged multifunctions of dynamic equilibrium.
$x_{\textrm{{fp}}}=(\lambda-1)/\lambda$ is a }\textit{\small fixed
point}{\small{} of $f_{\lambda}$. }}
\end{figure}

The labeling of the interdependent, interacting, stable points in
Fig. \ref{fig: mitosis} is in accordance with the following rule.
The interval $[0,1]$ is divided into two parts at $\frac{1}{2}$
with ``0'' corresponding to female $\text{\female}(\downarrow)$
and ``1'' to male $\text{\male}(\uparrow)$; the rationale in assigning
\female{}\textcolor{red}{\large{} }to $\mathbb{W}_{-}\left(\downarrow\right)$
and \male{} to $\mathbb{W}_{+}\left(\uparrow\right)$ being that emergence
of new structures is anchored in $\text{\female}(\downarrow)$ in
competitive-collaboration with self-organization in $\text{\male}(\uparrow)$,
through genetic variation, sexual reshuffling and natural selection
for biological life, and consumer culture selection and technological
innovation for economic life. At any stage of the iterative hierarchy
generated by the unfilled unstable points, the filled stable points
are labeled left to right according to the prescription that the female
``supply'' curve is positively sloped along the $x$-cost/variation/exergy
axis, while male ``demand'' is of negative slope. The dissipative
second-law $\mathbb{W}_{+}\left(\uparrow\right)$-engine runs as long
as its demand for evolutionary fuel is met by the induced concentrative
anti-second law $\mathbb{W}_{-}\left(\downarrow\right)$-pump with
necessary supply of variation for selection to work on in a causal-anticausal
feedback chain. The ``spin'' of $\mathbb{W}_{+}$ is taken to be
positive$\left(\uparrow\right)$ indicating natural dissipation, that
is inhibited by the unnatural increase in free-energy consequent gravitational
contraction sourced in negative$\left(\downarrow\right)$ $\mathbb{W}_{-}$. 

Hence the symbolic representation in a notation of unsegregated homologues
denoted by $\{h_{1}\parallel h_{2}\}$, segregated homologues by $h_{1}\parallel h_{2}$,
unsegregated sisters by $(s_{1},s_{2})$, and segregated sisters by
$s_{1},s_{2}$ --- a sequence of sisters being separated by a semicolon
; and a square bracket $[\cdots]$ denoting the tensor product blocks
in Fig. \ref{fig: Mendel} --- becomes\sublabon{equation} 
\begin{eqnarray}
N=1 & \mbox{P}_{1} & \{0\left(\downarrow\right)\parallel1\left(\uparrow\right)\}\label{eq: N.eq.1}\\
N=2 & \mbox{F}_{1} & \{00,01\parallel10,11\}\quad\mbox{(Fig. \ref{fig: Punnett-1} (ii)})\label{eq: N.eq.2}\\
N=3 & ?? & \{000,001;010,011\parallel100,101;110,111\}\quad\mbox{(Fig. \ref{fig: mitosis} (iii)})\label{eq: N.eq.3}\\
N=4 & \mbox{F}_{2} & {\normalcolor \{{\color{red}{\normalcolor [0000,0001;0010,0011]}}\,{\normalcolor {\color{blue}{\normalcolor [0100,0101;0110,0111]}}}\parallel}\nonumber \\
 &  & {\normalcolor {\normalcolor }\,\,\,{\color{yellow}{\normalcolor [1000,1001;1010,1011]}}\,{\color{green}{\normalcolor [}{\normalcolor 1100,1101;1110,1111]}}\}\quad\mbox{(Fig. \ref{fig: Punnett-2} (iii)})}\label{eq: N.eq.4}
\end{eqnarray}
\sublaboff{equation}\vspace{-0.2in}

\noindent for the self-organized, emergent levels of Fig. \ref{fig: mitosis},
the $N=4$ signature being that of the tensor product of Fig. \ref{fig: Mendel}.
The homologous units correspond to respective female-male contributions;
thus $\{010\parallel110\}$ and $\{0000\parallel1000\}$, $\{0101\parallel1101\}$
are examples of homologous coupling. However, as pointed out earlier,
the $N=1,\,2,\,3$-cycles of Fig. \ref{fig: mitosis} represent stable
states whereas the dihybrid matrix of Fig. \ref{fig: Mendel} correspond
to sex, meiosis, unstable gametes, and progeny zygotes considered
in Sec. \ref{sub: Meiosis-NegW}. Observe that the homologous components
$\female$, $\male$ lie on either side of the unstable fixed point
$x_{\textrm{{fp}}}=1-1/\lambda$ that yield these components through
bifurcation, and that there are no pure nonlinear sisters because
of the inevitable mixing of parental properties at every stage of
the process; genes replicate in the common background of the other
genes and their interactions, not in isolation. ``However independent
and free genes may be in their journey through generations, they are
very much \textit{not} free and independent agents in their control
of embrionic development. They collaborate and interact in inextricably
complex ways, both with each other, and with their external environment.
$\cdots$ The whole set of genes in a body constitutes a kind of genetic
climate or background, modifying and influencing the effects of any
particular gene'' \citep{Dawkins2006}. 

Observe that the $N=3$ case of (\ref{eq: N.eq.3}) does not correspond
to any progeny level. 

As a definite example $(N=2)$, the ``entangled'' holistic pattern
of Fig. \ref{fig: mitosis}(ii) clearly demostrates that the four
components of Eq. (\ref{eq: N.eq.2}) cannot be decoupled into Bell
states, being itself nonlinearly ``entangled'' rather than separated.
The various operations historically performed on the respective qubits
of the entangled pair to generate dense coding and teleportation $(N=3)$
for example, are not meaningful on the nonlinear holistic entities;
in fact it is possibly not significant to ascribe any specific qubit
to the individual members of the strings in Eq. (\ref{eq: N.eq.2}).
These suggestive differences between linear nonlocality generated
by externalities and nonlinear self-evolved complexity are the hallmarks
of departures of non-dynamic linear processes as exemplified by Punnett
squares and quantum entanglement and complex holism resulting from
the homeostasis of ``offer'' and ``confirmation'' adversaries
of demand and supply. 

Figure \ref{fig: mitosis} is a graphic representation of a possible
hypothetical correspondence of mitosis --- which along with meiosis
are among the most definitive attributes of Darwinian evolution ---
with self-organization and emergence. The unstable haploid gametes%
\footnote{\textbf{Gametes} are reproductive sex cells of haploid set of chromosomes.
The male reproductive sperm cell fuses with the female reproductive
egg or ovum cell to form a diploid fertilized zygote which then develops
into a new organism within the female environment. Gametes being haploid
are needed in the fertilization of a diploid zygote.%
} of the ova and sperm have only one of each of the 23 chromosomes
of the human genome and are, therefore, not stable diploids. Each
unstable {\large $\circ$}{\small{} }in (i), (ii), and (iv) is replaced
by two interlinked stable female $\female\left(\downarrow\right)$
and male $\male\left(\uparrow\right)$ states: the uncoupled \textit{sister}
{\large $\bullet-\circ-\bullet$} units of periodic points in (ii),
for example, can be considered to be replicated, genetically identical,
``daughter'' bifurcations of the respective ``parent'' {\large $\bullet$}
states of (i). The homologous constituents $0$ and $1$ remain on
their respective sides of the unstable fixed point $x_{\textrm{{fp}}}$:
left of $x_{\textrm{{fp}}}$ belongs to the female-pump, right to
male-engine. A stable diploid cell for some $\lambda<\lambda_{2}$,
destabilizes at $\lambda=\lambda_{2}$ to generate the homologous
pairs of 23 male $\male(\uparrow)$ and 23 female $\female(\downarrow)$
chromosomes in (a); each chromosome with one allele for every gene
and typically a repository of 1000 or more genes, occur in $\female-\male$
homologous pairs for every non-sex diploid cell in the body represented
in the figure by the dash-dot combination of the periodic genotypical
states linking the homologous units. 

In (ii), the environment changes sufficiently for $\lambda_{2}$ to
increase to $\lambda_{4}$ destabilizing the stable homologue $\female$($\downarrow$)-$\male$($\uparrow$)
resulting in bifurcation to two stable diploid daughters $00\,,\,10$,
$01\,,\,11$ of 46 chromosomes each. It is important to note the difference
in the dynamics of mitosis without cross-over, Fig. \ref{fig: mitosis},
and meiosis Fig. \ref{fig: meiosis-1}, with cross-over. In the former
the homologous fixed point pairs of $f^{(2n)}(x)$, like the $\{00\parallel10\}$
for example, occur for the same curvature profile of the graph of
$f^{(2n)}$ unlike in the odd-iterate case $f^{(2n+1)}$ of opposite
curvatures, Fig. \ref{fig: meiosis-1}. This significant difference
in the projected dynamics of the mitotic and meiotic cases is interpreted
as suggestive of a crossover of one with respect to the other. 

Periodic cycles are the ``eigenfunctions'' of the iterative \textit{generalized
nonlinear eigenvalue equation} $f^{(n)}(x)=x$ with iteration number
the ``eigenvalue'' $n$; unlike the linear case, however, these
composite cycles are not linearly superposed but appear as emergent,
self-organized, holistic entities. \textit{In this sense complex holism
represents a stronger form of ``entanglement'' than Bell's nonlocality}
While non-locality is a paradoxical manifestation of linear tensor
products, complex holism is a natural expression of the nonlinearity
of emergence and self-organization. Nature uses chaos as an intermediate
step in attaining states that would otherwise be inaccessible to it.
Well-posedness is an inefficient way of expressing a multitude of
possibilities as this requires a different input for every output;
instead nature chooses to express its myriad manifestations through
the multifunctional route \citep{Sengupta2003}. 

The transactional interpretation embodies --- through the ``offer''
and ``confirmation'' waves handshaking to complete an explicitly
nonlocal ``transaction'' --- the philosophy of the Pump-Engine dualism
of chanoxity. In this necessary antagonism between ``capital'' and
``culture'' representing bottom-up individualistic competition,
entropy decreasing order, and concentrative emergence and top-down
collective collaboration, entropy increasing disorder, and dissipative
self-organization, complex holism emerges as a dynamical homeostasis
of the win-win game in which neither participant wins and neither
loses.%
\footnote{In a \textbf{win-win game} all participants profit one way or other;
in conflict resolutions, it aims to accommodate all antagonists.%
}

\subsection{\textsf{\label{sub: Meiosis-NegW}Meiosis and the Negative World}}

\sublabon{figure}
\begin{figure}[!tbh]
\begin{centering}
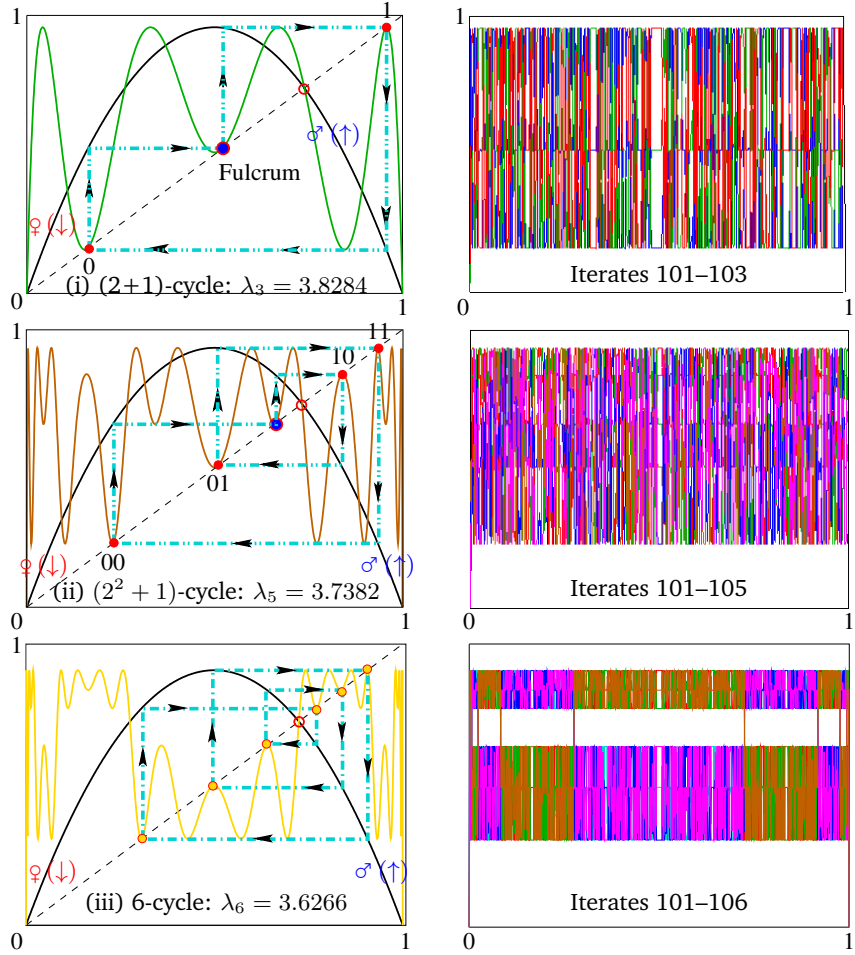
\par\end{centering}

\caption{\textbf{\small \label{fig: meiosis-1}Meiosis and Complex Evolution
(a): Logistics of meiosis.}{\small{} This diagram, the complement of
Fig. \ref{fig: mitosis} for all cycles not of the type $2^{N}$,
appears in the chaotic region $\lambda_{*}<\lambda<\lambda_{3}=1+\sqrt{8}$,
for both odd and even cycles. Odd $(2^{N}+1)$ cycles are especially
significant as the 3- and 5- cycles correspond to the generation of
the haploid gametes as illustrated in Fig. \ref{fig: meiosis-2} below.
The ``Extended Meiosis'' region $\lambda_{5}<\lambda<\lambda_{*}$
is where we believe the embryo and fetus develop leading to birth
at $\lambda_{*}$. Left to right 00, 01, 10, 11 are the gametes produced
in the 5-cycle of (ii). }}
\end{figure}

\begin{figure}[!tbh]
\begin{centering}
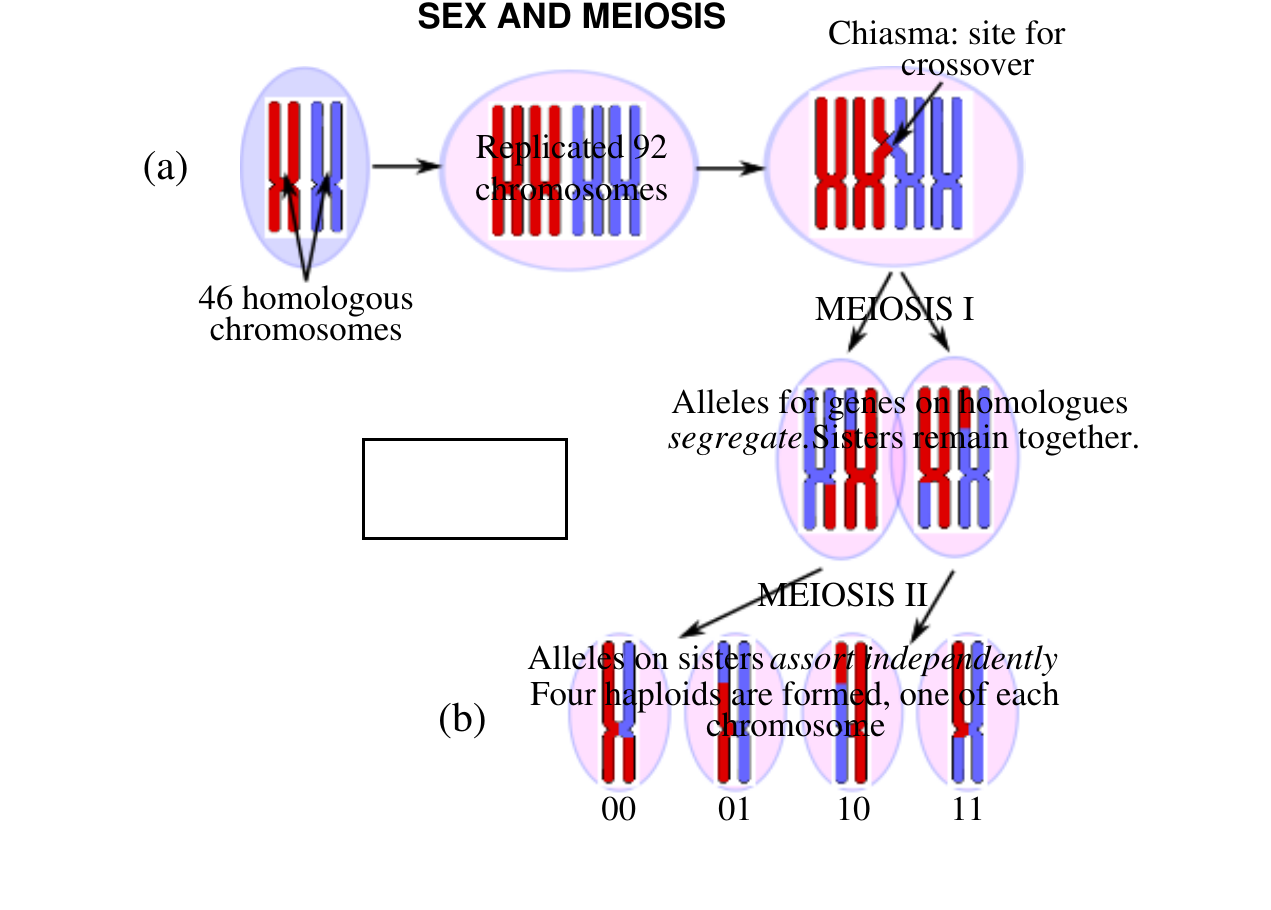
\par\end{centering}

\caption{\textbf{\small \label{fig: meiosis-2}Meiosis and Complex Evolution
(b).}{\small{} How the genes exchange according to Mendelian laws of
segregation and independent assortment of dyhybrid crossing during
meiosis. Let Red denote DOMINANT \textquotedblleft Female\textquotedblright ,
Green recessive \textquotedblleft Male\textquotedblright , replication
playing the role of two different traits like the round and yellow
seeds in Mendel's experiment. Gamete formation takes place in two
stages: MEIOSIS I: }\textit{\small Segregation}{\small . Organism
have two alleles of each gene, one from each parent. These alleles,
on homologous chromosomes, segregate separately into gametes so that
half of the gametes carry one allele and the other half the other
allele. MEIOSIS II: }\textit{\small Independent Assortment}{\small{}
}\textit{\small and cross-over}{\small . Alleles on sister chromatids
assort independently, i.e., segregation of alleles of one gene is
independent of the segregation of alleles of the other gene. Mendel's
Laws are clearly reductionist: in holistic reality independence of
segregation and assortment are unlikely to be valid as the existence
of a }\textit{\small unique}{\small{} fulcrum responsible for homologous
segragation as Fig. \ref{fig: meiosis-1} suggests. }}
\end{figure}
\sublaboff{figure}In this Section, we fill in the gaps in the Female-Male
rivalrous partnership As emphasized earlier, holism of severely stressed,
far-from-equilibrium systems, beyond the ambit of reductionism, does
not in any way negate its eminently successful analytical tools of
mainstream science. These methods --- so very successful under ``normal''
circumstances --- are simply inadequate in the ``revolutionary''
setting of stressed systems where the nonlinearity of mutual feedbacks
are essentially indispensable. This bottom-up synthesis does not falsify
the top-down analysis of reductionism: it merely suggests that there
is a contrapuntal process operating at a higher level, beyond that
of reductionism, as  in quantum theory beyond the dispensation of
classicalism. 

Among the most remarkable features of the appearance of higher forms
of life in $\mathbb{W}_{+}$ involving sexual reproduction is the
incredible self-organization of emergent structure appearing simply
from a fertilized egg, without any ``intelligent design'', except
for residence in the female uterus: although the zygote can be fertilized
outside the body in a ``test-tube'', it has to be transferred back
to the uterus to induce a successful pregnancy. This generation of
the order of life assumes special significance in our context because
order in the entropic world of $\mathbb{W}_{+}$ can occur only through
gravitational coalescence originating in $\mathbb{W}_{-}$ through
the intermediary of an induced pump. The female uterus therefore assumes
an important role of biological order in $\mathbb{W}_{+}$: it is
among the most significant expressions of $\mathbb{W}_{-}$ in $\mathbb{W}_{+}$. 

Period doubling bifurcations of type $2^{N},\, N=1,2,\cdots$, considered
in Fig. \ref{fig: mitosis} were useful in depicting self-organizing,
emergent, stable, diploid homologous states of $\text{\female\,-\,\male}$
origin. Due to the holistic nature of the dynamics involved, the process
is more than simple mitosis, involving the mixing of female-male strategies.
Figure \ref{fig: meiosis-1}, the meiotic counterpart of Fig. \ref{fig: mitosis},
illustrates the generation of gametes in the self-organizational structure
of natural evolution. Note that the 3-, 5-odd cycles of Fig. \ref{fig: meiosis-1}(i),
(ii), and the non-$2^{N}$ even 6-cycle (iii) differ from the 2-,
4-, 8-even cycles of Fig. \ref{fig: mitosis} in having the homologous
combinations separated by the additional fulcrum leading to the modification
of Eqs. (\ref{eq: N.eq.1}, \textit{b}) to represent gametes $0\parallel1$
and $00,01\parallel10,11$. The symbolic representations of Fig. \ref{fig: meiosis-1}
constitutes unstable gametes whereas the units of Fig. \ref{fig: mitosis}
more correctly represent evolved complex organisms, each in order-disorder
homeostasis, at the conclusion of a Meiosis I; this shuffling of the
genetic deck generate daughters that are distinct from each other
and from the parent. 

The meiotic cycles $f^{(i)}\!:i\neq2^{N}$ in Fig. \ref{fig: meiosis-1}\textit{,
b} occur for $\lambda>\lambda_{*}=3.5699457$ of the environment parameter
in the fully chaotic region that can be characterized by $\chi=0$
in the power-law  representation $f(x)=x^{1-\chi}$ where $\chi$,
the effective nonlinearity of $f(x)$, depends on $\lambda$ \citep{Sengupta2010-b};
observe in comparison, that the $2^{N}$ cycles in Fig. \ref{fig: mitosis}
are all in the complex domain $3\le\lambda<\lambda_{*}$. Hence the
time evolution of a natural system can be broken into two components,
the stable complex region $3\le\lambda<\lambda_{*}$ complemented
by the fully chaotic ``quantum'' domain $\lambda_{*}<\lambda\le\lambda_{3}$.
The chaotic region, in equivalence with $2<\lambda\le3$, is the $\mathbb{W}_{-}-\mathbb{W}_{+}$
``skin'': being the boundary of two homologous adversaries these
unstable states that by definition reside in the common neighbourhood
of $\mathbb{W}_{-}$ and $\mathbb{W}_{+}$ connect with the complex
region to generate $\mathbb{W}_{+}$-organizing, $\mathbb{W}_{-}$-emergent
``life''. This genotypic supply of individualistic capital ``gravity''
of $\mathbb{W}_{-}$ is the legacy of the dispersive entropic arrow
of the negative world \citep{Sengupta2010-c} as experienced in $\mathbb{W}_{+}$
of the phenotypes. Creation of life in the uterus --- and only in
the uterus --- from the zygote without any ``intelligent design''
bestows on the female the very special status of an envoy of gravitational
order of $\mathbb{W}_{-}$ in the entropic disorder of $\mathbb{W}_{+}$.
Biological life is supply dominated unlike the economic which is demand
driven. 

The clear distinction between the $2^{N}$ cycles of Fig. \ref{fig: mitosis}
and those of Fig. \ref{fig: meiosis-1} brings out this demarcation
forcefully: while the converged multifunction in the former are $\mathbb{W}_{+}$-stable
horizontal plots for $3\le\lambda<\lambda_{*}$ superimposed on $\mathbb{W}_{-}$-stable
verticals, those in the chaotic domain $\lambda>\lambda_{*}$ are
the more subdued $\mathbb{W}_{+}$-horizontals embedded in a far more
domineering and persistent background of $\mathbb{W}_{-}$-stables,
as illustrated by the converged multifunctions for large times. The
physics of Fig. \ref{fig: meiosis-1} portrays birth in dissipation-dictated
$\mathbb{W}_{+}$, coordinated and controlled by the gravitational
authority of $\mathbb{W}_{-}$: as $4>\lambda\rightarrow\lambda_{*}$,
haploid gametes formed in meiosis are fertilized by the injection
of sperm to provide the foundations of life, that mature, develop
and eventually decay in $\lambda<\lambda_{*}$. 

The value $\lambda_{*}$ of $\lambda$ is of decisive significance.
As observed in \citep{Sengupta2006}, numerical results suggest that
\[
\lim_{N\rightarrow\infty}\chi_{N}\overset{3\le\lambda\rightarrow\lambda_{*}}{\longrightarrow}1
\]
at the critical value of $\lambda_{*}=3.5699456$. Since $\chi=0$
gives the simplest linear relation in the effective power law $f(x)=x^{1-\chi}$\citep{Sengupta2006},
with\sublabon{equation} 

\begin{eqnarray}
\chi & = & 1-\frac{\ln\left\langle f(x)\right\rangle }{\ln\left\langle x\right\rangle },\qquad0\le\chi\le1,\label{eq: chi_1}\\
\left\langle x\right\rangle  & \triangleq & 2^{N}\overset{\lambda=\lambda_{*}}{\longrightarrow}\infty\label{eq: chi_2}\\
\left\langle f(x)\right\rangle  & \triangleq & 2f_{1}+{\textstyle \sum_{j=1}^{N}\sum_{i=1}^{2^{j-1}}}f_{i,i+2^{j-1}},\; N=1,2,\cdots,\nonumber \\
 & = & \{[(2f_{1}+f_{12})+f_{13}+f_{24}]+f_{15}+f_{26}+f_{37}+f_{48}\}\label{eq: chi_3}
\end{eqnarray}
\sublaboff{equation}for $(N=1)$, $[N=2]$, $\{N=3\}$, and
\begin{equation}
\chi_{N}=1-\frac{1}{N\ln2}\,\ln\left[2f_{1}+\sum_{j=1}^{N}\sum_{i=1}^{2^{j-1}}f_{i,i+2^{j-1}}\right]\label{eq: chi_N(c)}
\end{equation}
with 
\begin{equation}
\iota=\alpha=\chi,\qquad\lambda\in(3,\lambda_{*}:=3.5699456)\label{eq: iota=00003Dalpha=00003Dchi}
\end{equation}
in Regions (I), the measure of complexity \citep{Sengupta2006}, the
value $\chi=1$ indicates largest non-linearly emergent complexity
so that the logistic interaction is maximally complex at the transition
to the fully chaotic region. It is only in this region $3<\lambda<\lambda_{*}$
that a global synthesis of stability inspired self-organization and
instability driven emergence lead to the appearance of complex structures. 

What happens for $\lambda>\lambda_{*}$ in the fully chaotic region
where emergence persists for all times $N\rightarrow\infty$ with
no self-organization, indicates that on crossing the chaotic edge,
the system abruptly transforms to a state of effective linear simplicity.
This \textit{jump discontinuity} in $\chi$ demarcates order from
chaos, linearity from (extreme) nonlinearity, and simplicity from
complexity. This emergent but non-organizing region $\lambda>\lambda_{*}$
competes cooperatively with the complex domain $3\leq\lambda<\lambda_{*}$
where irreversibility generates self-organizing useful changes in
the internal structure of the system in order to attain the levels
of complexity needed in the evolution. While the state of eventual
evolutionary homeostasy appears only in $3\leq\lambda<\lambda_{*}$,
the relative linear simplicity of $\lambda>\lambda_{*}$ conceals
the resulting self-organizing thrust of the higher periodic windows
in this region, with the smallest period 3 appearing at $\lambda=1+\sqrt{8}=3.828427$. 

\begin{table}[!tbh]
\begin{centering}
\begin{tabular}{|r|>{\raggedright}m{2.6in}|}
\hline 
$\begin{array}{lclccclcc}
3 & \vartriangleright & 5 & \vartriangleright & \cdots & \vartriangleright & (2n+1)\cdot2^{0} & \vartriangleright & \cdots\\
3\cdot2 & \vartriangleright & 5\cdot2 & \vartriangleright & \cdots & \vartriangleright & (2n+1)\cdot2^{1} & \vartriangleright & \cdots\\
 & \vdots &  & \vdots &  & \vdots &  & \vdots\\
3\cdot2^{m} & \vartriangleright & 5\cdot2^{m} & \vartriangleright & \cdots & \vartriangleright & (2n+1)\cdot2^{m} & \vartriangleright & \cdots
\end{array}$ & $\begin{array}{l}
\mbox{(II): }\mathbb{W}_{-}:\lambda_{*}<\lambda\le\lambda_{3}.\mbox{ \textsc{Meiosis}, Fig. \ref{fig: meiosis-1},\,}b\\
Gonads\!:\mbox{Ovary, Testicle --- gametes.}\\
\mbox{Gamete compatibility. Cycles \ensuremath{\ge}\ 7 not of}\\
\mbox{the type \ensuremath{2^{n}}are unstable and do not occur.}
\end{array}$\tabularnewline
\hdashline$\begin{array}{cccccccccccccccccc}
\\
 & \vdots &  &  &  &  &  & \vdots &  &  &  & \vdots &  &  &  &  &  & \vdots\\
 &  &  &  &  &  &  &  &  & (\lambda=\lambda_{*})
\end{array}$ & $\begin{array}{l}
\mbox{\mbox{Sex}, }\lambda=\lambda_{*}.\, Fallopian\, tube\!:\mbox{\mbox{Ovulation,}}\\
\mbox{Fertilization. Zygote travels up fallopian}\\
\mbox{tube and embeds in the wall of uterus.}\\
\mbox{Pregnancy}.
\end{array}$\tabularnewline
\hdashline$\begin{array}{cccccccccccc}
\\
\cdots & \vartriangleright & 2^{N} & \vartriangleright & \cdots & \vartriangleright & 2^{3} & \vartriangleright & 2^{2} & \vartriangleright & 2 & \vartriangleright\\
\\
\end{array}$ & $\begin{array}{l}
\mbox{(I): }\mathbb{W}_{+}:\,3<\lambda<\lambda_{*}.\mbox{ \textsc{Mitosis}, Fig. \ref{fig: mitosis}}\\
Uterus\!:\mbox{Emergence of life.}\\
Birth\!:\mbox{Self-organization, Complexity.}
\end{array}$\tabularnewline
\hdashline$\begin{array}{cc}
\vartriangleright & 1\end{array}$ & $\mbox{(III)+(IV): }\mbox{ \ensuremath{3\ge\lambda\rightarrow0}\,\ Decease; Death.}$\tabularnewline
\hline 
\multicolumn{2}{|c|}{ Hence: ``Period Three Implies Chaos'' \citep{Li1975} in the sense
``period three implies all periods''.}\tabularnewline
\hline 
\end{tabular} \medskip{}

\par\end{centering}

\caption{{\small \label{tab: Sarkovskii}Sarkovskii ordering of natural numbers
and meosis-mitosis. The period doubling $\lim_{N\rightarrow\infty}2^{N}$
cycles converge at the chaotic limit $\lambda=\lambda_{*}=3.5699456$
from below as do the $\lim_{n,m\rightarrow\infty}(2n+1)2^{m}$ cycles
from above. As noted under $\mbox{(III)+(IV)}$, fixed-points are
non-dynamic and can only lead to degradation. }}
\end{table}

\sublaboff{table}

By the Sarkovskii ordering of natural numbers, there is embedded in
the fully chaotic region a backward arrow that induces connectivity
to lower periodic stability eventually terminating in the period doubling
sequence in $3\leq\lambda<\lambda_{*}$. According to this\textit{
Sarkovskii Theorem}%
\footnote{\textbf{Why Three?:} Why does period \textit{three} imply every other
period, hence chaos? Why is the physical space \textit{three} dimensional?
Why are codon groups of \textit{three} genetic symbols of amino acids
needed as the building blocks of proteins? And why are \textit{three
}components, $\female$, $\male$ and the environment, needed for
life? 

$\quad$Why is the binary system of two symbols inadequate in cases
such as these? %
} \citep{Sharkovskii1964}, if $f\!:\mathbb{R}\rightarrow\mathbb{R}$
is a continuous function having a $n$-periodic point, and if $n\vartriangleright m$
in ordering of all positive integers of Table \ref{tab: Sarkovskii}
then $f$ also has a $m$-periodic point; the ordering starts with
the odd numbers in increasing order, then 2 times the odds, 4 times
the odds, $\cdots$, and ends with the powers of two in decreasing
order, such that every positive integer appears only once in the list.
Sarkovskii's theorem does not entail the stability of all these cycles,
only that they exist, and that there is embedded in the fully chaotic
region a backward arrow that induces chaotic tunnelling to lower periodic
stability, eventually terminating with the period doubling sequence
in $3\leq\lambda<\lambda_{*}$. It is this reciprocal connectivity
between order and disorder that induces subsequent appearance of life
in $\mathbb{W}_{+}$. As in the transactional interpretation reflected
in Fig. \ref{fig: ChaNoXity-a}, this homeostasis is a consequence
of the collaboration between the forward arrow in time of the retarded
\textit{inverse} \textit{iterates} $f^{-i}$ dictated by the Second
Law of increasing entropy and symmetry in $\mathbb{W}_{+}$ and the
backward arrow in time of the advanced \textit{direct} \textit{iterates}
$f^{i}$ dictated by the reciprocal law of decreasing entropy and
symmetry of $\mathbb{W}_{-}$. Should for any reason this natural
$\lambda_{3}\rightarrow\lambda_{*}\rightarrow3\rightarrow0$ $\mathbb{W}_{+}$-forward
arrow be upset, the foundations of homeostasis of life will be disturbed
with resulting fatal consequences including cancer if the $\mathbb{W}_{-}$-specific
$3\rightarrow\lambda_{*}$ gains ascendency in subverting the natural
dialectics of $\mathbb{W}_{-}-\mathbb{W}_{+}$ relationship. 

The basic property that distinguishes meiosis from mitosis is the
crossover of homologous chromosomes in the former resulting in the
production of sperm and egg in the male and female gonads respectively.
The life cycle of Table \ref{tab: life-cycle} can be split into the
four sequential stages. \smallskip{}

\textbf{Stage I.} \textit{Meiosis; Formation of gametes:} Formation
of the gamete sperm and egg haploids in the male and female gonads.
A diploid somatic cell replicates, undergoes cross-over, followed
by segregation of the homologous and sister units in sequence to generate
the haploid sex cells. The sequence of division is important as mitosis
is essentially the segregation of replicated sister units to form
two separate diploid cells. 

\textbf{Stage II.} \textit{Fusion: Gametes$~\mapsto~$Zygotes:} In
the fallopian tube the male and female gametes from Stage I and sex
fuse as the zygote. This stage requires no addditional resources except
the environment of the fallopian and occurs in natural order as an
increase in entropy throughPackage textcomp Info: Setting pxr . The
eggs are born to be fertilized which the sperms hunting for; this
isPackagatural outcome of the coexistence of this predator-prey relationship
that occurs as a natural consequence of the second law. 

\textbf{Stage III.} \textit{Mitosis; Emergence of structures:} Embedding
of the high symmetry zygote in the wall of uterus followed by symmetry
breaking and emergence of structures in the uterus. 

\textbf{Stage IV.} \textit{Birth; Self-organization of emerged structures:}
For the first time in the cycle, outside the vehicle of the organism.
The emerged structures of Stage III undergo extensive self organization
following prolonged mitosis in this stage, finally leading to the
complex holism of human existence. \smallskip{}

Of the first three stages, gamete$\,\mapsto\,$zygote Stage II is
a $\mathbb{W}_{+}$-natural consequence of the predator-prey relationship
obtaining in the fallopian tube and should therefore be reproducible
in an artificailly induced environment, outside the confines of the
organismal vehicle. Stages I and III, by contrast, require active
collaboration of the free-energy of $\mathbb{W}_{-}$ if the gametes
in I and the baby in III are to be produced. These three stages can
hence be considered to collectively generalize to an \textit{extended
meiosis }ending with sister segregation of Stage III --- to include
the emergent fetal mitosis occuring in the uterus. This generalization,
together with Stage IV, completes the self-organized emergence of
gametic holism and life in $\mathbb{W}_{+}$.

\section{\label{sub: General Logistic}Generalized Demand, Supply, Logistic:
The Sink-Source Metaphor }

\begin{flushright}
\textsl{\FiveStarOpenCircled{} The activities of the financial markets
are often irrational. Prices go up for no apparent reason and then
suddenly the mood changes. What's about the latest spasm that has
convulsed bourses in Europe, Asia and North America is that the sell-off
is grounded in real and ever-more pressing concerns. Make no mistake,
something serious is going on here.\\The first cause for anxiety
is the global economy, and in particular the United States. The report
released on Thursday, August 18, by the Philadelphia Federal Reserve
covers only a small part of the Eastern U.S. but it has a good track
record for charting the ups and downs of the world's biggest economy.
The Philly Fed's barometer has just plunged deep into recession territory.
Two-and-a-half years ago, financial markets rallied strongly on the
assumption that the worst of the slump was over. There was relief
that Great depression 2 has been avoided. Now the talk is over a double-dip
recession. $\cdots$ \\At least then governments were in a position
to ride to the rescue. Today, governments are not seen as the solution
but as a part of the problem. What's more, the markets sense that
policymakers have run out of bullets to fire. They can't cut official
interest rates, they find it hard to justify more quantitative easing
when inflation is at current levels and almost every Western government
is currently trying to cut its budget deficit. \\Put all that together
and you get the full package: weak growth, weak banks, weak policy
response. That is not a good recipe for shares. Today's Tokyo Nikkei
market is at less than 25 percent of its level at the peak of the
stock market boom in the late 1980s.\hfill{}}\textsf{\textbf{\small Larry
Elliot}} \citep{Elliot2011}
\par\end{flushright}

\begin{figure}[!tbh]
\begin{centering}
{\renewcommand{\arraystretch}{1.25}%
\begin{tabular}{cc}
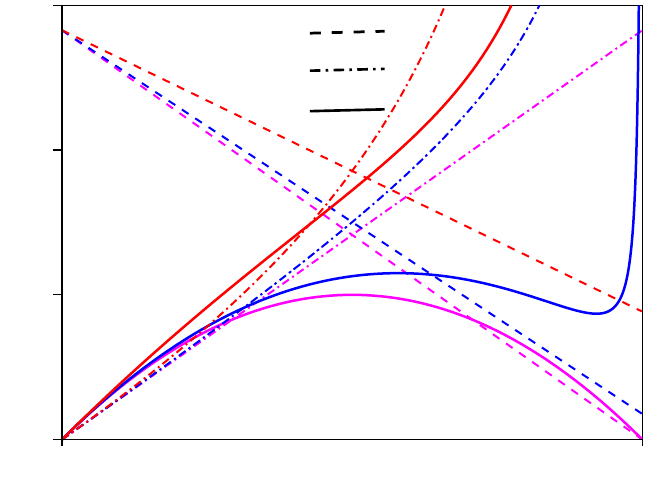 & 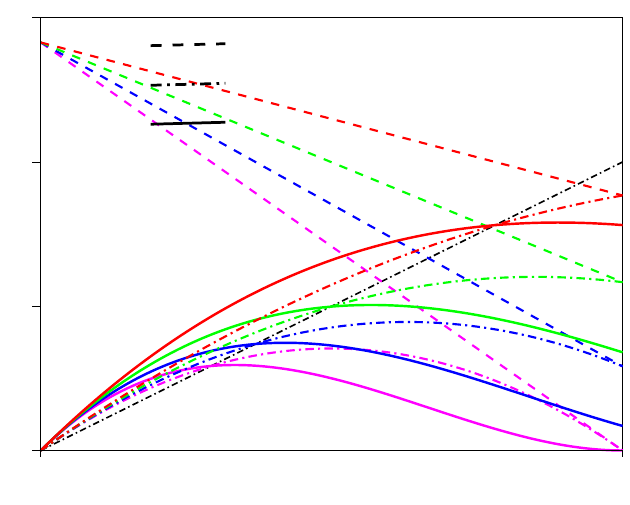\tabularnewline
\multicolumn{2}{c}{$(\iota q)Q=\left\{ \rho-2(1-\tau)\rho^{2}+(1-\tau)^{2}\rho^{3}\right\} Q_{h}^{2}$;$\quad Q_{h}=2,~\tau:=\frac{T_{c}}{T_{h}}\in[0,1)$}\tabularnewline
\end{tabular}}
\par\end{centering}

\caption{\label{fig: demand-supply}{\small Homeostasy of $Q=Q_{h}\left[\iota\left(1-\frac{T_{c}}{T_{h}}\right)+\frac{T_{c}}{T_{h}}\right]$
and $\iota q\triangleq(1-\iota)Q$: notice the collaborative-antagonism
inherent in this inverse balancing condition.} {\small The cubic interaction
$(\iota q)Q$ generates the required $\left(\left\uparrow ,\right\downarrow \right)$
behaviour with respect to the reversibility $\rho$. Supply is $\iota q$,
rather than $q$, because $q\rightarrow\infty$ as $\iota\rightarrow0$,
the condition of hot explosive death at maximum variation/concentration,
(i). This is unacceptable because as the entropic selection vanishes
with $\iota$, the variation free-energy and exergy is expected to
collaborate with this adversary and not defect to unlimited strength.
Thus $\iota q=Q=\left(\frac{T_{c}}{T_{h}}\right)Q_{h}$ iff $\iota=0$
at no evolution. The formal similarity of the generalized logistic
with the Ricker class $x_{t+1}=x_{t}e^{\lambda(1-x_{t}/K)}$ of source-sink
models in population dynamics \citep{Holt1997} is to be noted, the
exponential demand reducing to the linear $\lambda(1-x)$ in the logistic
case. }}
\end{figure}

\noindent To establish the comprehensiveness of this positive-negative
confrontational feedback in Nature, general characterizations of demand
sink and supply source with universal applicability is needed; in
fact ``the economic problem in modern economics is how to reconcile
the tension between scarcity and unsatiated wants'' \citep{Witztum2005}.
From the definitions of irreversibility $\iota$ and adaptibility
$\alpha$, the thermodynamic quantities (see Fig. \ref{fig: ChaNoXity-a})
\sublabon{equation}
\begin{eqnarray}
Q(T)\triangleq\left(\frac{T}{T_{h}}\right)Q_{h} & \!\!\!=\!\!\! & Q_{h}\left[\iota(T)\left(1-\frac{T_{c}}{T_{h}}\right)+\frac{T_{c}}{T_{h}}\right]\label{eq: Demand}\\
\iota q(T)\triangleq(1-\iota(T))Q & \!\!\!=\!\!\! & Q_{h}(1-\iota(T))\left[\iota(T)\left(1-\frac{T_{c}}{T_{h}}\right)+\frac{T_{c}}{T_{h}}\right]\label{eq: Supply}
\end{eqnarray}

\noindent for a given environmental input of ``fuel'' $Q_{h}$ and
adaptivity of the engine and pump $\alpha$, qualify as the generalized
``demand'' and ``supply'' in terms of the reversibility $\rho(T)=1-\iota(T)$.
Then by definition the non-linear cubic interaction
\begin{eqnarray}
(\iota q)Q & \!\!\!=\!\!\! & Q_{h}^{2}\left\{ \rho-2(1-\tau)\rho^{2}+(1-\tau)^{2}\rho^{3}\right\} ,\quad0\le\tau=\frac{T_{c}}{T_{h}}<1\label{eq: GenLog}\\
 & \!\!\!=\!\!\! & (Q_{h}^{2}\theta)\mu\theta(1-\theta),\quad\mu\triangleq\left(1-\tau\right)^{-1},\theta=\frac{T}{T_{h}}\nonumber 
\end{eqnarray}
 \sublaboff{equation}for the environmental parameter $\tau$ defines
a general supply-demand logistics for demand $Q$, supply $\iota q$
and an environmental scaling parameter $0\le\tau:=T_{c}/T_{h}<1$.
Pulliman \citep{Pulliam1988} for example argues that for many populations
a large fraction of individuals may occur in ``sink'' habitats,
where reproduction is insufficient to balance local mortality being
locally maintained by immigration from more-productive ``sources''
nearby: ``If this is commonly the case for natural populations, I
maintain that some basic ecological notions concerning niche size,
population regulation, and community structure must be reconsidered'',
invoking a BIDE --- Birth-Immigration-Death-Emigration --- discrete
group-theoretic habitat model for its exploration. While the supply
correspondence $S\Leftrightarrow\iota q(T):=\alpha(T)Q_{h}$ in this
positive-negative, auto-feedback loop is fairly obvious, the demand
analogy with $Q(T):=(T/T_{h})Q_{h}$ follows because the confrontation
of $Q$ and $\iota q$ bestows on the former a collective demand that
is met by individualistic supply $\iota q$ in a bidirectional loop
that sustains, and is sustained by each other, in the overall context
of the whole. This collective and collaborative consumer demand induces,
preserves, and nourishes the individualistic competitive supply $\iota q$
that constitutes the capitalist base of the firm: economic life is
demand induced. A complex holistic system is distinguished by the
fact it tends to maximise its survivality against second law entropic
cold death through homeostasy of competitive-collaboration. The plot
of this interaction between demand-supply, offer-confirmation, selection-mutation
is shown in Fig. \ref{fig: demand-supply} for different ratios $(T_{c}/T_{h})\in[0,1)$
that plays the role of the environmental parameter $\lambda$ in the
logistic interaction. It is reassuring to observe that the ``logistic''
obtains only if $0\le T_{c}<T_{h}$ --- the natural range for complex
life --- with $\iota qQ$ reducing to the linear $(1-\iota)$ at $T_{c}=T_{h}$
and increasing monotonically thereafter in $T_{c}>T_{h}$. 

For the precise correspondence, recall from Fig. \ref{fig: 2-phase}(iv)
that the product \sublabon{equation} 
\begin{equation}
\iota\alpha=\mu\theta(1-\theta)\quad\mu=(1-\tau)^{-1},\;\theta=T/T_{h}\label{eq: iota*alpha}
\end{equation}
as a standard logistic, $\alpha:=q/Q_{h}$, $T_{h}Q=Q_{h}T$, defines
the complete world $\mathbb{W}=(\mathbb{W}_{-})_{\iota\alpha<0}\oplus(\mathbb{W}_{+})_{\iota\alpha>0}$
in accordance with the sign of $\iota\alpha$. Hence our thermodynamic
logistic $\iota\alpha$ will reproduce the dynamic standard logistic
iff 
\begin{equation}
\frac{T_{c}}{T_{h}}=1-\frac{1}{\lambda},\label{eq: thermo-logistic(0)}
\end{equation}
a remarkable result that defines the thermodynamic environment in
terms of the fixed point $x_{\textrm{{fp}}}$ of the logistic map.
Finally,

\begin{equation}
\iota\alpha=\iota qQ\label{eq: thermo-logistic(1)}
\end{equation}
requires the heat extracted from the hot source $T_{h}$ to be dynamically
restricted to 
\begin{equation}
Q_{h}^{2}(\theta)=\theta^{-1},\quad Q_{h}(T)=\sqrt{\frac{T_{h}}{T}}\label{eq: Q_h}
\end{equation}
 --- a reminder of the (cause$\,\Leftrightarrow\,$effect) causality.
Clearly, this sustainable $T$ is that of Eq. (\ref{eq: T_pm(a)}).
The change of the independent variable from temperature $\theta$
in $\iota\alpha$ to fitness $\overline{W}(T)=\mu(1-\theta)=1-\iota(T)$
for $\iota qQ$ is a significant transition. Apart from rendering
the heat extracted from the hot source into a dynamical variable in
Eq. (\ref{eq: Q_h}), it downgrades the significance of reversibility/fitness/cost
in the dynamics of sustainable evolution. This prompts us, following
Fig. \ref{fig: ChaNoXity-a}, to consider the ratio of supply and
demand $\iota q/Q$ of the fitness $(1-\iota)$ as the sustainability
index that ensures continued supply of resources demanded by dissipation,
degradation and consumption in homeostatic dynamic equilibrium  defined
by evolution of the product $\iota qQ$, Eq. (\ref{eq: GenLog}).%
\footnote{\emph{Sustainable Development} is a pattern of growth in which resource
use aims to meet human needs while preserving the environment so that
these needs can be met not only in the present, but also for generations
to come, in satisfying ``the needs of the present without compromising
the ability of future generations to meet their own needs.'' \citep{Brundtland1987}%
} 

As a companion of Eq. (\ref{eq: CIndex}) of the complexity index
which required only the definitions of $\iota$ and $\alpha$, the
\emph{sustainability} \emph{index} is expected to depend also critically
on both the dynamics and thermodynamics of the processes, as in Eqs.
(\ref{eq: iota*alpha}, \ref{eq: thermo-logistic(0)}, \ref{eq: thermo-logistic(1)}).
Infact Eq. (\ref{eq: thermo-logistic(1)}) combined with Eq. (\ref{eq: iota=00003Dalpha})
can be presumed to represent sustainable holism with respect to%
\footnote{It is instructive to compare this with the Human Development Index
(HDI) of United Nations Development Programme in \emph{Human Development
Report 2007/2008: }the $\mbox{Dimension Index}=(\mbox{Actual Value}-\mbox{Minimum Value})/(\mbox{Maximum Value}-\mbox{Minimum Value})$,
from which the HDI is obtained as a simple linear average, is of the
form of thermodynamic ireversibility $\iota$. Our entropic measure
(\ref{eq: CIndex}) is of course vastly different from the numerically
averaged HDI, significantly it includes the growth component $(1-\iota)=\mbox{(\mbox{Maximum Value}-\mbox{Actual Value})/(\mbox{Maximum Value}-\mbox{Minimum Value})}$
that HDI does not. %
} \sublabon{equation}
\begin{eqnarray}
\theta & = & \frac{(1+\tau)+\sqrt{1+2\tau+9\tau^{2}-4\tau^{3}}}{2(2-\tau)}\label{eq: SI-1}\\
\iota & = & \frac{(1-\tau)(1-2\tau)+\sqrt{1+2\tau+9\tau^{2}-4\tau^{3}}}{2(2-\tau)(1-\tau)}\label{eq: SI-2}
\end{eqnarray}
\sublaboff{equation}with $\iota$ as of (\ref{eq: SI-2}), compare
Eq. (\ref{eq: T_plus(a)}, \ref{eq: iota(+/-)}).

These general conclusions are supported by investigations on evolutionary
stability of ``sink'' populations by Holt \citep{Holt1997}. By
identifying the exponential term in the discrete Ricker model $x_{t+1}=x_{t}e^{\lambda(1-x_{t}/K)}$
--- where $\lambda$ is the intrinsic growth rate and $K$ the carrying
capacity --- as the density dependent fitness $x_{t+1}/x_{t}$, the
demand-fitness function $W(x)=\exp[\lambda(1-x/K)]$ is taken to characterize
the ``source'' $x_{t}e^{\lambda(1-x_{t}/K)}$ in the \textit{source-sink
dynamics} of the model that is known to be stable for $\lambda<2$,
periodic for $2<\lambda<3.102$, and chaotic when $\lambda>3.102$.
Numerical investigations \citep{Holt1997} suggest that the evolutionary
persistence of sink populations (characterized by a constant fitness),
is a generic feature of systems with unstable dynamics in source habitats:
temporal variations in fitness in source favours the utilization of
sink habitats, concentrating individuals into the source of larger
growth rates. The qualitative prediction that species with locally
unstable population dynamics should tend to evolve generalized adaptations
across many habitats rather than specialization to any particular
one is manifest in the competitive-collaboration between the antagonistic
source and sink --- the source habitat with typical fitness $>1$
as the originator of order and emergence reflects $\mathbb{W}_{-}$:
``if numbers are not to increase in an unbounded fashion at high
densities'' the constraint of declining fitness suggests the logistic-like
Ricker model for the negative world in Holt. The sink habitat with
a long-term geometric growth-rate $<1$ reflects $\mathbb{W_{+}}$;
``this population ineluctably declines to lower densities and eventually
faces extinction'' by the inexorable entropic mandate of $\mathbb{W}_{+}$.
By contrast we treat the ``sink'' $\mathbb{W}_{+}$ on an equal
footing with ``source'' $\mathbb{W}_{-}$, the pair of antagonistic
cooperators generating the complexity of sustainable competitive-collaboration.

Modern individualistic, neo-classical mainstream economics, is an
orthodox static Newtonian equilibrium theory, where supply by the
firm equals the demand of the consumer in the mutation-selection balance
$sq^{2}=\mu_{+}$ \citep{Hamilton2009} where $s$ the selection coefficient
that models thermodynamic irreversibility $\iota$, is a measure of
the extent to which natural selection reduces the relative contribution
of a given genotype to the next generation, and $\mu_{+}$ is the
forward mutation rate of the allelic change $\mathbf{A}\overset{\mu_{+}}{\rightarrow}\mathbf{a}$,
$q$ being the frequency of the recessive allele $\mathbf{a}$. Linear
stability is central to this model that has come under severe strain
in recent times \citep{Bouchaud2008,Buchanan2009,Farmer2009,Ho2004,Shapiro2009}:
mainstream economics is concerned with the economy at a given point
in time, not in long-term development. The dynamics of historical
change of economics was alien to Adam Smith. The linear mathematics
founded in calculus with maximization and contraint-based optimization
seeking to maximize utility for the consumer and profit for firms
work with reasonable justification as long as its axioms of linearity
of people with rational preferences acting independently with full
and relevant information make sense. This framework of rationality
of economic agents of individuals or company working to maximize own
profits, of the invisible hand trickle-down effect transforming the
profit-seeking motive to collective societal benefaction, and of market
efficiency of prices faithfully reflecting all known information about
assets \citep{Bouchaud2008}, are relevant under severely restrictive
conditions. In reality, markets are rarely efficient, humans tend
to be over-focused in the short-term and blind in the long-term, and
errors get amplified, ultimately leading to collective irrationality,
panic and crashes. ``This picture suggests that by allowing individuals
to pursue their own interests, assuming that they are all rational,
the outcome of this decentralized system in which the government is
only needed to uphold law and order is both productive and allocative
efficient'' \citep{Witztum2005}, allocative efficiency implying
that the most preferred bundle for the consumer who represents social
preferences would be guaranteed by competitive pressures of the ``free-reign
of rational self-interest'' alone. Nevertheless, a careful analysis
of the pathway toward this conclusion demonstrates that the picture
of simultaneous maximization of utility and profit is obtained by
superimposing downward sloping demand characteristics onto the upward
sloping supply: how much to sell depends on the demand which reciprocally
depends on what is available. Social \emph{equity} as distributive
justice is consumptive hence entropically dissipative, profit is productive
hence generative; one without the other of collective pluralism, however,
is impotent on its own. The rational self-interest of the opposing
groups inhibit --- not reinforce --- each other to a meaningfully
useful eventuality. 

Several generations of economists ``have spent the last century elaborating
a system of thought that tries to explain the intricate relationships
of economic life with concepts invented to describe the motion of
planets. Because the intellectual superstructure of modern Western
economics was erected on the foundations of Newtonian physics, it
has become untenable'' \citep{Rothschild1995}. Free markets are
wild markets. Surprisingly, classical economics has no framework through
which to understand ``wild markets'' \citep{Bouchaud2008}. These
``perfect world'' models successfully forecast a few quarters ahead
in normal times but fail in the face of revolutionary changes \citep{Farmer2009},
as long as the influences independent of each other, the future is
derived from the past with no detectable backward arrow. But the recent
collapse, financial and societal, signals a systemic meltdown in which
interwined breakdowns have destabilized the system as a whole: ``there
has been a massive failure of the dominant economic model'' \citep{Bouchaud2008}.
Unable to explain ``the awesome complexities of real economic life
as experienced by workers and businesspeople, where history matters
and change is constant but largely unpredictable, Western economists
have barricaded themselves inside their obtuse mathematical models''
\citep{Rothschild1995}.

The competitive-collaboration of the engine and its self-generated
pump is identified as the tension between the consumer with its dispersive
collective spending engine (collaborative ``culture'') in conflict
with the individualistic resource generating pump (competitive ``capital'')
in mutual feedback cycles, attaining market homeostasis not through
linear optimization and equilibrium of intersecting supply-demand
profiles, but through nonlinear feedback loops that generate entangled
holistic structures: supply and demand in human societal metabolism
are rarely independent of each other. To take this into account, the
interactive feedback between the opposites of engine consumption and
pump production can be modelled as a \textit{product} of the supply
and demand factors that now, unlike in its static manifestation of
neo-classicalism, will evolve in time to induce dynamic equilibrium. 

What is nature's analogy of stressed, far-from-equilibrium thermodynamics
of open systems? The nonlinear, holistic ``mutation-selection balance''
of {\small exergy}-entropy confrontation depicted in Fig. \ref{fig: ChaNoXity-a}
suggests that economic profit (surplus) \sublabon{equation} 

\begin{equation}
\Pi(Y)=R-C(Y)\qquad(TS)\label{eq: profit-1}
\end{equation}

\noindent as the surplus value of reversible Carnot revenue $R$ and
total investment free-energy $C$, with $Y$ the output of the economy,
corresponds to irreversibility $\iota(T)$, Fig. \ref{fig: ChaNoXity-a}.
With the specific correspondences 

\begin{eqnarray}
\begin{array}{rcll}
R & = & W_{\textrm{{rev}}}={\displaystyle \left(1-\frac{T_{c}}{T_{h}}\right)}\, Q_{h} & (\mbox{Enthalpy, \ensuremath{U}Fixed)}\\
C(Y) & = & W(T)\triangleq(1-\iota)W_{\textrm{{rev}}}={\displaystyle \left(1-\frac{T}{T_{h}}\right)}\, Q_{h} & (\mbox{Free-Energy,\,\ensuremath{A})}\\
Y & = & T,
\end{array}
\end{eqnarray}
 \sublaboff{equation}$Q_{h}$ being the total infrastructural resources
needed for sustenance of complexity supporting the demand-supply interaction,
the $\iota=\alpha$ holism \citep{Sengupta2010-c} of Fig. \ref{fig: ChaNoXity-a}
is defined by \sublabon{equation} 
\begin{equation}
T=\frac{T_{h}(T_{h}+T_{c})+(T_{h}-T_{c})\sqrt{T_{h}^{2}+4T_{c}T_{h}}}{2(2T_{h}-T_{c})}.\quad(\mbox{Mutation-Selection balance})\label{eq: MutSel_Bal}
\end{equation}
This nonlinear mutation-selection free-energy-entropy balance requires
the very specific $R-C$ relationship 
\begin{equation}
C=\frac{3R-R\sqrt{5-4R}}{2(1+R)}\label{eq: cost}
\end{equation}
that solves the dynamic holistic problem, the profit being \sublaboff{equation}with
any unutilized profit, unavailable for the benefit of the system when
the complex inheritance $T$ of Eq. (\ref{eq: MutSel_Bal}) is not
achieved in (\ref{eq: profit-1}) for $\Pi\triangleq TS$, only increasing
the entropy of the universe, Fig. \ref{fig: ChaNoXity-a}, leading
to decease and implosive cold-death --- the pursuit of greater profit
margins leads to more intensive exploitation of natural resources
which in turn leads to ecological and environmental degradation. Economic
growth is an important component of development, but it cannot be
a goal in itself, nor can it go on indefinitely \citep{IUNC1991}:
holism is Nature's way of reinventing surplus economic energy of profit/benefit
as new information, with the economic selection coefficient $s$ (irreversility
$\iota$) being $\Pi/R$. Adopting the convention that the maximum
entropy forward state of dissipation, degradation, and waste comprises
biologically less fit while its opposite of enforced constructivism,
usefulness and order defines fitness, the significance of this analysis
is that Nature discards the high-entropy \textquotedblleft bad\textquotedblright{}
to make way for the low-entropy \textquotedblleft good\textquotedblright{}
to achieve a holistic mutation-selection balance for revolutionary,
far-from-equilibrium cases beyond Darwinian reductionism, in its dynamical
quest of life. Paradoxically either on its own spells doom: tragic
heat death of the commons --- ``Freedom in a commons brings ruin
to all'' according to Hardin \citep{Hardin1968} writing on population
increase leading to insatiable demand --- or obliteration of cold
death. Indeed, ``the simplest summary of this analysis $\cdots$
is this: the commons is justifiable only under (near-equilibrium)
conditions of low-population density. As the population has increased
(to far-from-equilibrium) the commons has had to be abandoned in one
aspect after another'' \citep{Hardin1968}; only a judicious intermingling
of the opposites can support and sustain Life. Nature is in fact a
delicately balanced nonlinear complex of \textquotedblleft capital\textquotedblright{}
and \textquotedblleft culture\textquotedblright{} representing the
arrows of individualism and collectivism. 

As an example of the destructive potential of the \textquotedblleft stabilize,
privatize, liberalize\textquotedblright{} mantra of ``capital''
sans ``culture'', the debate now is not on whether the broader interpretation
of a neoliberal manifesto enshrined in the Washington Consensus \citep{Williamson2002}
--- seen as a shift from state-led dirigisme to market-oriented policies
with focus on GDP growth --- ``is dead or alive, but over what will
replace'' this ``damaged brand''%
\footnote{The many areas conspicuous by their absence in the Washington Consensus
revolve around significant market failures that markets on their own
cannot repair but require active interventionist policies. They were
not part of Consensus-style agenda because the Consensus relies on
well-functioning markets to solve the development challenges and ``viewed
any state interference in the economy with suspicion'', being ``essentially
contemptuous of equity concerns\textquotedblright{} \citep{Birdsall2010}.
One of the paradoxical consequences of the 2008\textendash 09 financial
crisis has been noted to be that ``Americans and Britons will finally
learn that open capital markets combined with unregulated financial
sectors is a disaster in the waiting. $\cdots$ Implicit in the Reagan-Thatcher
doctrine was the belief that markets were an acceptable substitute
for efficient government. The crisis demonstrated that unregulated
or poorly regulated markets can produce extraordinary costs'' with
the prospect that ``historians may well point to the financial crisis
as the end of American economic dominance in global affairs.'' \citep{Birdsall2011}%
}. Infact, with the Consensus in its broad sense considered to have
survived till the global financial crisis of 2008, ``occasionally,
the reader has to remind himself that the book he is holding in his
hands is not some radical manifesto, but a report prepared by the
seat of orthodoxy in the universe of development policy'', writes
Rodrik \citep{Rodrik2006} of the World Bank\textquoteright s important
marker \emph{Economic Growth in the 1990s: Learning from a Decade
of Reform} \citep{Bank2005}, ``a rather extraordinary document insofar
as it shows how far we have come from the original Washington Consensus.
There are no confident assertions here of what works and what doesn\textquoteright t
\textemdash{} and no blue-prints for policymakers to adopt. The emphasis
is on the need for humility, for policy diversity, for selective and
modest reforms, and for experimentation. $\cdots$ There are contending
interpretations of what has gone wrong and how to move forward. But
the mere fact that such views have been put forward in an official
World Bank publication is indicative of the changing nature of the
debate and of the space that is opening up within orthodox circles
for alternative visions of development policy.'' 

Among these alternative wisdoms, the Beijing Consensus of sustainable
development \citep{Beijing2004} is beginning to replace the ``widely-discredited''
Washington Consensus. In this new strategy, ``innovation to reduce
the friction-losses of reform'', corresponding to our entropy reducing
individualistic pump $P$, is the antagonistic partner of ``chaos-management''
beyond per-capita GDP measures of quality-of-life, sustainability,
equality and equity that represent the adversity of entropy-increasing,
collaborative engine $E$ in ``China\textquoteright s new approach
to development stressing chaos management'' through complexity and
holism, ``driven by a desire to have equitable, peaceful high-quality
growth using economics and governance to improve society''. Thus
Cao Baijun \citep{Baijun2007} summarizing on China's entry in a critical
era of rapid development that has resulted in tensions in its economy
and society, feels that in order to develop China's economy in a sound
and stable manner it is necessary, among others, to deal with the
following: first the focus should shift from the pursuit of economic
growth and pure development to quality economic growth and sustainable
development, second a balance between economic growth and social development
should strive toward establishing a society founded on people's well-being
and commitment to social equality and harmony, and third a resource-saving
society aiming at a balance between nature and man leading to ecologically
sustainable development. A recall of Figs. \ref{fig: ChaNoXity-a},
\emph{b} establishes that these goals comprise a possible implementation
of the collaborative antagonism of individualism and collectivism
on which the holism of sustainable development is founded, with the
fitness $(1-\alpha)$ serving as an index of sustsinability corresponding
to an appropriately defined logistic $\iota qQ$. 

In the dynamical two-phase mixture $T$ portrayed in Eqs. (\ref{eq: rigged},
\ref{eq: T_pm(a)}, \textit{b}) and Fig. \ref{fig: ChaNoXity-a},
the canonical morphisms can be identified with demand $Q$ for the
entropy-increasing arrow, and supply $\iota q$ for entropy-decreasing
feedback in Eq. (\ref{eq: rigged}). Their logistic coupling $\iota qQ$
achieves the proper ``transactional'' interaction via bi-directional
``handshake'' of supply and demand leading to the homeostasis $T$
of $\mathcal{H}$, that for Eq. (\ref{eq: rigged}) translates to
$\int\delta(x-a)f(x)=f(a)$, for $\delta\in\Psi^{\times}$ in $\mathbb{W}_{-}$
and $f\in\Psi$ in $\mathbb{W}_{+}$. The generalized function $\delta(x)$
is a constant in $\mathbb{W}_{+}$ for $x\neq0$ and a constant in
$\mathbb{W}_{-}$ at $x=0$; thus for example the Poisson kernel 
\[
\delta_{\varepsilon}(x)=\frac{\varepsilon}{\pi(x^{2}+\varepsilon^{2})}\underset{\varepsilon\rightarrow0}{\overset{\mathbf{G}}{\longrightarrow}}0\in\begin{cases}
\begin{array}{l}
\mathbb{W}_{+},\quad\left|x\right|>0\\
\mathbb{W}_{-},\quad x=0.
\end{array}\end{cases}
\]
\textit{converegs graphically} to the Dirac delta $\delta(x)$ \citep{Sengupta2003}.
Graphical convergence to multifunctions like $\delta(x)$ enlarges
the space of functions to correspondences for successful portrayal
of chaos and its complex derivatives. 

As a classic example of the demand-supply feedback dynamics in $\mathbb{W}_{-}-\mathbb{W}_{+}$,
consider the Case elementary singular eigenfunction superposition
solution \sublabon{equation} 
\begin{equation}
\Phi(x,\mu)=a(\nu_{0})e^{-x/\nu_{0}}\phi(\mu,\nu_{0})+a(-\nu_{0})e^{x/\nu_{0}}\phi(-\nu_{0},\mu)+\int_{-1}^{1}a(\nu)e^{-x/\nu}\phi(\mu,\nu)d\nu;\label{eq: MonoNeutron(a)}
\end{equation}
of the \textit{linear} monoenergetic neutron transport equation 
\begin{equation}
\mu\frac{\partial\Phi(x,\mu)}{\partial x}+\Phi(x,\mu)=\frac{c}{2}\int_{-1}^{1}\Phi(x,\mu^{\prime})d\mu^{\prime},\qquad0<c<1,\,-1\leq\mu\leq1\label{eq: MonoNeutron(b)}
\end{equation}

\noindent \sublaboff{equation}for $\Phi(x,\mu)\triangleq e^{-x/\nu}\phi(\nu,\mu)$
and $\int_{-1}^{1}\phi(\nu,\mu)d\mu=1$, where 
\[
\begin{array}{l}
\left.\begin{array}{l}
\phi(\mu,\nu_{0})={\displaystyle \frac{c\nu_{0}}{2}\frac{1}{\nu_{0}-\mu}}\\
{\displaystyle \frac{c\nu_{0}}{2}\ln\frac{\nu_{0}+1}{\nu_{0}-1}}=1;
\end{array}\right\} \quad\mid\pm\nu_{0}\mid>1\\
\left.\begin{array}{l}
\phi(\mu,\nu)={\displaystyle \frac{c\nu}{2}\mathcal{P}\frac{1}{\nu-\mu}}+\lambda(\nu)\delta(\nu-\mu)\\
\lambda(\nu)=1-{\displaystyle \frac{c\nu}{2}\ln\frac{1+\nu}{1-\nu}},
\end{array}\right\} \quad\nu\in[-1,1]
\end{array}
\]
and the conjugate Poisson kernel 
\[
\mathcal{P}_{\varepsilon}(x)=\frac{x}{x^{2}+\varepsilon^{2}}\underset{\varepsilon\rightarrow0}{\overset{\mathbf{G}}{\longrightarrow}}\begin{cases}
\begin{array}{l}
{\displaystyle \frac{1}{x}}\in\mathbb{W}_{+},\quad\left|x\right|>0\\
0\in\mathbb{W}_{-},\quad x=0.
\end{array}\end{cases}
\]
converges graphically to the multifunction principal value. Then 
\begin{equation}
\Phi_{\varepsilon}(x,\mu)=a(-\nu_{0})e^{x/\nu_{0}}\phi(-\nu_{0},\mu)+a(\nu_{0})e^{-x/\nu_{0}}\phi(\mu,\nu_{0})+\sum_{i=-N,\neq0}^{N}a(\nu_{i})e^{-x/\nu_{i}}\phi_{\varepsilon}(\mu,\nu_{i})\label{Eqn: DiscSpect}
\end{equation}

\noindent is the discretized spectral solution \citep{Sengupta1984,Sengupta1988}
of Eq. (\ref{eq: MonoNeutron(b)}), where 
\begin{align*}
\phi_{\varepsilon}(\nu,\mu) & =\frac{c\nu}{2}\frac{\nu-\mu}{(\mu-\nu)^{2}+\varepsilon^{2}}+\lambda_{\varepsilon}(\nu)\frac{\varepsilon}{\pi\left((\mu-\nu)^{2}+\varepsilon^{2}\right)}\underset{\varepsilon\rightarrow0}{\overset{\mathbf{G}}{\longrightarrow}}\phi(\nu,\mu)\\
\lambda_{\varepsilon}(\nu) & ={\displaystyle \frac{\pi}{\tan^{-1}(1+\nu)/\varepsilon+\tan^{-1}(1-\nu)/\varepsilon}\left(1-\frac{c\nu}{4}\ln\frac{(1+\nu)^{2}+\varepsilon^{2}}{(1-\nu)^{2}+\varepsilon^{2}}\right)\underset{\varepsilon\rightarrow0}{\overset{\mathbf{G}}{\longrightarrow}}\lambda(\nu)}
\end{align*}
are the regularizations in $\mathbb{W}_{+}$ of the corresponding
singular multifunctions in $\mathbb{W}_{-}$. The excellent numerical
results reported in \citep{Sengupta1988} bear testimony to the fruitfulness
and reality of $\mathbb{W}_{-}-\mathbb{W}_{+}$ collaboration-in-adversity
in structuring the complexities of nature.

\subsection{\noindent Climate Change and Global Warming}

\noindent Equations (\ref{eq: T_plus(a)}), (\ref{eq: CIndex}), (\ref{eq: iota(+/-)})
show that for a given $\tau$ of varying $T_{c}$ and $T_{h}$, $\theta$
and $\iota$ --- therefore the complexity index $\sigma_{\textrm{{C}}}$
--- remains unaltered, although the absolute $T$ --- a more faithful
representation of sustainability --- indeed does change. 

\emph{Climate change} involves change in $\sigma_{\textrm{{C}}}$
through variation of $\tau$ of Eq. (\ref{eq: iota(+/-)}) consequent
unrestrained human and related interferences in the ecology leading
to far-from-equilibrium instability of the environment, forced out
of its naturally defining infinite sink-source characteristics. \emph{Global
warming} is characterized by varying $\tau$ and $\theta$ in the
context of Eq. (\ref{eq: SI-1}, \ref{eq: SI-2}) with the \emph{sustainability}
\emph{index} given, formally as in (\ref{eq: CIndex}), by 
\begin{equation}
\sigma_{\textrm{{S}}}=\sqrt{\iota(1-\iota)}\label{eq: SIndex}
\end{equation}
with more heat needing to be entropically dissipated out of the source
in accordance with (\ref{eq: Q_h}) in order that $T$ remains constant.
Invariance of $T$ under $(T_{c}\rightarrow0)\Leftrightarrow(T_{h}\rightarrow\infty)$
of Eq. (\ref{eq: reciprocal_a}) conflicts with the constancy of $\tau$.
This antagonistic posture of environmental  health can be met by extracting
more heat from the hot source according to Eq. (\ref{eq: Q_h}) maintaining
$\left(T/T_{h}\right)$ constant, in order that the (engine)demand-supply(pump)
relationship $\iota qQ$ evolves as the standard logistic $\iota\alpha$
of Eq. (\ref{eq: iota*alpha}), simultaneously satisfying the complexity
condition $\iota(T)=\alpha(T)$ of Eq. (\ref{eq: iota=00003Dalpha}).
If, however, the non-equilibrium disturbances fail to be restored
to the equilibrium conditions, monitoring the environment through
enforced entropic dissipation and cooling such that $\tau$ remains
constant, non-sustainable global warming with climate change will
inevitably follow. 

Ability to reduce the dialectical dynamics of any living system to
a confrontation between a generalized demand (eg. growing population
and consumerism) and a generalized supply (eg. limited resources of
the earth) and their nonlinear encounter is the basis of a successful
win-win negotiation between the antagonists that makes life possible.

\subsection{\noindent \label{sub: Fixed-Periodic}Fixed and Periodic Points:
The Bottomline}

What is the basis of the unflattering observation of Rothschild \citep{Rothschild1995}
that the ``awesome complexities of real economic life as experienced
by workers and business people'' have led economists to barricade
themselves ``inside their obtuse mathematical models''? The reality
of the simple Mendel's pea plant experiments --- summarized in Sec.
\ref{sec: Conclusion} below --- without the distinguishing features
of entanglements, bears testimony of the efficacy of Punett squares
and utility matrices --- under normal circumstances. In these non-revolutionary
situations the simple dynamics governed by the \textit{fixed} \textit{points}
$f(x)=x$ of (not necessarily linear) mappings suffice: the \textit{best
response mapping}
\[
\beta(s)\triangleq\{\tau:u(\tau,s)=\max_{t\in S}u(t,s)\},
\]
where $u$ is the utility/payoff function (see Sec. \ref{sec: Punnett})
of any finite two-player symmetric game has a Kakutani fixed point
that is a Nash equilibrium for the game, for example. In general the
stable fixed points of the replicator-mutator equations are evolutionary
stable states. In far-from equilibrium, exclusively nonlinear instances,
however, the intricate complexities of the \textit{periodic $p$oints
}
\begin{equation}
f^{(n)}(x)=x,\qquad n>1\label{eq: periodic}
\end{equation}
consisting of the stable fixed points of the $n^{\textrm{{th}}}$
iterate $f^{(n)}$ of $f$ and their limit cycles determine complex
homeostasy in the extended, graphically converged, multifunctional,
limit space: understandably invariants of the future --- as much as
those of the present --- determine homeostasy of an evolving dynamical
system. The $n$ eigenfunctions of Eq. (\ref{eq: periodic}) combine
non-linearly through equivalence classes as in Figs. \ref{fig: mitosis}
and \ref{fig: meiosis-1} to generate the homeostasy of self-organized
altruism from the emergent selfish individualism of the eigenvectors.
This is how altruist collectivism reciprocates selfish individualism
that in turn is sustained by the generosity sourced in the inevitable
entropic dissipation --- meaningful only under the exergic concentration
of $\mathbb{W}_{-}$-individualism --- in $\mathbb{W}_{+}$. Entropic
altruism is natural in $\mathbb{W}_{+}$; exergic individualism emerges
only to sustain it. 

An example of the revolutionary mutation-selection balance is illustrated
in Fig. \ref{fig: ChaNoXity-a}. The equilibrium condition $T$ representing
the graphically converged $2^{N}$ chain-dashed limit cycles of Figs.
\ref{fig: mitosis} and \ref{fig: meiosis-1} is to be compared with
the well-known equilibrium frequency under the fixed-point condition
$\Delta q\equiv q_{t+1}-q_{t}=0$ \citep{Lyubich2001}, with $q$
a sum of the conflicting contributions of decreasing entropic-selection
and increasing exergic-mutation. Then for genotypes $\mathbf{AA}$,
$\mathbf{Aa}$ and $\mathbf{aa}$, with \citep{Hamilton2009} \sublabon{equation}
\begin{eqnarray}
\Delta q_{\textrm{{NS}}} & \hspace{-0.2cm}=\hspace{-0.2cm} & -\frac{s(1-q)q^{2}}{1-sq^{2}}\label{eq: Mut-Sel(a)}\\
\Delta q_{\textrm{{Mut}}} & \hspace{-0.2cm}=\hspace{-0.2cm} & \mu_{+}(1-q),\nonumber 
\end{eqnarray}
the fixed point requirement $\Delta(q_{\textrm{{NS}}}+q_{\textrm{{Mut}}})=0$
of 
\begin{equation}
q_{\textrm{{eq}}}=\sqrt{\frac{\mu_{+}}{\mu_{+}s+s}}\label{eq: Mut-Sel(b)}
\end{equation}
\sublaboff{equation}reduces to the classical result $\sqrt{\mu_{+}/s}$
--- with $s=0$ in the denominator of (\ref{eq: Mut-Sel(a)}) for
the mean relative fitness , Fig. \ref{fig: ChaNoXity-a} --- for $\mathbf{A}\overset{\mu_{+}}{\underset{s}{\rightleftarrows}}\mathbf{a}$.
In comparison the periodic-cycle, complex-holistic, chain-dashed attractors
of Figs. \ref{fig: mitosis} and \ref{fig: meiosis-1} asymptotically
lead to the global, multifunctional phenotypes over many (essentially
infinite) generations as shown in the respective figures. 

Clearly the intricacies of nonlinear periodicity and chaos can lead
to startingly new possibilities beyond the reach of fixed-point reductionism.

\section{\label{sec: Punnett}Punnett Square and Economic Payoff: The Mixed
Nash Equilibrium }

Game Theory --- the formal study of conflict and collaboration ---
is designed to address situations in which the outcome of a person's
decision depends not just on how he individually operates but also
on the collective behaviour of the other interdependent members he
interacts with: this in essence constitutes \textit{competitive-collaboration}
of unity-in-diversity, perforce linear in the game theoretic setting.
The basic principle of biology and their economic correspondences
that mutations are more likely to succeed when they improve the reproducibility
of the concerned organism holds only as far as it goes --- genotype
fitness is not an individualistic attribute, it depends collectively
on all other (non-mutant) parts, with linear (quantum) non-locality
and non-linear (complex) holism being the inevitable consequences.
In the linear setting therefore, equilibrium concepts like Nash equilibrium,
Pareto optimality, tensor products should be relevant in the study
of Punnett squares --- a shorthand way of determining the probability
of having a certain type of offspring given the parents' genotypes
--- culminating in a global appreciation of the homeostatic holism
of which these are but linear manifestations. 

In this Section our aim is a formalization of the reductionist techniques
of linear non-locality that share common features with the checkerboard
methods of Punnett square, a summary of every possible combination
of one maternal allele with one paternal allele used by biologists
to determine the probability of an offspring having a particular genotype,
and the game-theoretic economic payoff (utility) matrix of a (normal-form)
game $\mathscr{G}=\left\langle P,\mathbf{S},\mathbf{u}\right\rangle $
of a finite set $N=\{n\}_{n=1}^{N}$ of $N$ diploid \textit{players
$P$,} $\mathbf{S}=\{S_{n}\}_{n=1}^{N}$ a set of \textit{pure strategies
$S_{n}=\{1,2,\cdots,k_{n}\}$ }available to player $n$, $\mathbf{s}=\{s_{n}\in S_{n}\}_{n=1}^{N}$
a pure strategy profile of the players, and $\mathbf{u}=\{u_{n}\!:\prod_{m=1}^{N}S_{m}\rightarrow\mathbb{R}\}_{n=1}^{N}$
the real-valued economic \textit{utility (payoff) functions }for the
players. 

Our starting point is the realization, following Figs. \ref{fig: Punnett-1}
and \ref{fig: Punnett-2}, that the Punnett square can also be expressed
in the language of economic utility/payoff matrices, in the global
backdrop of non-locality motivated by tensor products of Fig. \ref{fig: Punnett-2}
under these formal correspondences: 

\noindent \begin{center}
{\renewcommand{\arraystretch}{1.25}%
\begin{tabular}{|c|c|}
\hline 
Biology & Game Theory\tabularnewline
\hline 
\hline 
Female \female($\downarrow$) & Player 1\tabularnewline
\hline 
Male \male($\uparrow$) & Player 2\tabularnewline
\hline 
\multirow{2}{*}{{\renewcommand{\arraystretch}{1.00}$\begin{array}{c}
\mbox{Allele/}\\
\mbox{Gamete}
\end{array}$}$\begin{cases}
\mbox{Spin} & \!\!\!\!(\downarrow)\\
\mbox{Spin} & \!\!\!\!(\uparrow)
\end{cases}$} & Strategy 1\tabularnewline
\cline{2-2} 
 & Strategy 2\tabularnewline
\hline 
Genotype fitness & Utility/Payoff\tabularnewline
\hline 
\end{tabular}}
\par\end{center}

\noindent The utility matrix corresponding to Fig. \ref{fig: Punnett-1}
(ii) then becomes

\begin{center}
{\renewcommand{\arraystretch}{1.25}%
\begin{tabular}{|c|c||>{\centering}m{1.3cm}|>{\centering}m{1.25cm}|}
\hline 
\multicolumn{2}{|c||}{Player {\large \male }: Engine} & \multicolumn{2}{c|}{{\large \male{}} Allele Strategy}\tabularnewline
\cline{3-4} 
\multicolumn{2}{|c||}{Player {\large \female }: Pump} & $0\left(\downarrow\right)$ & $1\left(\uparrow\right)$\tabularnewline
\cline{1-2} 
\multicolumn{2}{|c||}{{\large{} \female{}} Allele Frequency} & $q$ & $(1-q)$\tabularnewline
\hline 
\hline 
$0\left(\downarrow\right)$  & $p$ & $\left\downarrow ,\right\downarrow $ & $\left\downarrow ,\right\uparrow $\tabularnewline
\hline 
$1\left(\uparrow\right)$  & $(1-p)$ & $\left\uparrow ,\right\downarrow $ & $\left\uparrow ,\right\uparrow $\tabularnewline
\hline 
\multicolumn{1}{c}{} & \multicolumn{1}{c|}{} & \multicolumn{2}{c|}{Offspring genotypes}\tabularnewline
\cline{3-4} 
\end{tabular}}
\par\end{center}

\noindent With all payoffs positive because of emergence and self-organization,
it is clear that in addition to any pure strategy Nash equilibria,
there is a mixed strategy equilibrium \sublabon{equation} 
\begin{equation}
(p,q)=\left(\frac{u_{22}^{(2)}-u_{21}^{(2)}}{(u_{11}^{(2)}+u_{22}^{(2)})-(u_{12}^{(2)}+u_{21}^{(2)})},\,\frac{u_{22}^{(1)}-u_{12}^{(1)}}{(u_{11}^{(1)}+u_{22}^{(1)})-(u_{12}^{(1)}+u_{21}^{(1)})}\right)\label{eq: MixedStrat}
\end{equation}
for a genotype utility matrix $\mathbf{U}=\{(u_{mn}^{(1)},u_{mn}^{(2)})\}_{m,n=1}^{2}$
where the allele frequency $p\in(0,1)$ is total number of $(\downarrow)$
alleles in the population as a fraction of $2N$ is the probability
of Player 1 choosing strategy $(\downarrow)$ (and $1-p$ of choosing
$(\uparrow)$) and $q\in(0,1)$, $1-q$, are the probabilities of
Player 2 doing the same. 

For symmetric games --- like Prisoner's Dilemma and Hawk-Dove, when
agents do not have distinct roles and the payoffs are independent
of their identities --- $(u_{mm}^{(1)}=u_{mm}^{(2)})_{m=1,2}$, $(u_{mn}^{(1),(2)}=u_{nm}^{(2),(1)})_{m\neq n}$,
the equilibrium allelic/gamete strategy reduces to $q=p$ with 
\begin{equation}
p=\frac{u_{22}^{(1)}-u_{12}^{(1)}}{(u_{11}^{(1)}+u_{22}^{(1)})-(u_{12}^{(1)}+u_{21}^{(1)})},\label{eq: SymStrat}
\end{equation}
for respective genotype payoffs $u_{11}^{(1)}q+u_{12}^{(1)}(1-q)$
of Player 1 (Pump), and $u_{11}^{(2)}p+u_{21}^{(2)}(1-p)$ for Player
2 (Engine). In an anti-coordination games without uncorrelated asymmetry
defined by $u_{11}^{(1)}=u_{11}^{(2)}=\mathcal{C}$, $u_{22}^{(1)}=u_{22}^{(2)}=\mathcal{T}$,
$u_{12}^{(1)}=u_{21}^{(2)}=\mathcal{W}$, $u_{12}^{(2)}=u_{21}^{(1)}=\mathcal{L}$,
the mixed strategy pair 

\begin{equation}
(p,q)=\left(\frac{\mathcal{W}-\mathcal{T}}{(\mathcal{W}-\mathcal{T})+(\mathcal{L}-\mathcal{C})},\frac{\mathcal{W}-\mathcal{T}}{(\mathcal{W}-\mathcal{T})+(\mathcal{L}-\mathcal{C})}\right)\label{eq: p=00003Dq strategy}
\end{equation}
\sublaboff{equation}is a Nash equilibrium, provided $p=q\in(0,1)$,
with an expected genotype fitness utility 
\[
\frac{\mathcal{WL}-\mathcal{TC}}{(\mathcal{W}-\mathcal{T})+(\mathcal{L}-\mathcal{C})}.
\]

Consider the Pump-Engine system as a two-player game between players
\female{} Capital and \male{} Culture each with allele strategies
$(\uparrow)$ derived from male parental Engine Culture and $(\downarrow)$
from female Pump Capital, Fig. \ref{fig: Punnett-1} (b). In this
hybrid, Capital-Culture discoordination $\mathbb{W}_{-}-\mathbb{W}_{+}$
game %
\footnote{\noindent In \textit{anti-coordination games,} $u_{11}^{(1)}<u_{21}^{(1)}$,
$u_{22}^{(1)}<u_{12}^{(1)}$, $u_{11}^{(2)}<u_{12}^{(2)}$ , $u_{22}^{(2)}<u_{21}^{(2)}$
, competing with different strategies for the rivalrous and non-excludable
shared resource is mutually beneficial for the players: sharing comes
at a cost of negative externality. In \textit{coordination games,
}$u_{11}^{(1)}>u_{21}^{(1)}$, $u_{22}^{(1)}>u_{12}^{(1)}$, $u_{11}^{(2)}>u_{12}^{(2)}$
, $u_{22}^{(2)}>u_{21}^{(2)}$ , collaborating the resource is beneficial
for all; the resource is non-rivalrous, sharing creating positive
externalities. A \textit{discoordination} game $u_{11}^{(1)}>u_{21}^{(1)}$,
$u_{22}^{(1)}>u_{12}^{(1)}$, $u_{11}^{(2)}<u_{12}^{(2)}$ , $u_{22}^{(2)}<u_{21}^{(2)}$
combines both for simultaneous competition and collaboration at the
linear reductionist level. %
} with the players conscious of their identities and strategies, the
non-symmetric Pump-Engine utility compared to the symmetric Hawk-Dove,
the indicated payoffs are interpreted as follows.
\begin{enumerate}
\item $(\left\downarrow ,\right\downarrow )$: $\female(\downarrow)$ dominates
$\male(\downarrow)$
\item $(\left\downarrow ,\right\uparrow )$: $\male(\uparrow)$ dominates
$\female(\downarrow)$
\item $(\left\uparrow ,\right\downarrow )$: $\male(\downarrow)$ dominates
$\female(\uparrow)$
\item $(\left\uparrow ,\right\uparrow )$: $\male(\uparrow)$ dominates
$\female(\uparrow)$
\end{enumerate}
\noindent The male authority of 2., 3., 4. reflects the eventual entropic
real world dissipation in $\mathbb{W}_{+}$ over {\small exergy} moderation
of 1. from $\mathbb{W}_{-}$. Without 1. however, there would be entropic
cold death exclusively: no life, only the structureless frozen global
symmetry of the cosmic inevitability second law. This implied inhibition
of entropic dissipation is the source of complexity, holism, and life. 

\noindent \begin{center}
{\renewcommand{\arraystretch}{1.25}%
\begin{tabular}{c}
\begin{tabular}{|c|c||c|c|}
\hline 
\multicolumn{2}{|c|}{\noun{Dove }{\large \male{}}} & \multicolumn{2}{c|}{{\large \male{}}}\tabularnewline
\cline{3-4} 
\multicolumn{2}{|c|}{\noun{Hawk }{\large \female{}}} & $(\downarrow)$ & $(\uparrow)$\tabularnewline
\hline 
\hline 
\multirow{2}{*}{{\large \female{}}} & $(\downarrow)$ & $\mathcal{C},\mathcal{C}$ & $\mathcal{W},\mathcal{L}$\tabularnewline
\cline{2-4} 
 & $(\uparrow)$ & $\mathcal{L},\mathcal{W}$ & $\mathcal{T},\mathcal{T}$\tabularnewline
\hline 
\multicolumn{4}{|c|}{$(\downarrow)$: Hawk, $(\uparrow)$: Dove}\tabularnewline
\hline 
\end{tabular}$\;\;$%
\begin{tabular}{|c|c||c|c|}
\hline 
\multicolumn{2}{|c||}{\noun{Prisoner}} & \multicolumn{2}{c|}{{\large \male{}}}\tabularnewline
\cline{3-4} 
\multicolumn{2}{|c||}{\noun{Dilemma}} & $(\downarrow)$ & $(\uparrow)$\tabularnewline
\hline 
\hline 
\multirow{2}{*}{{\large \female{}}} & $(\downarrow)$ & $\mathcal{L},\mathcal{L}$ & $\mathcal{W},\mathcal{C}$\tabularnewline
\cline{2-4} 
 & $(\uparrow)$ & $\mathcal{C},\mathcal{W}$ & $\mathcal{T},\mathcal{T}$\tabularnewline
\hline 
\multicolumn{4}{|c|}{$(\downarrow)$: Defect, $(\uparrow)$: Cooperate}\tabularnewline
\hline 
\end{tabular}\tabularnewline
\begin{tabular}{|c|c||c|c|}
\hline 
\multicolumn{2}{|c||}{\noun{Culture }{\large \male{}}} & \multicolumn{2}{c|}{{\large \male{}}}\tabularnewline
\cline{3-4} 
\multicolumn{2}{|c||}{\noun{Capital }{\large \female{}}} & $(\downarrow)$ & $(\uparrow)$\tabularnewline
\hline 
\hline 
\multirow{2}{*}{{\large \female{}}} & $(\downarrow)$ & $\mathcal{T},\mathcal{L}$ & $\mathcal{C},\mathcal{W}$\tabularnewline
\cline{2-4} 
 & $(\uparrow)$ & $\mathcal{C},\mathcal{W}$ & $\mathcal{L},\mathcal{T}$\tabularnewline
\hline 
\multicolumn{4}{|c|}{$(\downarrow)$: Capital, $(\uparrow)$: Culture}\tabularnewline
\end{tabular}\tabularnewline
\hline 
\multicolumn{1}{|c|}{$\mathcal{W>T>L>C}$.\textsc{$\quad$For PD: }Negative Payoff\textsc{;}
years in jail}\tabularnewline
\hline 
\end{tabular}}
\par\end{center}

It is significant that the payoffs for P-E is distinguished by the
absence of Pump resource $\mathcal{W}$ in the row entries and the
absence of Engine resource $\mathcal{C}$ in the column entry. This
curious feature of the P-E realization as observed earlier in \citep{Sengupta2010-c},
represents the essence of competitively collaborating unity-in-diversity:
the dispersion of $E$ is proportional to the domain $T-T_{c}$ of
$P$, and the concentration of $P$ depends on $T_{h}-T$ of $E$.
Thus an increase in $\iota$ can occur only at the expense of $P$
which opposes this tendency; reciprocally a decrease in $\iota$ is
resisted by $E$. The induced pump $P$ prevents the entire internal
resource $T_{h}-T_{c}$ from dispersion at $\iota=1$ by defining
for some $\iota<1$ a homeostatic temperature $T_{c}<T<T_{h}$, with
$E$ and $P$ interacting with each other in the spirit of competitive-collaboration
at the induced inheritance $T$. This inverse dependency of the two-phase
Pump-Engine system characterizes its collaborative-competitiveness
and directly contributes to the emergence of self-organization. 

Our variant of the Hawk-Dove game based on the reality that the Dove,
true to the Engine it models, is a scavenger that cannot be sustained
without supply of resource from the Hawk Pump --- which in turn with
its explosive concentration of $\mathbb{W}_{-}$ free-energy that
has no place in the entropic world of $\mathbb{W}_{+}$ except in
association with agents that can make use of it --- is not viable
in isolation of the Dove, has no pure Nash Equilibrium. The equilibrium
\[
(p,q)=\left(\frac{\mathcal{W}-\mathcal{T}}{2\mathcal{W}-(\mathcal{T}+\mathcal{L})},\,\frac{\mathcal{L}-\mathcal{C}}{\mathcal{T}+\mathcal{L}-2\mathcal{C}}\right)
\]
of probabilities from Eq. (\ref{eq: p=00003Dq strategy}) is a mixed
Nash equilibrium for the Capital-Culture game with payoffs 
\[
\left(\frac{\mathcal{TL}-\mathcal{C}^{2}}{\mathcal{T}+\mathcal{L}-2\mathcal{C}},\,\frac{\mathcal{W}^{2}-\mathcal{TL}}{2\mathcal{W}-(\mathcal{T}+\mathcal{L})}\right).
\]
for \female{} Female and \male{} Male respectively; note again that
while the pump $p$ is independent of $\mathcal{C}$ and the engine
$q$ of $\mathcal{W}$, the \textit{utility} of the Female bears the
expected inverse independence of $\mathcal{W}$ and the Male Engine
of $\mathcal{C}$. 

Generally for an even ordered $I\times I$ utility matrix with probabilities
$\{p_{i}:\sum p_{i}=1\}$ and $\{q_{j}:\sum q_{i}=1\}$ for the allele
strategy/gamete frequency of Female and Male respectively, the homozygous
sum of the diagonal products $\sum p_{i}q_{i}$ represents the irreversibility
$\iota$ responsible for ensuring entropic selection of fitness, while
the heterozygous $\sum_{i\neq j}p_{i}q_{j}$ of the off-diagonal terms
represents variation $(1-\iota)$ of dynamic evolution. This suggests
a correspondence of thermodynamic irreversibility with the diagonal
terms of the utility matrix, and of reversibility with the off-diagonal
terms. Quantum decoherence of the heterozygous component is a mechanism
by which open quantum systems of superpositions interact with their
environment to generate spontaneous suppression of interference and
appearance of the definiteness of classical objectivity. This irreversible
decay of the off-diagonal terms is the basis of decoherence that effectively
bypasses ``collapse'' of the state to one of its eigenstates on
performing a measurement. Non-local entanglement and interference,
are more pronounced and pervasive in nonlinear complexity than in
linear, isolated and closed, quantum systems. Complex homeostasy of
holism that cannot be represented by the classical objectivity of
wave-function collapse or of decoherence is a reminder of the limitations
of linear characterization of severely nonlinear, stressed, far-from-equilibrium,
systems. In this sense the mixed Nash equilibria --- which respects
both (linear) interaction and interference and is more non-local than
either decoherence or wave-function collapse of quantum systems can
admit --- is a more faithful portrayal of the objective reality of
complexity and holism. 

Heterozygosity of an organism, as considered in ChaNoXity, is a continuous
parameter determined by the homeostasy of $T$ which arises from the
mulifunctional graphical limit of $(\left\uparrow \right\uparrow )\oplus(\left\uparrow \right\downarrow )\oplus(\left\downarrow \right\uparrow )\oplus(\left\downarrow \right\downarrow )$.

\subsection{The Central Dogma: Limitations of Linear Reductionism}

DNA contains the codes for manufacturing various proteins. According
to the central dogma of molecular biology, the one-way flow of information
DNA~$\rightarrow$~RNA~$\rightarrow$~Protein is the basis of
all life on Earth: ``once information has passed into protein, it
cannot get out again'' \citep{Crick1970}, back to the nucleic acid.
The 3 major classes of biopolymers --- nucleic acids DNA, RNA and
protein --- allow 9 possible reductionist transfer of information
as shown in Table \ref{tab: Central-Dogma}: normal \textit{general
transfers} {\Large \smiley{}} that can occur in all cells, restricted
\textit{special transfers} {\Large \frownie{}} do not occur in most
cells but may occur in special circumstances as in virus-infected
cells and in the laboratory, and forbidden \textit{unknown} \textit{transfers}
\CrossOpenShadow{} \citep{Crick1970}. 

\begin{table}[!tbh]
\noindent \begin{centering}
{\renewcommand{\arraystretch}{1.5}%
\begin{tabular}{|c||c|c|c|}
\hline 
 & DNA & RNA & Protein\tabularnewline
\hline 
\hline 
DNA & {\Large \smiley{}} & {\Large \smiley{}} & {\Large \frownie{}}\tabularnewline
\hline 
RNA & {\Large \frownie{}} & {\Large \frownie{}} & {\Large \smiley{}}\tabularnewline
\hline 
Protein & \CrossOpenShadow{} & \CrossOpenShadow{} & \CrossOpenShadow{}\tabularnewline
\hline 
\end{tabular}}
\par\end{centering}

\caption{\label{tab: Central-Dogma}\textbf{\small The Central Dogma}{\small{}
of molecular biology states that mRNA is transcribed faithfully from
DNA and is translated faithfully into protein: $\mbox{DNA}\protect\overset{\textrm{replication}}{\longrightarrow}\mbox{DNA}\protect\overset{\textrm{transcription}}{\longrightarrow}\mbox{mRNA}\protect\overset{\textrm{translation}}{\longrightarrow}\mbox{Protein}$,
as deterministic copy-paste, one-to-one, faithful transfers. The zygote
self-replicates into two cells that divides into two and the process
of mitosis continues. A gene is a functional unit on a chromosome
which directs the synthesis of a particular protein. Humans have 23
pairs of chromosomes, each with two non-identical copies one derived
from each parent. }}
\end{table}

The above reductionist approach suffers the same individualistic limitations
considered earlier: the entropy increasing free-energy utilization
of genetic information through bidirectional feedback leading to the
forward DNA~$\rightarrow$~RNA~$\rightarrow$~Protein process
does not forbid, rather it actually seeks, the antagosistic dissipation-concentration
collaboration inside the female body. It is quite remarkable that
this replication-transcription-translation mechanism requires the
$\mathbb{W}_{-}$-environment of the uterus as the source and sustenance
of locally induced gravitational concentration of genetic information
stored in the DNA. According to this view all the components of Table
\ref{tab: Central-Dogma} participate collectively and competitively
in enabling the forward generation of proteins, the other arrows being
hidden from direct observation under normal circumstances. Without
them however, the information content of the DNA might not have been
there in the first place, the observed forward arrow being the consequence
of mutual antagonism of entropy and {\small exergy}. This bidirectional
information flow in complex biological systems --- rather than the
uni-directionality of classical dogma --- appear to support the recent
finding of widespread differences between DNA sequences and their
corresponding RNA transcripts in human cells \citep{Li2011} demonstrating
that these differences result in proteins that do not precisely match
the genes that encode them, and that mRNA and proteins --- not simply
the DNA --- might hold the key to understanding the genetic basis
of molecular biology. The ``mad cow disease'' (BSE) for example,
have been recorded to be transmitted even after the infectious media
was treated by means that normally destroy genetic material, DNA and
RNA. When the medium was treated by agents that only destroy proteins
and leave nucleic acids intact, the infection was however blocked.
This immediately indicates that BSE is actually transmitted by proteins.

Molecular geneticists studying the genetic material have over the
last few decades been turning up evidence that increasingly contradicts
the Central Dogma. There is an immense amount of necessary cross talk
between genes and the environment, that not only changes the function
of the genes but also the structure of the genes and genomes. Thus
Shapiro \citep{Shapiro2009} believes that the following lessons from
current molecular discoveries ``are likely to lead us to a significant
reformulation of our basic assumptions about the organization and
role of the genome in phenotypic expressions, heridity, and evolution''. 
\begin{itemize}
\item There is no unidirectional flow of information from one class of biological
molecule to another. Many types of molecules participate in information
transfer. In particular, genomic functions are inherently interactive
because isolated DNA is virtually inert and probably never exists
in that state in a cellular context. DNA cannot replicate or segregate
to daughter cells by itself. 
\item Classical atomistic-reductionist concepts are no longer tenable. Each
process involves multiple molecular components and one region of the
genome may be important for more than one process. Heredity thus has
to reflect the inherently systemic and distributed nature of genome
organization.
\item The post-central dogma discoveries relate to the importance of multivalent
and combinatorial techniques. The mobility and interaction of different
submolecular domains are increasingly apparent. It is of great biological
significance that multivalent operations provide the potential for
feedback, regulation, and robustness that simple mechanical devices
lack. 
\item Genomic change arises from natural genetic engineering, not from accidents.
Realization that DNA change is a biochemical process opens up new
ways of thinking about the role of natural genetic engineering in
normal life cycles and the potential for nonrandom processes in evolution. 
\item Informatic-entropic rather than mechanistic processes control cell-functions. 
\item Feedback signals play a central role in cell operations. The use of
signals is critical for basic functions like homeostatic regulation,
adaption to changing conditions, cellular differentiation, and multicellular
morphogenesis. Unpredictable signals in biological processes generates
an inescapable indeterminacy that contradicts the central dogma and
other reductionist statements of genetic determinism. 
\end{itemize}
The general correspondence of the above with the foundations of chanoxity
are all too evident to require further elaboration, all of which goes
against the basic tenets of central dogma of linear, mechanistic control.
Instead, ``layers upon layers of chaotic complexity are coordinated,
it seems, by mutual agreement, in an incredibly elaborate, exquisite
dance of life that dances itself freely and spontaneously into being''
\citep{Ho2004}; \citep{Doolittle2000,Isambert2009,Koonin2009,Koonin2009a,Purushotham2010}.
To reflect this new realism, our use of the terms ``variation''
includes sexual genetic recombination, gene flow, HGT, and ``selection''
admits random genetic drift available on demand in extreme circumstances:
while an organism's phenotype is obviously determined by its genotype
and the environment, the new dialectics requires higher forms of homeostasis
to be an expression of a win-win game between the two that itself
generates and sustains each other.

\subsection{Entropy-Exergy Antagonism --- Protein Folding}

These considerations should have some important bearing in the understanding
of protein folding in hydrophobic media. The hydrophobic effect is
fundamentally based on the tendency of polar water molecules to exclude
non-polar molecules leading to the segregation of water and non-polar
substances and apparent repulsion between water and hydrocarbons.
The hydrophobic effect is an important force providing the main impetus
for protein folding, formation of the lipid bilayer, insertion of
membrane proteins to the nonpolar lipid environment and protein small
molecule interactions. 

Depending on the polarity of the side chain, amino acids vary in their
hydrophilic or hydrophobic character. These properties are important
in protein structure and protein-protein interactions. The importance
of the physical properties of the side chains arises from its influence
on amino acid residue interactions with other structures, both within
and between proteins. The distribution of hydrophilic and hydrophobic
amino acids determines the tertiary structure of the protein, and
their physical location on the outside of the proteins influences
their quaternary structure. Hydrophilic and hydrophobic interactions
of the proteins need not rely only on the side chains of amino acids.
By various post-translational modifications other chains can be attached
to the proteins, forming hydrophobic lipoproteins or hydrophilic glycoproteins. 

\begin{wrapfigure}{o}{0.5\columnwidth}%
\begin{centering}
\includegraphics[scale=0.95]{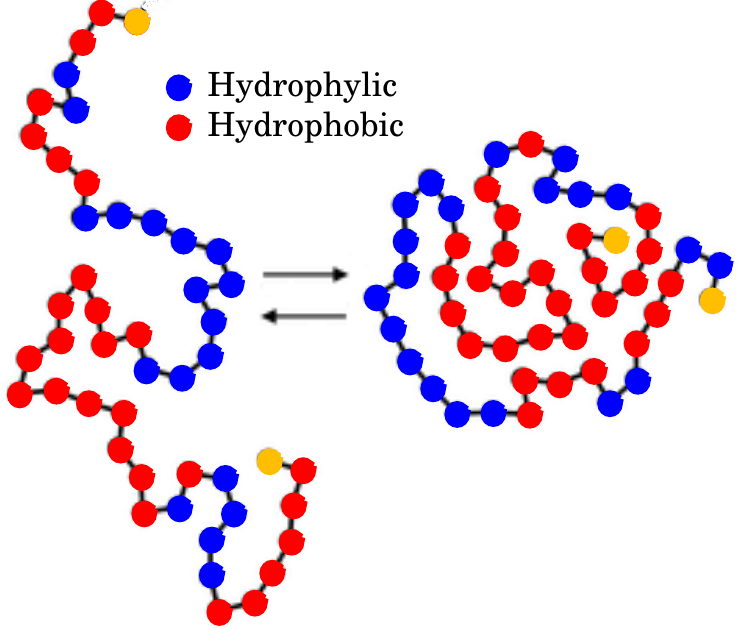}
\par\end{centering}

\caption{\textcolor{green}{\small \label{fig: protein-folding}}{\small Hydrophilic
entropic dispersion cohabitating with hydrophobic}\textcolor{red}{\small{}
}{\small free-energy concentration is the guiding principle of protein
folding. In this (entropy$\:\rightleftarrows\:$exergy) confrontation,
neither of the adversaries win, neither lose: no one dominates and
none is recessive. The genotype$\:\rightarrow\:$phenotype map is
typically non-injectively non-invertible --- a necessary condition
for emergence and self-organization. }}
\end{wrapfigure}%
Protein folding, Fig. \ref{fig: protein-folding}, is the physical
process by which a polypeptide folds into its characteristic and functional
three-dimensional structure from a random coil. Each protein exists
as an unfolded polypeptide random coil when translated from a sequence
of mRNA to a linear chain of amino acids, the amino acids interacting
with each other to produce the well-defined three-dimensional folded
protein. Folded proteins have a hydrophobic core in which side chain
packing stabilizes the folded state, and the hydrophilic charged or
polar side chains occupy the solvent-exposed surface where they interact
with surrounding water. Minimizing the number of hydrophobic side-chains
exposed to water is an important driving force behind the folding
process. For hydrophobic collapse the free-energy increases with corresponding
decrease in entropy in the interior, in the hydrophilic surface exterior
the opposite happens preventing thereby total collapse of the interior.
The folded protein represents homeostasy of the entropy and free-energy/exergy
adversaries. In this new form of dynamism, individuals retain their
identity in contributing to the collective that sustains them as much
as their nurture and support keeps the commune from disintegrating.
While this may not be the best option for either, without the collective
pluralism, neither would be better off. Recall that a misfolded protein
can be a serious liability rather than an asset; the correct three-dimensional
structure is essential. Failure to fold into the intended shape usually
produces inactive proteins with different undesirable properties including
toxic prions. Several neuro-degenerative diseases can result from
incorrect misfoldings, and many allergies may be caused if the immune
system fails to produce the required antibodies. 

A \textit{micelle} is an aggregate of surfactant molecules dispersed
in a liquid colloid. A typical micelle in aqueous solution forms an
aggregate with the hydrophilic head regions in contact with surrounding
solvent, separating the hydrophobic single tail regions in the micelle
centre. Inverse micelles have the headgroups at the centre with the
tails extending out. Micelles are approximately spherical in shape
This process of micellization is part of the phase behavior of many
lipids according to their polymorphism. Micelles form spontaneously
because of a balance between hydrophilic entropy and hydrophobic free-energy.
In water, the hydrophobic effect is the driving force for micelle
formation, despite the fact that assembling surfactant molecules together
reduces their entropy. At very low concentrations of the lipid, only
monomers are present in true solution. As the concentration of the
lipid is increased, a point is reached at which the unfavorable entropy
considerations, derived from the hydrophobic end of the molecule,
become dominant. At this point, the lipid hydrocarbon chains of a
portion of the lipids must be sequestered away from the water. Therefore,
the lipid starts to form micelles, thereby constituting a natural
mechanism of forming nanoparticles. Targeted drug delivery to tumors,
for example, piggybacking such natural vehicle is an attractive possibility;
however given the holistic nature of the human complex system, it
is not clear how much of this ``targeting'' can remain outside the
mutual influence of non-targeted organisms. This opens up the domain
of Darwinian or evolutionary medicine, the applications of evolutionary
theory to health and disease that provides a complementary approach
to the mechanistic explanations that dominate medical science, particularly
medical education \citep{Nesse2008}.

\section{\label{sec: Conclusion}Conclusion: Irreducible Complexity Without
Intelligent Design}

\noindent \begin{flushright}
\textsf{\textsl{\small \FiveStarOpenCircled{}}}\textsl{ A scientific
revolution happens when the paradigm (of normal science) breaks down.
In normal periods you need only people who are good at working with
the technical tools --- the master craftspeople. During revolutionary
periods you need seers, who can peer ahead into the darkness. $\cdots$
We are in a revolutionary period but are using the inadequate tools
of normal science. We are horribly stuck and need real seers, badly.
$\cdots$ Do you want a revolution in science? Let in a few revolutionaries.
The payoff could be discovering how the universe works.}\hfill{}\textsf{\textbf{\small Lee
Smolin}}\textit{,} The Trouble with Physics: The Rise of String Theory,
the Fall of a Science, and What comes Next\textit{,} Houghton Mifflin
Company, New York, (2006).
\par\end{flushright}

\noindent The new world-view of complex holism calls for a renovation
of the way science is done: ``We are'', undoubtedly, ``in a revolutionary
period but are using the inadequate tools of normal science''. The
new science has a distinctive mathematics of nonlinearity, multiplicity,
non-smoothness and equivalence classes leading to a new physics of
stand-off between selfish individualism and altruist collectivism,
and an interpretative philosophy where both the adversaries win and
both lose \citep{Sengupta2010-c}. Of course, ``this doesn't mean
that atomism or reductionism are wrong, but it means that they must
be understood in a more subtle and beautiful way than before''; indeed
``no one is saying the second law of thermodynamics is wrong, just
that there is a contrapuntal process organizing things at a higher
level''. Just as the advent of quantum mechanics did not signal the
demise of Newton, complexity and holism simply ventures beyond the
linear pathway of reductionism in exploring the theme that ``the
geometry of spacetime is a beautiful expression of the idea that the
properties of any one part of the world are determined by its relationships
and entanglement with the rest of the world''. 

Indeed, ``the real blow to the idea that the choice of which laws
govern nature is determined only by mechanisms acting at the smallest
scales came from the dramatic failure of string theory''. String
theory, ``a contender for the theory of everything (TOE), a manner
of describing the known fundamental forces and matter in a mathematically
complete system'', however ``has yet to make testable experimental
predictions, leading some to claim that it cannot be considered a
part of science'', seeks to be the supreme manifesto in reductionism
in its quest for the ultimate: Matter$\:\rightarrow\:$Molecule$\:\rightarrow\:$Atom$\:\rightarrow\:$Proton/Neutron/Electron$\:\rightarrow\:$Quark$\:\rightarrow\:$String.
The very strong peer sentiments expressed above, however, lends credence
to the possible roadmap of adventure beyond reductionism --- do strings
represent the genes of Nature? --- and ChaNoxity with its mathematically
self-contained and consistent view, can provide a global view of the
unexplored possibilities that lie beyond reductionism. 

This terrain beyond reductionism is synthetic rather that analytic
and is distinguished by bumps, blockades, multifunctions and jumps,
contrasted with the ubiquitous uniqueness, smoothness, and continuity
of Newton. Negative temperature and specific heat and decreasing symmetry
generators of structure, run counter to the established laws of increasing
symmetry and entropy that characterize the real world $\mathbb{W}_{+}$.
The collaborative manifestation of Female $\mathbb{W}_{-}$ on Male
$\mathbb{W}_{+}$ is through the entropy reducing effect of ``gravity'':
this is the source of Schroedinger's neg-entropy and the initiator
of the $\mathbb{W}_{-}-\mathbb{W}_{+}$ homeostasis of life. 

Although the New Science does not negate reductionism, its paradigmatic
shift in methodology calls for a fresh beginning to proceed beyond
its rich analytical legacy. This is easier said than done: the multifaceted
inertia can be expected to be not only academic but social, cultural,
economic and political. ``In the West, those who hold to a view of
a theistic God, including the Christian fundamentalists of such power
in the United States, find themselves in a cultural war with those
who do not believe in a transcendent God, whether agnostic or atheistic.
This war is evidenced by the fierce battle over Intelligent Design
being waged politically and in the court systems of the United States.
While the battleground is Darwinism, the deeply emotional issues are
more fundamental. These include the belief of many religious people
that without God's authority, morality has no basis. Literally, for
those in the West who hold to these views, part of the passion underlying
religious conviction is the fear that the very foundations of Western
society will tumble if faith in a transcendent God is not upheld''
\citep{Kauffman2007}. 

The attentive reader has probably not failed to observe that our perception
of complexity is in remarkable apparent consonance with that of Irreducible
Complexity of ``a single system composed of several well-matched
interacting parts that contribute to the basic function, wherein the
removal of any one of the parts causes the system to effectively cease
functioning'' \citep{Behe1996} leading to the discredited syllogism
``Whenever complex design exist, there must have been a designer;
nature is complex; therefore nature must have had an intelligent designer''
that ``everyone understands to be God'' \citep{Jones2005}. Our
explorations beyond reductionism, into the virgin territory of its
holistic neighbour, does not advance the inevitability of ``creation
science'' or ``scientific creationism'' that ``is simply not science''
because it depends on ``supernatural intervention''%
\footnote{Unless we interpret $\mathbb{W}_{-}$ to represent the elusive invisible
hand of God.%
}. The complex entities of Figs. \ref{fig: mitosis}, \ref{fig: ChaNoXity-a},
\ref{fig: meiosis-1} for example do not appear instantaneously, they
evolve in time. Once formed, however, they are indeed ``irreducible''
in the sense that a disturbance in any of the organs $\bullet-\circ-\bullet$
in Fig. \ref{fig: mitosis} will severely compromise the entire system.
As explicated in Looijen \citep{Looijen2000}, the whole is more than
the sum of the parts in the sense that the ``whole has emergent properties
which its component parts do not possess, neither separately nor when
simply added together nor in other partial combinations. It is only
in the specific combination in which the parts occur in the whole,
resulting from their specific mutual interactions, that the emergent
properties of the whole appear''. In fact the entropic, dissipative,
implosive world $\mathbb{W}_{+}$ is reductionist, the exergic, gravitationally
concentrative, explosive world $\mathbb{W}_{-}$ --- from its lack
of structures and patterns --- is holistic, and it is the complexity
of $\mathbb{W}_{-}-\mathbb{W}_{+}$ that represents a respectable
``handshake'' of the two antagonists: the reductioist $\mathbb{W}_{+}$
demands that holistic supply $\mathbb{W}_{-}$ rescues it from entropic
eventuality in return for its own survival from volcanic exergic concentration. 

The central issue, however, is likely to be the fundamental notion
of competitive-collaboration. That collectiveness is not a mere by-product
of selfish individualism as Adam Smith would have us believe %
\footnote{``The supposed omniscience and perfect efficacy of a free market
with hindsight looks more like propaganda against communism than plausible
science. In reality, markets are not efficient, humans tend to be
over-focused in the short-term and blind in the long-term, and errors
get amplified, ultimately leading to collective irrationality, panic
and crashes. Free markets are wild markets. Surprisingly, classical
economics has no framework to understand 'wild' markets. $\cdots$
The recent financial collapse was a systemic meltdown, in which interwined
breakdowns $\cdots$ conspired to destabilize the system as a whole.
We have had a massive failure of the dominant economic model'' \citep{Bouchaud2008}.%
} is really at the centre of the controversy: selfish individualism
and altruist collectivism share a common platform for mutual benefit
with neither superseding the other: if Male and Female failed to collaborate,
the family would be dysfunctional. Female Capital cannot survive without
the cushion of Male Culture which, in its turn, would starve to death
without the supply of essentials by the former. In this world of mutualism,
profit as the source of surplus economic energy does not in itself
represent stored exergy. The reciprocal feedback of collective culture
is indispensible for transformation of entropic profit/benefit to
exergic information. In the metaphor of Richard Dawkins \citep{Dawkins2006},
collective altruism of the \textit{Selfish Gene} against their own
self-interest leads to unselfish action by the organisms, much like
the win-win game of Fig. \ref{fig: ChaNoXity-a}, \textit{b. }The
gene-centric view of evolution holds that those genes whose phenotypic
effects successfully promote their own propagation will be favorably
selected in detriment to their competitors, thereby producing adaptations
for the benefit of genes and reproductive success of the organism.
However, genes not being directly visible to natural selection, the
unit of selection is the phenotype: mutational allelic differences
in genes generate phenotypes differences --- the raw material for
natural selection --- that indirectly acts upon the genes. The genes
are expressed in successive generations in proportion to the selective
value of their phenotypic effects, thereby completing the genotype$\;\rightleftarrows\;$phenotype
win-win contract. This antagonism of individualistic selfishness and
collective altruism are necessary components of this contract. 

What is Life? This remarkable query of Schroedinger of 1944 that \textquotedblleft must
surely rank among the most influential of scientific writings in this
century\textquotedblright , having led to epochal discoveries subsequently,
has been rather difficult to define: the ability to reproduce, often
considered a crucial ingredient, would imply for example that a mule
was never alive! Considering the non-dead to be alive, however, and
taking \textquotedblleft dead\textquotedblright{} as the second law
equilibrium state of maximum entropy, far-from-equilibrium complex
holism naturally constitutes \textquotedblleft life\textquotedblright{}
and hence the \textquotedblleft cohabitation of opposites\textquotedblright .
This is the essence of evolutionary existence in Nature, the apparent
defiance of the cosmic determinism of the second law implying an anti-second
law arrow that establishes a bottom-up pump competitively collaborating
with the second law top-down engine. The universality of this philosophy
leads, as we have endeavoured to demonstrate, in a natural way to
a comprehensive view of evolutions in Nature. 

This abstract and non-specific vision of life emphasises the emergent,
self-organizing character of evolving open systems. Availability of
exergy embodied in the $\mathbb{W}_{-}$-pump generated by a dissipating
$\mathbb{W}_{+}$-engine as a defense against its eventual second-law
entropic death comprises the sustaining immunity of life: the evolving
two-phase mixture of collective cooperating engine and individualistic
competing pump induces the top-down-bottom-up homeostasy of the ``living''. 

At the same time in accordance with Sec. \ref{sub: Fixed-Periodic},
we must not lose sight of the the significance of the boundary, the
``skin'' of $\mathbb{W}_{-}\mbox{-}\mathbb{W}_{+}$, of quantum
entanglement. This is a region of linear reductionist first order
representation of holism, of near-to-equilibrium phenomena. In this
domain of the EPR non-locality of Alice and Bob, non-holistic structures
like bacteria, algae and plants need not go through the complex of
chanoxity to exist: as demonstrated in Sec. \ref{sub: near-to-equilibrium}
a quantum non-local, Punnett square like representation is sufficient
for these lower structures. The wondrous manifesto of complex holism
establishes itself fully in humans with relatively higher complexity
$\chi$ (Fig. \ref{fig: 2-phase}) --- reflective of the rigours of
competitive-collaboration of the nonlinear qubit --- born the most
helpless struggling-to-survive infants. The organs must go through
a painstaking process of holistic self-organization in $\mathbb{W}_{+}$
for a long period before attaining functionality --- compare with
the lesser animals and plants of linear Punnett squares and pay-off
matrices distinguished by smaller values of $\chi$, where a new-born
is nearly biologically-ready for the intricate maneuvers of existence. 

\noindent \begin{flushright}
\textsf{\textsl{\small \FiveStarOpenCircled{}}}\textsl{ When you have
eliminated the impossible, whatever remains, however improbable, must
be the truth.}\\\hfill{} \textsf{\textbf{\small Sir Arthur Conan
Doyle}}
\par\end{flushright}{\small \par}

\noindent \textbf{Acknowledgement. }It is a pleasure to record my
exposure to ``enthalpy-entropy'' antagonism by Dr Keshab Gangopadhyay
of Nems/Mems Works LLC, Columbia USA at the Second World Congress
on Cancer (September 2010, Kerala India) which sowed the seeds of
the present Adventure. 

\bibliographystyle{plain}
\bibliography{osegu}

\end{document}